\documentclass[a4paper,fleqn,usenatbib,useAMS]{mnras}
\usepackage[T1]{fontenc}
\usepackage{ae,aecompl}

\usepackage{longtable}
\usepackage{rotating}
\usepackage{graphicx}	% Including figure files
\usepackage{amsmath}	% Advanced maths commands
\usepackage{amssymb}	% Extra maths symbols
\usepackage{multirow,bigdelim,array,pdflscape,booktabs,psfig}
\newcolumntype{C}[1]{>{\centering\arraybackslash}m{#1}}

\usepackage{float}
\usepackage{array}
\usepackage{makecell}
\usepackage{color,natbib}
\def\apjl{ApJ}%
\def\apjs{ApJS}%
\def\aap{A\&A}%          % Astronomy and Astrophysics
\usepackage[normalem]{ulem}

\newcommand{\kms}{$\rm{\,km \,s}^{-1}$}

\title[LeMMINGs: 1.5-GHz pc-scale radio structures of nearby galaxies]{LeMMINGs. I. The eMERLIN
  legacy survey of nearby galaxies. 1.5-GHz parsec-scale radio
  structures and cores}

\author[R. D. Baldi et al.]{R.~D.~Baldi$^{1}$\thanks{E-mail: r.baldi@soton.ac.uk},
D.~R.~A. Williams$^{1}$,
I.~M. McHardy$^{1}$,
R.~J. Beswick$^{2}$,
M.~K. Argo$^{2,3}$,
\newauthor B.~T. Dullo$^{4}$,
J.~H. Knapen$^{5,6}$,
E. Brinks$^{7}$,
T.~W.~B. Muxlow$^{2}$,
S. Aalto$^{8}$,  A. Alberdi$^{9}$,
\newauthor G.~J. Bendo$^{2,10}$, 
S. Corbel$^{11,12}$,
R. Evans$^{13}$,
D.~M. Fenech$^{14}$,
D.~A. Green$^{15}$,
\newauthor 
H.-R. Kl\"{o}ckner$^{16}$,
E. K\"{o}rding$^{17}$, 
P. Kharb$^{18}$,
T.~J. Maccarone$^{19}$,
I. Mart\'i-Vidal$^{8}$,
\newauthor C.~G. Mundell$^{20}$, 
F. Panessa$^{21}$, 
A.~B. Peck$^{22}$,
M.~A. P\'erez-Torres$^{9}$,
D.~J. Saikia$^{18}$,
\newauthor P. Saikia$^{17,23}$, F. Shankar$^{1}$, R.~E. Spencer$^{2}$, 
I.~R. Stevens$^{24}$, P. Uttley$^{25}$ and
J. Westcott$^{7}$\\
$^{1}$ School of Physics and Astronomy, University of Southampton, Southampton, SO17 1BJ, UK\\
$^{2}$ Jodrell Bank Centre for Astrophysics, School of Physics and Astronomy, The University of Manchester, Manchester, M13 9PL, UK\\
$^{3}$ Jeremiah Horrocks Institute, University of Central Lancashire, Preston PR1 2HE, UK\\
$^{4}$ Departamento de Astrofisica y Ciencias de la Atmosfera, Universidad Complutense de Madrid, E-28040 Madrid, Spain\\
$^{5}$ Instituto de Astrofisica de Canarias, Via Lactea S/N, E-38205, La Laguna, Tenerife, Spain\\
$^{6}$ Departamento de Astrofisica, Universidad de La Laguna, E-38206, La Laguna, Tenerife, Spain\\
$^{7}$ Centre for Astrophysics Research, University of Hertfordshire, College Lane, Hatfield, AL10 9AB, UK\\
$^{8}$ Department of Space, Earth and Environment, Chalmers University of Technology, Onsala Space Observatory, 43992 Onsala, Sweden\\
$^{9}$ Instituto de Astrofisica de Andaluc\'ia (IAA, CSIC); Glorieta de la Astronom\'ia s/n, 18008-Granada, Spain\\
$^{10}$ ALMA Regional Centre Node, UK \\
$^{11}$ Laboratoire AIM (CEA/IRFU - CNRS/INSU - Universit\'e Paris Diderot), CEA DSM/IRFU/SAp, F-91191 Gif-sur-Yvette, France\\
$^{12}$ Station de Radioastronomie de Nan\c{c}ay, Observatoire de Paris, PSL Research University, CNRS, Univ. Orl\'{e}ans, 18330 Nan\c{c}ay, France\\
$^{13}$ School of Physics and Astronomy, Cardiff University, Cardiff, CF243AA, UK \\
$^{14}$ Department of Physics \& Astronomy, University College London, Gower Street, London WC1E 6BT, UK\\
$^{15}$ Astrophysics Group, Cavendish Laboratory, 19 J.~J.~Thomson Avenue, Cambridge CB3 0HE, UK\\
$^{16}$ Max-Planck-Institut f\''{u}r Radioastronomie, Auf dem H\''{u}gel 69, 53121 Bonn, Germany \\
$^{17}$ Department of Astrophysics/IMAPP Radboud University, Nijmegen, The Netherlands\\
$^{18}$ National Centre for Radio Astrophysics - Tata Institute of Fundamental Research, Postbag 3, Ganeshkhind, Pune 411007, India\\
$^{19}$ Department of Physics, Box 41051, Science Building, Texas Tech University, Lubbock, TX 79409-1051, US\\
$^{20}$ Department of Physics, University of Bath, Claverton Down, Bath, BA2 7AY, UK\\
$^{21}$ INAF -- IAPS Rome, Via Fosso del Cavaliere 100, I-00133 Roma, Italy\\
$^{22}$ Gemini North Operations Center, 670 N. A'ohoku Pl, Hilo, HI 96720  USA \\
$^{23}$ New York University Abu Dhabi, PO Box 129188, Abu Dhabi, UAE\\
$^{24}$ School of Physics and Astronomy, University of Birmingham, Edgbaston, Birmingham B15 2TT, UK\\
$^{25}$ Anton Pannekoek Institute, University of Amsterdam, Science Park 904, 1098 XH Amsterdam, The Netherlands}

\pubyear{2018}

\begin{document}
\label{firstpage}
\pagerange{\pageref{firstpage}--\pageref{lastpage}}
\maketitle

\begin{abstract}
 %250 words limit, now 253 

  We present the first data release of high-resolution ($\leq0.2$
  arcsec) 1.5-GHz radio images of 103 nearby galaxies from the Palomar
  sample, observed with the eMERLIN array, as part of the LeMMINGs
  survey.  This sample includes galaxies which are active (LINER and
  Seyfert) and quiescent (H{\sc ii} galaxies and Absorption line
  galaxies, ALG), which are reclassified based upon revised
  emission-line diagrams. We detect radio emission $\gtrsim$ 0.2 mJy
  for 47/103 galaxies (22/34 for LINERS, 4/4 for Seyferts, 16/51 for
  H{\sc ii} galaxies and 5/14 for ALGs) with radio sizes typically of
  $\lesssim$100 pc. We identify the radio core position within the radio
  structures for 41 sources. Half of the sample shows jetted
  morphologies. The remaining half shows single radio cores or
  complex morphologies. LINERs show radio structures more
  core-brightened than Seyferts.  Radio luminosities of the sample
  range from 10$^{32}$ to 10$^{40}$ erg s$^{-1}$: LINERs and H{\sc ii}
  galaxies show the highest and the lowest radio powers respectively,
  while ALGs and Seyferts have intermediate luminosities.

  We find that radio core luminosities correlate with black hole (BH)
  mass down to $\sim$10$^{7}$ M$_{\odot}$, but a break emerges at
  lower masses. Using [O~III] line luminosity as a proxy for the
  accretion luminosity, active nuclei and jetted H{\sc ii} galaxies
  follow an optical fundamental plane of BH activity, suggesting a
  common disc--jet relationship. In conclusion, LINER nuclei are the
  scaled-down version of FR~I radio galaxies; Seyferts show less
  collimated jets; H{\sc ii} galaxies may host weak active BHs and/or
  nuclear star-forming cores; and recurrent BH activity may account
  for ALG properties.

  %Although jetted galaxies
  %have black hole (BH) masses larger than $\sim$10$^{6}$ M$_{\odot}$,
  %radio core luminosities correlate with BH masses down to
  %$\sim$10$^{7}$ M$_{\odot}$, but the relationship is less clear at
  %lower masses.
 
  %Using the [O~III] luminosity as an indicator of the AGN bolometric
  %luminosity, we find: i) LINERs are the scaled-down versions of FR~I
  %radio galaxies, implying a similar central engine; ii) Seyferts show
  %less collimated jets than LINERs, powered by high-accretion-rate
  %BHs; iii) H{\sc ii} galaxies and ALGs are a mixed population of galaxies
  %which either host weakly active or silent BHs, or powered by SF. All
  %jetted AGN follow the optical fundamental plane of BH activity,
  %suggesting a scale invariance between accretion and the jet
  %launching mechanism.

\end{abstract}

\begin{keywords}
  galaxies: active -- galaxies: jet -- galaxies: nuclei -- galaxies:
  star formation -- radio continuum: galaxies
\end{keywords}

%%%%%%%%%%%%%%%%%%%%%%%%%%%%%%%%%%%%%%%%%%%%%%%%%%

%%%%%%%%%%%%%%%%% BODY OF PAPER %%%%%%%%%%%%%%%%%%

%\usepackage{newtxtext}                  % Good fonts
%\usepackage[slantedGreek]{newtxmath}    %   "    "   (slanted Greek)
\section{Introduction}

The advent of large-area, high-sensitivity, sub-arcsecond resolution
surveys enables the investigation of large samples of extragalactic
radio sources in order to understand the interplay and dominance of
the two astrophysical processes which dominate the emission from our
Universe: star-formation (SF) and accretion, i.e. the growth of
galaxies and of super-massive black holes (SMBH). SF is fundamental
for the formation and evolution of galaxies whilst accretion, the
phenomenon responsible for powering Active Galactic Nuclei (AGN),
provides a major power source in the distant Universe, dominating the
emission across most wavelengths in the form of QSOs (Quasi Stellar
Objects) and blazars. Conversely, in the local Universe,
low-luminosity AGN (LLAGN, defined by \citealt{ho97a} as AGN with
H$\alpha$ luminosity $\lesssim$ 10$^{40}$ erg s$^{-1}$) and SF similarly
contribute to illuminating the galaxies and the surrounding media.

Observational and theoretical studies support the co-evolution of the
SMBH and host galaxy properties (e.g., the BH mass $-$ stellar
velocity dispersion, or $M_{\rm BH}-\sigma$ relationship,
\citealt{ferrarese00,tremaine02,dimatteo05,dimatteo08,shankar16}) and
also the connection between SF and nuclear activity in LLAGN
(e.g. \citealt{kauffmann07,lamassa13}), where the faint radiation from
the AGN allows the detection and the study of SF.

LLAGN generally differ from luminous AGN because they are powered by
either low accretion rate SMBHs and/or radiatively inefficient
accreting SMBHs (e.g. \citealt{ho99,maoz07,panessa07}), resulting in
undetectably low optical to X-ray luminosities. LLAGN are important
because: i) they numerically dominate the local Universe, outnumbering
QSOs, and representing the most common mode of SMBH accretion; ii)
they are similar to quiescent galaxies, where the central SMBH is not
necessarily active, allowing the study of nuclear triggering
mechanisms; iii) the nuclear emission does not swamp the emission of
the host galaxy, allowing for the discrimination of the accretion
process from SF; iv) they are in a regime of low accretion rate, an
optimal condition to investigate the transition of the radiative
efficiency of accretion discs; v) they can be associated with lower BH
masses, crucial for constraining the scaling relations and the
cosmological evolutionary models for SMBHs and galaxies
\citep{barausse17}; vi) they sample the low end of the luminosity
function, important for a complete census of local SMBH activity.

The best way to study nuclear accretion and to disentangle it from the
emission associated with star-forming regions is by observing nearby
galaxies. The proximity enables high linear resolution and high
sensitivity observations, allowing detections of lower luminosity
sources, comparable to Sgr~A*. Therefore, the local Universe provides
a unique opportunity to explore LLAGN and their interplay with their
host galaxies at the low end of galaxy and AGN luminosity functions.

%The understanding of LLAGN would benefit from a multi-band approach,
%since observations at different frequencies give additional
%constraints on the origin of the emission, either from SF or AGN. This
%is the case, when cross-matching large surveys in radio, infrared,
%optical, and X-ray bands\footnote{For example, Faint Images of the
%  Radio Sky at Twenty centimetres survey, FIRST; \citep{becker95}, the
%  Wide-field Infrared Survey Explorer, WISE; \citep{wright10}, the
%  Sloan Digital Sky Survey, SDSS; \citep{stoughton02}, and the
%  XMM-Newton serendipitous survey, 3XMM; \citep{rosen16}.}: an
%enormous quantity of information about the local Universe can be
%exploited for investigating the accretion properties of LLAGN in
%comparison with more luminous AGN (e.g.,
%\citealt{best05a,baldi10b,asmus15,mingo16}).

Although all bands are useful to fully constrain the origin of the
emission from LLAGN, radio observations can offer the best single
diagnostic of SF and nuclear accretion. Indeed, SF and AGN emit in the
cm--mm band, where the two processes can be detected at low luminosity
regime, since the galaxy contributes negligibly. Furthermore,
long-baseline radio arrays provide high-angular resolution and high
sensitivity, which are essential to detect and isolate the
low-brightness nuclear emission with respect to the dominant galaxy
emission. X-ray observations can also probe BH accretion at the galaxy
centre. However, due to the cost and resolution limitation of the
X-ray satellites and the great range of earth-based sensitive radio
telescopes available to astronomers (e.g, VLA, VLBI, LOFAR, ALMA,
eMERLIN), deep radio interferometric observations of local galaxies
are ideal for probing small spatial scales of often $\lesssim$500 pc
in the innermost central regions and identifying the emission from
SMBHs.

High-resolution measurements of radio emission, flux density and its
variability, compactness of the source, radio spectral index,
brightness temperatures and polarisation are unique diagnostic tools
for determining the nature of the radio emission. Synchrotron
radiation generated in magnetic fields and free--free emission are
typically the primary physical origin of the radio emission in local
LLAGN. In addition, LLAGN are more efficient than QSOs at launching
radio jets, whose synchrotron emission in some cases dominates the
total output \citep{koerding08}. The detection of jets or high
brightness temperature radio cores represent a valid method of the
identification of an active SMBH at the galaxy centre, in the absence
of other AGN signatures such as high-ionization emission lines or an
X-ray nucleus.

In the last half century, several interferometric radio surveys at 1.4
and 5 GHz with a resolution of $\sim$1 arcsec, have been conducted on
local LLAGN
(e.g. \citealt{ekers73,wrobel91a,sadler84,sadler89,slee94,ulvestad89,vila90,sadler95,ulvestad01a,gallimore06})
revealing a large fraction of radio cores, not only in elliptical but
also in bulge-dominated spiral galaxies. LLAGN typically have compact
flat-radio-spectrum cores, interpreted as the synchrotron
self-absorbed base of the jet which fuels larger-scale radio
emission. Flat-spectrum radio sources can also result from thermal
emission due to ionized gas in H{\sc ii} regions or from free--free
non-thermal emission, occurring in compact nuclear star-bursts
\citep{condon91} with low-brightness temperatures ($<$10$^{5}$ K). In
addition, LLAGN exhibit a large variety of radio morphologies: single
core, core--jet, Fanaroff--Riley morphologies (FR,
\citealt{fanaroff74}), double lobes and complex structures. The
presence of composite radio structures may indicate the contribution
of SF in the observed radio emission.

%Since LLAGN tend to be radio louder than luminous AGN, it is still
%historically debating weather the distribution of radio loudness of
%active nuclei is dichotomous, bimodal or continuous
%(e.g. \citealt{white00,ho01c,laor03,cirasuolo03,terashima03,sikora07}). Whether
%a radio-loud AGN which shows collimated relativistic jets from parsec
%to kpc-Mpc scales is intrinsically different from radio-quiet AGN,
%weak and confined to kpc scales, is one of the big questions of modern
%astrophysics. The relatively few available high-resolution and deep
%studies of radio-quiet and loud LLAGN lead to mixed results
%(e.g. \citealt{nagar02,anderson05,wrobel06,giroletti09,panessa13}).

Unambiguously determining the physical nature of the radio emission in
nearby galaxies is more than just of mere phenomenological
interest. Understanding how radio core emission is associated with
accretion-powered nuclei was the goal of several observational studies
(e.g.,
\citealt{ho99b,hardcastle00,giroletti09,capetti06,nagar05}). Theoretical
papers and numerical simulations which explored the accretion
disc--jet relationship in the case of galactic nuclei
(e.g. \citealt{begelman84,
  lovelace96,falcke99,beckwith08,massaglia16}), suggested that LLAGN
have nuclear and jet properties, similar to powerful AGN
\citep{ulvestad01b}. Compact nuclear radio emission and jets are
expected from the accretion discs in the form of advection-dominated
accretion flows (ADAF, \citealt{narayan98}) at low accretion rates
($<$10$^{-3}$ of the Eddington rate). At higher rates, accretion takes
the form of a geometrically thin, optically thick disc (standard disc,
\citealt{shakura73}) with a hot, X-ray emitting corona, typically
associated with luminous Seyferts. Many radio studies have reported
jet structures associated with Seyferts (e.g.,
\citealt{ulvestad84,kukula95,kukula99,thean00,orienti10,kharb17}). However,
there are indications that by either increasing the accretion rate or
the radiative efficiency of a standard disc, the fraction of radio
jets observed in sources optically classified as Seyferts decreases
from low-power Seyferts to QSOs
\citep{ho01a,ulvestad01a,blundell01,blundell03,heywood07}. This radio
behaviour of Seyferts generally suggests that jet production in
radiatively efficient discs is still viable, but it is less efficient
than in ADAF discs
\citep{kukula99,nagar99,giroletti09,king11}. However, since jets are
universally observed across all types of AGN, it is still under debate
whether the accretion-jet symbiosis is scale invariant, regardless of
the accretion mode, e.g., in the so-called `fundamental plane of BH
activity' (FPBHA)
\citep{merloni03,falcke04,kording06b,panessa07,plotkin12,bonchi13}.
This plane is believed to unify all active BHs at different mass
scales (from galactic BHs to AGN), indicating similar physical
mechanisms in accretion and jet production from accreting compact
objects.

A systematic, deep radio study of a large and complete sample of LLAGN
at milli-arcsecond resolution and $\mu$Jy beam$^{-1}$ sensitivity
promises to provide insights on the radio properties of LLAGN. While
previous studies focused on only known active galaxies, a
comprehensive radio study of inactive galaxies is necessary to
investigate low-level SMBH accretion activities, akin to that of
Sgr~A*. With these intentions, a new radio survey has been designed,
based on the capabilities of the {\it enhanced Multi-Element Radio
  Linked Interferometer Network} (eMERLIN). The project is named the
{\bf L}egacy {\bf eM}ERLIN {\bf M}ulti-band {\bf I}maging of {\bf
  N}earby {\bf G}alaxies {\bf s}urvey (LeMMINGs, PI: McHardy and
Beswick; \citealt{beswick14}).  The target of this survey is the
Palomar nearby galaxy sample \citep{filippenko85}, optically selected,
with no radio constraint bias.  In this paper we present results from
an interim sample of 103 galaxies, essentially a random sub-set from
the Palomar sample which demonstrates the data and imaging quality and
shows the progress of the Legacy Programme. This sample includes LLAGN
and quiescent galaxies with a broad distribution of nuclear and host
properties. In addition, excellent ancillary data that cover wide
wavelength ranges, i.e., from infrared to X-ray are available for the
LeMMINGs sample of galaxies. Such a multi-band survey will unveil the
nature of the central engines in LLAGN.

This paper is organised as follows: in Section~\ref{sample} we present
the LeMMINGs project and sample; the observations and calibration of
the radio data are explained in Section~\ref{observations}. The radio
properties of the sample are presented in Section~\ref{radioresults}
and the connection between the radio and optical properties to study
the SMBH activity are examined in Section~\ref{optical-radio}. We
discuss the results for the different optical classes in
Sections~\ref{discussion} and we provide an overview of the
accretion--radio connection for LLAGN and draw conclusions in
Section~\ref{overview}. Appendix \ref{app} shows the radio images
of the radio detected galaxies and presents several tables listing
their properties.

\vspace{-0.2cm}
\section{The sample and the legacy}
\label{sample}

The sample investigated by the LeMMINGs survey is a subset of the
Palomar sample, which itself includes nearby galaxies with $\delta >
0^{\circ}$ and $B_{T} \leq$12.5 mag from the Revised Shapley--Ames
Catalog of Bright Galaxies (RSA; \citealt{sandage81}) and the Second
Reference Catalogue of Bright Galaxies (RC2;
\citealt{devaucouleurs76}). The Palomar sample was observed by
\citet{ho95} with the Hale 5m telescope at the Palomar Observatory
\citep{filippenko85} to carry out an intense spectroscopic campaign.

The Palomar galaxies have been optically classified, based on the
ratios of several optical narrow lines (i.e., H$\beta$, [O~III],
H$\alpha$, [N~II]) extracted from a long-slit spectra on a nuclear
region of $2\times4$ arcsec$^{2}$. Several optical classes are
identified based on position in emission-line diagnostic diagrams
similar to the methodology of \citet{veilleux87}. \citet{ho97a}
distinguish H{\sc ii}, Seyfert, LINER, and Transition galaxies among
the objects which show prominent emission lines.

H{\sc ii} galaxies ($\sim$42 per cent of the Palomar sample) are
considered star-forming galaxies where the emission lines are ascribed
to young massive stars. Seyferts and LINERs ($\sim$11 and 19 per
cent of the Palomar sample, respectively) correspond to two active
types of the central SMBHs. The Transition galaxies ($\sim$ 14 per
cent) are sources with composite characteristics of LINER nuclei
contaminated by H{\sc ii} emission regions (e.g. \citealt{ho93}).  The
Transition class is not commonly used since its selection is more
restrictive than the other optical classes (see
Section~\ref{optclass}). Therefore we prefer to remove this class of
objects and convert the Transition galaxies into LINERs or H{\sc ii}
galaxies based on different line diagnostics. The rest of the sample
($\sim$14 per cent of the Palomar sample) shows no significant optical
emission lines and are typically in elliptical and lenticular
galaxies. These sources are named Absorption-line galaxies (ALG). The
lack of an emitting-line nucleus can be explained either by the
inactivity of the SMBH or the non-detection of a very weak active
SMBH, whose emission lines are dominated by an old stellar population
(similar to low-power radio galaxies, \citealt{buttiglione09}), or
diluted by a non-thermal power-law continuum (similar to BL~Lacs,
\citealt{scarpa97,capetti10}), or dimmed by extinction through the host
galaxy \citep{goulding09}. However, as already suggested by
\citet{ho03} and pointed out by several authors
(e.g. \citealt{best05a,baldi10b}), the spectro-photometric properties
of a large fraction of ALGs can be reconciled with the picture of
LINERs.

The active galaxies in the Palomar sample have also been
observed at low and high resolution (arcsec and mas) at different
cm-band frequencies (VLA,
\citealt{nagar00,nagar01,nagar02,filho00,nagar05,ho01a}; VLBA/VLBI,
\citealt{falcke00}, Ulvestad \&
\onecolumn
\begin{center}
%\begin{longtable}{lC{1.85cm}C{2cm}C{0.3cm}ccC{1.8cm}C{1.7cm}C{0.8cm}cC{0.5cm}}
\begin{longtable}{lccC{0.4cm}C{0.35cm}cC{0.4cm}ccC{0.35cm}C{0.35cm}C{0.2cm}}
\caption[Properties of the  sample.]{Radio properties of the sample.} 
\label{radioprop} \\

%This is the header for the first page of the table...
\hline 
\hline
Name & RA & DEC & D    &LEM & phase cal &  Q & Beam & PA & rms & det & morph \\
 (1) & (2)& (3) & (4)  & (5) & (6)      & (7)& (8) & (9) & (10)& (11) & (12) \\     
\hline	
\endfirsthead

%This is the header for the remaining page(s) of the table...
\multicolumn{3}{c}{{\tablename} \thetable{} -- Continued} \\[0.5ex]
\hline
\hline
Name & RA & DEC & D    &LEM & phase cal &  Q & Beam & PA & rms & det & morph \\
(1) & (2)& (3) & (4)  & (5) & (6)      & (7)& (8) & (9) & (10)& (11) & (12) \\     
\hline 

\endhead

%This is the footer for all pages except the last page of the table...
\hline
  \multicolumn{12}{c}{{Continued on Next Page}} \\
\endfoot

%This is the footer for the last page of the table...
 \\[-1.8ex] 
\endlastfoot

%Now the data...
NGC~7817  & 00 03 58.899 &  +20 45 08.42 & 31.5 &04 &  0004+2019 & ++&  0.21$\times$0.13 & 33.4  & 82  &  U &  \\
IC~10 	  & 00 20 23.16  &   +59 17 34.7 & 1.3  &04 &  0027+5958 & ++&  0.21$\times$0.11 & 88.2  &  89 & U  &  \\
NGC~147   & 00 33 12.120 &  +48 30 31.46 & 0.7  &04 &  0039+4900 & ++& 0.24$\times$0.14 & -63.2  &  73 & I    &  A \\
NGC~185   & 00 38 57.970 &  +48 20 14.56 & 0.7  &04 &  0039+4900 & ++& 0.17$\times$0.14 & -84.3  &   82 & U &  \\
NGC~205   & 00 40 22.075 &  +41 41 07.08 & 0.7  &04 &  0038+4137 & ++& 0.20$\times$0.13 &  79.8  & 86 &  U &  \\
NGC~221   & 00 42 41.825 &  +40 51 54.61 & 0.7  &04 &  0038+4137 & ++&  0.36$\times$0.15  & -34.7   & 82 & U &  \\
NGC~224   & 00 42 44.330 &  +41 16 07.50 & 0.7  &04 &  0038+413  & ++& 0.19$\times$0.14 & -71.4  & 79 & U &  \\
NGC~266   & 00 49 47.793 &  +32 16 39.81 & 62.4 &03 &  0048+3157 & +& 0.48$\times$0.31 &  -40.0 & 95  & I   & A  \\
NGC~278   & 00 52 04.300 &  +47 33 01.93 & 11.8 &04 &  0039+4900 & ++& 0.29$\times$0.14 &  -55.2 & 50 & I    & A \\
NGC~315   & 00 57 48.883 &  +30 21 08.81 & 65.8 &03 &  0048+3157 & +& 0.12$\times$0.12 &  45.0 &  5500   &  I   & A \\
NGC~404   & 01 09 27.021 &  +35 43 05.27 & 2.4  &03 &  0112+3522 & +& 0.43$\times$0.26 & -46.8 &  61 & U &   \\
NGC~410   & 01 10 58.872 &  +33 09 07.30 & 70.6 &03 &  0112+3522 & +&  0.52$\times$0.32 & -46.85 & 140 & I   & A  \\
NGC~507   & 01 23 39.950 &  +33 15 22.22 & 65.7 &03 &  0112+3522 & +& 0.46$\times$0.35 &  -67.8 & 180 & I    & A \\
NGC~598   & 01 33 50.904 &  +30 39 35.79 & 0.7  &03 &  0137+3122 & +& 0.32$\times$0.13  &  12.1 &  103 & U & \\
IC~1727   & 01 47 29.890 &  +27 20 00.06 & 8.2  &03 &  0151+2744 & +& 0.48$\times$0.17 & -8.8  &  98 & U & \\
NGC~672   & 01 47 54.523 &  +27 25 58.00 & 7.5  &03 &  0151+2744 & +& 0.50$\times$0.23  & -16.3 &  118 & U & \\
NGC~697   & 01 51 17.568 &  +22 21 28.69 & 41.6 &03 &  0152+2207 & +& 0.43$\times$0.13  &  13.7 &  95 &  U & \\
NGC~777   & 02 00 14.90  &  +31 25 46.00 & 66.5 &02 &  0205+3219 & m& 0.23$\times$0.15 &  7.6  &  77 & I    & A \\
NGC~784   & 02 01 16.932 &  +28 50 14.13 & 4.7  &03 &  0151+2744 & +& 0.48$\times$0.22 &  -29.3 &  131 & U & \\
NGC~812   & 02 06 51.497 &  +44 34 22.48 & 108.8&02 &  0204+4403 & m& 0.21$\times$0.16 &  -76.5 & 79 & U & \\
NGC~818   & 02 08 44.510 &  +38 46 38.11 & 59.4 &02 &  0204+3649 & m& 0.21$\times$0.17 &   54.9 & 72 & U  & \\
NGC~841   & 02 11 17.334 &  +37 29 49.96 & 59.5 &02 &  0204+3649 & m& 0.21$\times$0.18 &   50.3 & 81 &  U & \\
NGC~890   & 02 22 01.008 &  +33 15 57.94 & 53.4 &02 &  0226+3421 & m& 0.22$\times$0.18 &   42.1 & 75 &  U & \\
NGC~891   & 02 22 32.907 &  +42 20 53.95 & 9.6  &02 &  0222+4302 & m& 0.20$\times$0.17 & 65.6   & 64 &  unI & \\
NGC~925   & 02 27 16.913 &  +33 34 43.97 & 9.4  &02 &  0226+3421 & m& 0.22$\times$0.18 &   42.7 & 75 & U & \\
NGC~959   & 02 32 23.945 &  +35 29 40.80 & 10.1 &02 &  0226+3421 & m& 0.20$\times$0.15  &  43.4  &  81 & U & \\
NGC~972   & 02 34 13.34  &  +29:18:40.57 & 21.4 &02 &  0237+2848 & m& 0.24$\times$0.17 & 37.2  & 69 & I   & B \\
NGC~2273  & 06 50 08.663 &  +60 50 44.50 & 28.4 &27 &  0707+6110 & ++& 0.20$\times$0.16 &  48.1 & 930 & I   & E \\
NGC~2342  & 07 09 18.089 &  +20 38 09.22 & 69.5 &27 &  0700+1709 & ++& 0.18$\times$0.18 &    0.0 & 65 & I  & E \\
NGC~2268  & 07 14 17.441 &  +84 22 56.18 & 34.4 &25 &  0702+8549 & +& 0.20$\times$0.20 &   0.0 &  91 &  U & \\
UGC~3828  & 07 24 35.784 &  +57 58 02.98 & 46.8 &27 &  0707+6110 & ++& 0.19$\times$0.16 & 50.3 &  71  & I  & C \\
NGC~2276  & 07 27 14.485 &  +85 45 16.20 & 36.8 &25 &  0702+8549 & +&  0.20$\times$0.20 &   0.0 &  110 &  U & \\
NGC~2300  & 07 32 20.486 &  +85 42 31.90 & 31.0 &25 &  0702+8549 & +& 0.20$\times$0.20 & 0.0 &   60  & I  & A \\
UGC~4028  & 07 50 49.918 &  +74 21 27.79 & 52.7 &25 &  0749+7420 & +& 0.20$\times$0.20 & 0.0  & 100   & I   & C  \\
NGC~2500  & 08 01 53.225 &  +50 44 13.56 & 10.1 &27 &  0808+4950 & ++& 0.20$\times$0.16 & 48.6 & 93 & U  & \\
NGC~2543  & 08 12 57.955 &  +36 15 16.23 & 32.9 &27 &  0815+3635 & ++& 0.24$\times$0.16 & 35.0  & 77 & U & \\
NGC~2537  & 08 13 14.643 &  +45 59 23.25 & 9.0  &27 &  0806+4504 & ++& 0.21$\times$0.16 &  40.3 & 92 & U & \\
NGC~2541  & 08 14 40.073 &  +49 03 41.18 & 10.6 &27 &  0808+4950 & ++& 0.20$\times$0.16 & 48.9  & 87 & U & \\
NGC~2639  & 08 43 38.093 &  +50 12 19.94 & 42.6 &27 &  0849+5108 & ++& 0.20$\times$0.19 & 68.0 &  130 & I   & C \\
NGC~2634  & 08 48 25.433 &  +73 58 01.62 & 30.2 &25 &  0930+7420 & +& 0.20$\times$0.20 & 0.0 & 82  & I   & A \\
NGC~2681  & 08 53 32.739 &  +51 18 49.35 & 13.3 &27 &  0849+5108 & ++& 0.19$\times$0.16 &  51.8 & 920 & I  & C \\
IC~520 	  & 08 53 42.275 &  +73 29 27.32 & 47.0 &25 &  0930+7420 & +& 0.20$\times$0.20 &    0.0 &  102 & U & \\
NGC~2655  & 08 55 37.731 &  +78 13 23.10 & 24.4 &25 &  0919+7825 & +& 0.20$\times$0.20 & 0.0  &  120  & I  & E \\
NGC~2715  & 09 08 06.196 &  +78 05 06.57 & 20.4 &25 &  0919+7825 & +& 0.20$\times$0.20 &   0.0 &  93 & U & \\
NGC~2748  & 09 13 43.037 &  +76 28 31.23 & 23.8 &25 &  0930+7420 & +& 0.20$\times$0.20 &   0.0 &  90 & U& \\
NGC~2841  & 09 22 02.655 &  +50 58 35.32 & 12.0 &27 &  0929+5013 & ++& 0.19$\times$0.16 & 52.7 & 74   & I  & C \\
NGC~3184  & 10 18 16.985 &  +41 25 27.77 & 8.7  &17 &  1020+4320 & ++& 0.34$\times$0.28 &   32.4 &  108 & U &  \\
NGC~3198  & 10 19 54.990 &  +45 32 58.88 & 10.8 &17 &  1020+4320 & ++& 0.20$\times$0.20 &   0.00 &  114  & I & A \\
NGC~3294  & 10 36 16.267 &  +37 19 28.52 & 26.7 &17 &  1033+3935 & ++& 0.37$\times$0.29 &  -56.4 & 119 & U  &  \\
NGC~3319  & 10 39 09.533 &  +41 41 12.74 & 11.5 &17 &  1033+3935 & ++& 0.45$\times$0.33 &  -49.0 & 98 &  U &  \\
NGC~3414  & 10 51 16.242 &  +27 58 29.88 & 23.9 &17 &  1102+2757 & ++& 0.22$\times$0.14 &  20.60  & 60  & I & C  \\
NGC~3430  & 10 52 11.415 &  +32 57 01.53 & 26.7 &17 &  1050+3430 & ++& 0.20$\times$0.20 &    0.0 & 100  & I  & E \\
NGC~3432  & 10 52 31.132 &  +36 37 07.60 & 7.8  &17 &  1050+3430 & ++& 0.47$\times$0.28 &  -16.7 & 110  & I  & A  \\
NGC~3583  & 11 14 10.979 &  +48 19 06.16 & 34.0 &18 &  1110+4817 & ++& 0.17$\times$0.13 & -59.4  & 79 & U &  \\
NGC~3600  & 11 15 51.980 &  +41 35 28.96 & 10.5 &18 &  1110+4403 & ++& 0.17$\times$0.14 &  -55.8 & 77  & U &  \\
NGC~3652  & 11 22 39.056 &  +37 45 54.14 & 33.5 &17 &  1130+3815 & ++& 0.20$\times$0.20  & 0.0 &  89 & U  &  \\ 
NGC~3665  & 11 24 43.630 &  +38 45 46.05 & 32.4 &17 &  1130+3815 & ++& 0.20$\times$0.20 & 0.0  &  61 & I  & B  \\
NGC~3675  & 11 26 08.584 &  +43 35 09.30 & 12.8 &18 &  1110+4403 & ++& 0.17$\times$0.14  & -63.3  & 80  & I  & A  \\
NGC~3726  & 11 33 21.174 &  +47 01 44.73 & 17.0 &18 &  1138+4745 & ++& 0.18$\times$0.13 &  -59.0  &  63 & U &  \\
NGC~3877  & 11 46 07.782 &  +47 29 40.20 & 91.6 &18 &  1138+4745 & ++& 0.18$\times$0.13 &   -54.7 &  66 &  U &  \\
NGC~3893  & 11 48 38.207 &  +48 42 38.84 & 17.0 &18 &  1153+4931 & ++& 0.18$\times$0.13 &  -59.4 & 58  & U &  \\
NGC~3938  & 11 52 49.453 &  +44 07 14.63 & 17.0 &18 &  1155+4555 & ++& 0.19$\times$0.14& -54.2  &  74 & I & A \\
NGC~3949  & 11 53 41.748 &  +47 51 31.62 & 17.0 &18 &  1153+4931 & ++& 0.18$\times$0.13 & -56.5  & 71 & U &  \\
NGC~4013  & 11 58 31.417 &  +43 56 49.28 & 17.0 &18 &  1155+4555 & ++& 0.19$\times$0.14 & -56.4  &  70 & unI &  \\
NGC~4051  & 12 03 09.686 &  +44 31 52.54 & 17.0 &18 &  1155+4555 & ++& 0.18$\times$0.14 &  -58.6  &  71 & I & C  \\
NGC~4914  & 13 00 42.967 &  +37 18 55.20 & 62.4 &10 &  1308+3546 & ++& 0.21$\times$0.19 & 28.4 & 71 & U &  \\
NGC~5005  & 13 10 56.312 &  +37 03 32.19 & 21.3 &10 &  1308+3546 & ++& 0.21$\times$0.19 & 27.0 & 93  & I  & D  \\
NGC~5055  & 13 15 49.329 &  +42 01 45.44 & 7.2  &10 &  1324+4048 & ++& 0.20$\times$0.19 & 68.1 & 84 & U &  \\
NGC~5112  & 13 21 56.503 &  +38 44 04.23 & 20.5 &10 &  1324+4048 & ++& 0.20$\times$0.19 & 34.4 & 104 & U &  \\
NGC~5194  & 13 29 52.698 &  +47 11 42.93 & 7.7  &10 &  1335+4542 & ++& 0.18$\times$0.17 & 82.7& 101 & I  & D  \\
NGC~5195  & 13 29 59.590 &  +47 15 58.06 & 9.3  &10 &  1335+4542 & ++& 0.20$\times$0.18 &  82.8 & 80 & I & C \\
NGC~5273  & 13 42 08.386 &  +35 39 15.26 & 21.3 &09 &  1340+3754 & m& 0.25$\times$0.17 & 32.4 & 57  & unI  &  \\
NGC~5297  & 13 46 23.694 &  +43 52 19.34 & 37.8 &10 &  1335+4542 & ++& 0.20$\times$0.19 & 85.9 & 102 & U &  \\
NGC~5353  & 13 53 26.741 &  +40 16 59.24 & 37.8 &09 &  1405+4056 & m& 0.18$\times$0.14 & 29.6  & 180 & I  & A \\
NGC~5371  & 13 55 40.023 &  +40 27 42.37 & 37.8 &09 &  1405+4056 & m& 0.20$\times$0.15 & 35.0 &  80 & U &  \\
NGC~5377  & 13 56 16.670 &  +47 14 08.03 & 31.0 &10 &  1358+4737 & ++& 0.20$\times$0.18 & -86.5 & 91 & I & C  \\
NGC~5383  & 13 57 04.980 &  +41 50 45.60 & 37.8 &09 &  1405+4056 & m& 0.20$\times$0.15 & 35.0 & 85 & U &  \\
NGC~5395  & 13 58 37.939 &  +37 25 28.49 & 46.7 &09 &  1340+3754 & m& 0.23$\times$0.15 & 30.8 & 77 & U &  \\
NGC~5448  & 14 02 50.072 &  +49 10 21.53 & 32.6 &10 &  1358+4737 & ++&  0.19$\times$0.18 &  -86.6 & 86 & I & C \\
NGC~5523  & 14 14 52.310 &  +25 19 03.41 & 21.5 &09 &  1419+2706 & m& 0.26$\times$0.14 & 26.5 & 79 & U   &  \\
NGC~5557  & 14 18 25.708 &  +36 29 37.28 & 42.6 &09 &  1426+3625 & m& 0.24$\times$0.15 & 30.6 & 90 & U   &  \\
NGC~5660  & 14 29 49.801 &  +49 37 21.40 & 37.2 &07 &  1439+4958 & ++& 0.29$\times$0.17 &   13.4 & 77 & U  &  \\
NGC~5656  & 14 30 25.514 &  +35 19 15.98 & 42.6 &09 &  1426+3625 & m& 0.23$\times$0.15 & 29.6 & 86 &  U   &  \\
NGC~5676  & 14 32 46.853 &  +49 27 28.11 & 34.5 &07 &  1439+4958 & ++& 0.35$\times$0.19  & 23.2 & 85  & unI  &  \\
NGC~5866  & 15 06 29.561 &  +55 45 47.91 & 15.3 &07 &  1510+5702 & ++& 0.18$\times$0.18  & 0.0   & 97 & I & D \\
NGC~5879  & 15 09 46.751 &  +57 00 00.76 & 16.8 &07 &  1510+5702 & ++& 0.18$\times$0.18 & 0.0  & 71  & I  & A  \\
NGC~5905  & 15 15 23.324 &  +55 31 01.59 & 44.4 &07 &  1510+5702 & ++&  0.28$\times$0.19 &  18.6 &  75 &  U &  \\
NGC~5907  & 15 15 53.687 &  +56 19 43.86 & 14.9 &07 &  1510+5702 & ++&  0.28$\times$0.19 &  18.6 &  90 & unI  &  \\
NGC~5982  & 15 38 39.778 &  +59 21 21.21 & 38.7 &07 &  1559+5924 & ++& 0.26$\times$0.19 & 20.2 &  78 &  U &  \\
NGC~5985  & 15 39 37.090 &  +59 19 55.02 & 39.2 &07 &  1559+5924 & ++& 0.26$\times$0.19 & 22.1 & 57 & I  & D  \\
NGC~6015  & 15 51 25.232 &  +62 18 36.11 & 17.5 &07 &  1559+5924 & ++& 0.23$\times$0.18 &  25.4 & 67 & unI &  \\
NGC~6140  & 16 20 58.162 &  +65 23 25.98 & 18.6 &07 &  1623+6624 & ++& 0.23$\times$0.18 & 23.2 &  75 & U &  \\
NGC~6702  & 18 46 57.576 &  +45 42 20.56 & 62.8 &05 &  1852+4855 & m& 0.19$\times$0.16 & 56.7 & 74 & I  & C \\
NGC~6703  & 18 47 18.845 &  +45 33 02.33 & 35.9 &05 &  1852+4855 & m& 0.19$\times$0.15 & 53.6  & 66 & I & A \\
NGC~6946  & 20 34 52.332 &  +60 09 13.24 & 5.5  &05 &  2010+6116 & m& 0.19$\times$0.15 & -84.1 & 57 & I & E \\
NGC~6951  & 20 37 14.118 &  +66 06 20.02 & 24.1 &05 &  2015+6554 & m& 0.20$\times$0.15 & 88.5  & 55 & I & C \\
NGC~7217  & 22 07 52.368 &  +31 21 33.32 & 16.0 &05 &  2217+3156 & m& 0.18$\times$0.18 &  0.0  & 75  & I & C \\
NGC~7331  & 22 37 04.102 &  +34 24 57.31 & 14.3 &05 &  2253+3236 & m& 0.15$\times$0.15 &    0.0 &  81 & U &  \\
NGC~7332  & 22 37 24.522 &  +23 47 54.06 & 18.2 &05 &  2238+2749 & m& 0.15$\times$0.15 &    0.0 & 86 &  U &  \\
NGC~7457  & 23 00 59.934 &  +30 08 41.61 & 12.3 &05 &  2253+3236 & m& 0.15$\times$0.15 &    0.0 & 92 & U &  \\
NGC~7640  & 23 22 06.584 &  +40 50 43.54 & 8.6  &05 &  2312+3847 & m& 0.15$\times$0.15 &   0.0 & 81 &  U &  \\
NGC~7741  & 23 43 54.375 &  +26 04 32.07 & 12.3 &04 &  2343+2339 & ++&  0.19$\times$0.14 &  36.0  & 70 & U &  \\
NGC~7798  & 23 59 25.503 &  +20 44 59.59 & 32.6 &04 &  0004+2019 & ++& 0.20$\times$0.14 & 35.9  & 31 &  I  & C \\
 
%NGC~5548  & 14 17 59.513 &  +25 08 12.45 &09 &  J1419+2706 & & 67 	& SA0/a        &SEYFERT   &291 	& 8.78  \\
%NGC~2685  & 08 55 34.750 &  +58 44 03.87 &25 &  J0930+7420 &  {c} 	& 16.2 	& SB0 pec      &SEYFERT    &93.8 	& 6.81  \\

\hline
\hline
\end{longtable}
\end{center}
Column description: (1) source name; (2)-(3) RA and DEC position
(J2000) from NED, using optical or infrared images; (4) distance (Mpc)
from \citet{ho97a}; (5) LeMMINGs observation block; (6) phase
calibrator name; (7) raw data and calibration quality: `++' = very
good; `+' = good; `m' = moderate; (8) restoring beam size in arcsec in
full resolution map; (9) PA angle (degree) in full resolution map; (10) rms in
full resolution map in $\mu$Jy beam$^{-1}$; (11) radio detection
status of the source: `I' = detected and core identified; `U' =
undetected; `unI' = detected but core unidentfied; (12) radio
morphological class: A = core/core--jet; B = one-sided jet; C =
triple; D = doubled-lobed ; E = jet+complex.  \twocolumn

\noindent
Ho 2001b,
\citealt{nagar02,falcke02,anderson04,filho04,nagar05,filho04,nagar05,panessa13};
MERLIN \citealt{filho06}). However, an exhaustive sub-arcsecond
sub-mJy radio study of the Palomar sample, unbiased towards the
inactive galaxies, is still missing and is the goal of the LeMMINGs
survey.

%eMERLIN is an exceptional instrument because its baselines, which are
%intermediate between JVLA and VLBA/I, and its wide-band correlators
%offer milli-arcsecond resolution and $\mu$Jy sensitivity
%simultaneously. Therefore, when applied to a complete sample of local
%galaxies, eMERLIN's capabilities give unprecedented views of AGN and
%SF on parsec scales, at unparallelled sensitivity. 

The LeMMINGs project focuses on a declination-cut subset of the
Palomar galaxies, with declination $>$ 20$^{\circ}$ (280 targets, more
than one half of the original Palomar sample), to ensure proper
visibility ($u{-}v$) coverage at higher latitudes, accessible to the
eMERLIN array.  This legacy survey has been granted 810 hours of
eMERLIN time in total, representing the second largest of the eMERLIN
legacy surveys. The aim of this project is to detect pc-scale radio
components associated with either SF regions or AGN. The survey is
being carried out at 1.5 and 5 GHz using a two-tiered approach: the
`shallow' and the `deep' sample. The former consists of short
observations of a large sample of 280 objects for a total of 750
hours. The latter consists of six scientifically interesting targets
observed for longer exposure times for a total of 60 hours to reach a
higher sensitivity. The observations of these six sources include the
Lovell telescope and are published as separate papers: M~82
\citep{muxlow10}, IC~10 \citep{westcott17}, NGC~4151
\citep{williams17}, NGC~5322 \citep{dullo18}, M~51b
(Rampadarath et al, submitted), and NGC~6217 (Williams et al, in
prep.).

In this work, we present the first release\footnote{The eMERLIN radio
  images will be released through the Centre de Donn{\'e}es
  astronomiques de Strasbourg (CDS) website to store and distribute
  the astronomical data in early 2018.}  of the eMERLIN results of the
`shallow sample' at 1.5 GHz, focusing on the central galaxy region
(0.73 arcmin$^{2}$) to explore the nuclear radio properties. In this
first data release, we publish the data for 103 galaxies which are
randomly selected from the original sample of 280 targets and are
observed and calibrated until the end of the first semester of
2017. In the next data release, we will complete the sample and
perform a detailed multi-wavelength study, which will provide better
statistics from greater numbers of sources per class, critical to draw
final conclusions on the LeMMINGs sample.

Although the 103 sources (listed in Table~\ref{radioprop}) are
randomly selected from the Palomar sample, all the optical classes (30
LINERs, 7 Seyferts, 16 Transition galaxies, 37 H{\sc ii} galaxies, and
13 ALGs, according to \citealt{ho97a}) are included. However, we will
revise the optical classifications based on the new updated diagnostic
schemes (Section~\ref{optclass}). The sample encompasses all galaxy
morphologies (E, S0, Sa-b-c-d-m, I0, Pec). The galaxies extend in
distance out to 120 Mpc, but with a median distance of $\sim$20 Mpc,
similar to the original Palomar sample.

%Ho et al sample
%30 LINER
%16 TRANS
%7 SEYFERT
%37 HII
%13 ALG

%It is important to note that, unlike most previous radio surveys, which
%mostly focused on active galaxies, we observe all optical classes,
%being unbiased towards nuclear classifications to unveil the true
%relationships between BHs and host galaxies.

%Only \citet{ulvestad02} have completed a survey
%of a well-defined sub-sample of 40 H{\sc ii} type nuclei in the Palomar
%sample, but they found that none of them has a compact radio nucleus
%at the flux levels of those in LLAGN in the sample. 

\section{Observations and Data Reduction}
\label{observations}

The LeMMINGs shallow sample was observed with eMERLIN, an array
consisting of seven radio telescopes spread across the UK. The
configuration of eMERLIN is identical to that of MERLIN, but with a
wider bandwidth and greater sensitivity. The observations were made at
L band (20 cm, 1.5 GHz) without the Lovell telescope during
2015--2017. The sample was split into blocks of 10 galaxies (plus
calibrators) per observation, clustered on the sky in intervals of
right ascension to minimise the slewing of the telescopes. We use a
`snapshot imaging' mode: each observation has a time-on-source of
$\sim$48 minutes per target, with the observations spread out over
several hour angles to improve the $u{-}v$ coverage. The observational
set-up used a total bandwidth of 512 MHz split into 8 intermediate
frequencies (IFs), each with a width of 64 MHz and consisting of 128
channels, centred on 1.5 GHz between 1.244 -- 1.756 GHz.

In this work we provide the observations from eleven observing blocks:
2, 3, 4, 5, 7, 9, 10, 17, 18, 25, and 27, which include 110
targets. Several phase calibrators were used across the datasets,
filling the gaps between the target pointings. The target-phase
calibrator loop usually lasted 10-11 min, with 3-4 min on the phase
calibrator and 7 min on the target, for a total of 7 visits to fully
cover the $u{-}v$ plane. The phase calibrators were selected from the
VLBA calibrator lists \citep{beasley02} to be unresolved on eMERLIN
baseline scales. The bandpass calibrator (OQ~208) and the flux
calibrator (3C~286) were typically observed at the end of each run.

%2: mancano core baseline for some sources
%3: No Defford baselines, flux/BP cal data copied from LEM4 to save Darnhall baselines 
%4: ok
%5: low qualitity data
%7: ok
%9: manca antenna Darnhall
%10: ok
%17: ok
%18: ok
%25:Defford baselines flagged: No flux calibrators on Defford baselines, only one snap snot one Defford baselines
%27: ok

\subsection{Data calibration}

The data reduction was carried out using \verb'AIPS' (the Astronomical
Image Processing Software from NRAO) software, following the
calibration steps detailed in the eMERLIN cookbook (Belles et al
2015).

The large separation between the eMERLIN telescopes yield projected
baselines in the $uv$-plane of up to $\sim$1300~k$\lambda$ for our
datasets. This results in high-resolution maps (beam FWHM $\sim$150
mas), which are necessary to study the pc-scale nuclear emission of
the sources.  Spurious, low-level radio frequency interference (RFI)
can affect the observations of our targets, which are expected to be
intrinsically weak in radio. First, we excised bright RFI, using the
\verb'SERPENT' auto-flagging code \citep{peck13} written in
ParselTongue \citep{kettenis06} principally for use in the eMERLIN
pipeline \citep{argo15}. Next, we carried out a further inspection of
the data with the tasks \verb'SPFLG' and \verb'IBLED' to remove any
remaining RFI not recognised by the automatic flagger. The first and
the last 10 channels in each IF were flagged because they are in the
least sensitive areas of the passband and contribute mostly noise.  We
estimate that the flagged data is generally $\sim$30 per cent of a
full dataset, but represents a higher portion in those runs where
antennas were missing from the observations.

No pipeline was used for the data reduction. We began the calibration
procedures by fitting the offsets in delays using the task
\verb'FRING' before calibrating for phase and gain.  We followed
standard calibration procedures to determine the complex gain
solutions from the flux calibrator to apply later to the
data. Bandpass solutions were calculated with \verb'BPASS' and applied
to the data, followed by imaging the phase calibrator. We performed
several rounds of phase-only self-calibration on the phase calibrators
and a final round of self-calibration in phase and amplitude. The
phase solutions are stable within $\pm$20 degrees independent of
baseline. The amplitude solutions show typical variations within a
range of 10-20 per cent. The solutions from self-calibration were then
applied to the target field. For some scans, the calibration failed
due to lack of sufficiently long exposure times on the phase
calibrator caused by antenna problems and we deleted the affected
calibrator-target scans for some objects.

%At the end of the calibration, we find that the observing blocks 4, 7,
%10, 17, 18, and 27 have standard quality data, while block 5 shows
%poor but still valid calibration solutions. For block 3 and 25 there
%are nearly no baselines with the Defford telescope, while the Darnhall
%antenna is missing for block 9. For block 2, the core baselines (Mk2,
%Pickmere and Darnhall antennas) are partially missing for some
%scans. In summary, we fail the calibration for seven observed objects
%(NGC~783, NGC~2685, NGC~3395, NGC~5033, NGC~5354, NGC~5548, NGC~7080).
%Hence, those sources miss in the final sample, which now consists of
%103 galaxies. Table~\ref{list} displays the final list of observed
%sources with their associated phase calibrators and observing blocks.

At the end of the calibration, we find that observing blocks 4, 7, 10,
17, 18, and 27 have standard quality data, while block 5 shows poor
but still valid calibration solutions. During observing blocks 2, 3, 9
and 25 some eMERLIN antennas partially participate in observations due
to a combination of weather and technical issues (weak flux
calibrators on short scans, antenna dropouts in the observing run, or
low-elevation observations). In summary, adequate data was acquired
for all but seven of the observed galaxies (NGC~783, NGC~2685,
NGC~3395, NGC~5033, NGC~5354, NGC~5548, NGC~7080). The exclusion of
these seven galaxies resulted in a final sample of 103 galaxies
presented in this paper. Table~\ref{radioprop} displays the final list
of sources with their associated phase calibrators, observing blocks
and their calibration quality.

After a further inspection of the target data to check the quality and
remove possible bad scans, the calibrated datasets were imaged (Stokes
I) using the standard multi-scale, multi-frequency synthesis algorithm
\verb'IMAGR', with a cell-size of 0.05 arcsec and natural
weighting. Because of the short observations with the snapshot imaging
mode and the large bandwidth, the H{\sc i} absorption line, which
falls in the covered radio spectral window, was expected not to
significantly contribute and, hence, it was not considered later in
the calibration and in the imaging process. We performed a few rounds
of self-calibration in phase and a final one in phase and amplitude,
using 1-2 min of integration time and using at least 3-$\sigma$
detections for solutions. However, only sources with flux densities
higher than 5 produced adequate solutions.  This procedure improved
the signal-to-noise of the final maps and reduced the phase and gain
solution scatter. Bright sources in the fields were mapped separately
while imaging the targets to reduce the noise in the field. At this
stage of the project, we focused only on the central region of fields
within an area of at least 1024$\times$1024 pixels around the target
position from NED (optical or infrared images), which corresponds to
an area of 0.73 arcmin$^{2}$ (within the primary beam). Mapping the
entire primary-beam field will be addressed in future papers. Several
images were created in different resolutions (i) to study the
morphology of each source, picking up the low-brightness diffuse
emission, (ii) to increase the detection probability in case no source
appears in the full-resolution map, and (iii) to detect any possible
background sources or off-nuclear components.  Lower-resolution maps
were obtained using different values of the \verb'UVTAPER' parameter
in the \verb'IMAGR' task, ranging between 750 k$\lambda$ to 200
k$\lambda$. This parameter specifies the width of the Gaussian
function in the $uv$-plane to down-weight long-baseline data
points. Smaller units of this parameter correspond to a shallower
angular resolution, i.e. a larger beam size, up to a factor 10 larger
than ones reached in full resolution(i.e. $\lesssim$1.5 arcsec).

For sources which have very elliptical restoring beams (with
major-minor axis ratio higher than $\sim$2), we create a further map
using a circular beam with a radius (typically 0.18--0.2 mas,
presented in Appendix \ref{app}) corresponding to the area equivalent to
the elliptical beam. This approach helps us to identify possible
genuine radio morphological elongations of the sources in the
direction of the major axis of the beam.

We present in the Appendix \ref{app} the full and low-resolution maps
of the detected sources (Figures~\ref{maps1} and \ref{maps2}). The
uv-tapered maps are typically obtained with $u{-}v$ scales of 750
k$\lambda$, corresponding to a resolution of 0.4 arcsec, double that
of the full-resolution images. The large dynamic range of the radio
maps highlight their quality, which both depend on the phase and
amplitude solutions of the phase calibrators and the self-calibration
on the target (if performed). For a small fraction of sources (in
blocks 2, 3, 9, and 25), the image quality is modest, but sill
adequate for the purposes of our survey.

\section{Radio results}
\label{radioresults}

In this Section we present the results related to the 103 radio images
of the LeMMINGs legacy survey, which are released in this paper.

\subsection{Radio maps and source parameters}

Since the goal of this work is to study the radio emission associated
with the SMBH, we search for detected radio sources in the innermost
region of the optical galaxy extent (within a radius of 5 kpc at 20
Mpc) in full resolution and at multiple tapering scales. This choice
favours the selection of radio sources co-spatial with the optical
galaxy centre.

With a typical beam size of 0.15-0.2 arcsec at full resolution, each
radio map of our sample reveals either single or multiple components.
We adopt a consistent method in extracting source parameters
(position, deconvolved size, peak flux density, integrated flux
density, beam size and PA), by fitting a two-dimensional Gaussian
model with the task \verb'JMFIT'.  This procedure works for compact,
unresolved components. For extended resolved structures we measure the
flux density within interactively defined boundaries with \verb'IMEAN'
for rectangular boxes or \verb'TVSTAT' for irregularly shaped regions,
following the emitting area of the sources. We always repeat the
measurements several times using similar regions to derive an average
value. The rms is measured in a region around the central source if
present or at the optical centre, using a rectangular box with
\verb'IMSTAT'.  We also apply this method to the uv-tapered maps to
search for detected components. We consider the components detected if
their flux densities were above 3$\sigma$ above the typical rms
measured around the source.  Additional components not present in full
resolution may appear in the uv-tapered map because the larger beam
allows for low-brightness diffuse emission to emerge above the rms in
the image. In this case we derive a 3$\sigma$ upper-limit flux density
for the undetected counterparts at full resolution.

In general, the final natural-weighted full-resolution maps reach a
median rms of 70 $\mu$Jy beam$^{-1}$, with a standard deviation of
0.54. Searching components within the central galaxy region, we detect
significant radio emission ($\gtrsim$0.2 mJy) at the galaxy centre for
47 out of the 103 sources ($\sim$46 per cent). For such sources, we
also extract the flux densities of their counterparts in the
uv-tapered maps. The main parameters of the full-resolution maps (beam
size, PA, rms) are listed in Table~\ref{radioprop}. The source
parameters for the detected sources in full and low resolution are
listed in Table~\ref{tabdet} and Table~\ref{tabsfr} in Appendix
\ref{app}. For the remaining 56 objects, no significant radio emission
has been detected in the core region of their fields both in full and
in low-resolution images, which are essentially consistent with the
3$\sigma$ rms level.

The detected sources show a large variety of radio morphologies from
pc-scale unresolved single components to extended and complex
shapes. The low-resolution maps typically show radio structures more
extended than those revealed from full-resolution maps. The notable
cases are the maps of NGC~5005 and NGC~5194: kpc-scale double lobes
come out when reducing the resolution. The sizes of the radio
structures vary from 150 mas for the unresolved sources to 26 arcsec
for NGC~5005 with a median size of 1.5 arcsec. This interval
corresponds to a physical linear range of $\sim$10-1700 pc (median
$\sim$100 pc), depending on the galaxy of the sample. In
Table~\ref{tabdet} and \ref{tabsfr} we list the radio sizes for the
resolved sources.

We perform a further step in the analysis, i.e. identifying the radio
cores. We define cores in our sources as the unresolved radio centre
of the sources, which may represent either a possible jet base or an
unresolved central star-forming region, where the SMBH resides.  The
identification of the radio core in our maps depends on the specific
morphology and on the proximity to the optical centre, taken from the
NED catalogue (using the HST optical centroid in the best-case
scenario). The core is generally recognised as an unresolved component
at the centre of a symmetric structure (i.e. twin jets). In case of
radio asymmetry, the optical centre and the brightness of the radio
components (with the core brighter than the other components, i.e. in
the case of one-sided jets) aid with pinpointing the radio core.
Nevertheless, the distance between the optical galaxy centre and the
closest possible radio component is the main criterion for a core
identification.

Previous VLA observations on the Palomar sample detected and
identified radio cores with a resolution of $\sim$1 arcsec (e.g.,
\citealt{nagar02,filho00,nagar05}), generally matching the optical
galaxy centres. Since our eMERLIN observations are 5 times higher in
resolution, the VLA 1 arcsec-beam size can be used to roughly estimate
the optical-radio offset we expect to find for our sample. Such an
offset corresponds to an average area with a radius of $\sim$100 pc
for our sample. Indeed, the typical radio-optical separation observed
in our sample is 0.3 arcsec, largely within the VLA resolution. To be
precise, for 38 out of the 47 detected sources, radio emission is
detected within 1 arcsec from the optical centre. Only three sources
(NGC~147, UGC~4028, and NGC~3432) show a radio core at 1.5 arcsec from
the optical centre. For these 41 (38+3) objects we assume that the
radio cores are correctly identified (see Section~\ref{core-ident})
and we call them `identified' hereafter. For the remaining 6 objects
out of 47 detected (NGC~189, NGC~4013, NGC~5273, NGC~5676, NGC~5907,
and NGC~6015) the identification of their radio cores is questionable
either because of the lack of a clear core within 2 arcsec from the
optical centre or the presence of multiple components of ambiguous
nature. Furthermore, another two sources (NGC~2342 and NGC~3198),
whose radio core has already been identified, show an unexpected
additional radio source in the central field, possibly related to a
background AGN. We call these 8 (6+2) sources `unidentified' hereafter
(see Section~\ref{unidentified}). The status of the radio
detection/core-identification is listed in column 11 of
Table~\ref{radioprop}.

\subsection{Identified sources}
\label{core-ident}

Figure~\ref{maps1} presents the full and low-resolution radio maps
(source characteristics in Table~\ref{tabdet}; radio contours and
restoring beam in Table~\ref{contours}) of the 41 detected galaxies,
where we identify the radio cores within 1.5 arcsec from the optical
centre.

To test the possibility that the radio cores are associated with the
galaxies and not with background objects, we calculate the probability
of detecting a radio source above the survey detection limit of 0.2
mJy within a given sky area. For this purpose, we consider the source
count distribution obtained with VLA observations at 1.4 GHz of the
13$^{th}$ XMM-Newton/ROSAT Deep X-ray survey area \citep{seymour04}
over a 30-arcmin diameter region as no eMERLIN wide-field survey
exists at this frequency in the literature. Based on this source
density distribution, we expect to randomly detect one radio source
out of 103 galaxies within a circular area of radius 3.7 arcsec. This
result suggests that the radio cores, identified within the
radio-optical offset of 1.5 arcsec observed for the 41 galaxies, are
likely to be associated with the target sources.

The radio nuclei are detected at full resolution for all, except five
targets (NGC~147, NGC~3430, NGC~3675, NGC~5985, and NGC~6702) for
which only the low-resolution maps reveal the radio cores.  The core
peak flux densities for most of the sample fall below $\sim$1 mJy
beam$^{-1}$ on a scale of 0.2 arcsec, reaching values up to $\sim$500
mJy beam$^{-1}$ for NGC~315. Assuming that a source is considered
unresolved if its deconvolved size is smaller than the beam
size at the FWHM for both major and minor axes, $\theta_{\rm M}$ and
$\theta_{\rm m}$, most of these central components appear unresolved as
cores. In fact, the peak flux densities of the core components are
typically consistent with their integrated flux densities within a
factor of $\sim$2. The cores, which appear slightly resolved, typically
show a protrusion associated with a core--jet morphology
\citep{conway97}.

Most of the identified sources (25/41) show clear extended radio
structures.  In this paper we preferentially favour the interpretation
that this extended radio emission originates from radio jets, based on
their morphologies. This is because the possibility that the radio
emission from a galaxy disc is limited by the properties of our
eMERLIN observations (long baselines and snapshot imaging), which are
sensitive to compact radio emission like in jets, but insensitive to
diffuse low-brightness emission, as expected by a galaxy disc
\citep{brown61,kennicutt83} (see Section~\ref{anysfg} for discussion).

We classify the identified sources into five radio structures based on
both the full- and low-resolution maps (see column 12 of
Table~\ref{radioprop}). The {\it core/core--jet} morphology (16
galaxies marked as A) consists of sources which show bright cores with
a possible elongation in one direction (core--jet morphology) and
radio components aligned in the same direction of the possible jet: an
example is NGC~266. The {\it one-sided jets} (two galaxies marked as
B) are sources which clearly show a jet extended asymmetrically: a
clear example is NGC~3665. The {\it triple-sources} (14 galaxies marked
as C) are sources which show three aligned components in
full-resolution maps, which may turn into {\it twin symmetric jets} in
low-resolution maps: for example, UGC~3828. Large {\it double-lobed}
sources (4 galaxies marked as D) are objects which show two large
radio lobes at large scales either in full or in low resolution maps,
e.g. NGC~5005. The last class is represented by `{\it jet+complex}' (5
galaxies marked as E) which reveal a possible jet but additional
components emerge making the overall radio appearance complex: an
example is NGC~2273. The radio sources which show jet-like
morphologies (one-sided, two-sided, triple, double-lobed sources) are
hereafter called `jetted' and those without a clear jet,
`non-jetted'. Note that the radio classifications can be equivocal
since the radio structures are generally characterised by low
brightness.

\subsection{Unidentified sources}
\label{unidentified}

Eight radio sources have been classified as `unidentified'. Six of
them do not show an unequivocal radio core, further than at least 2
arcsec from the optical centre. The remaining two are unknown radio
sources detected in the central field of NGC~2342 and
NGC~3198. Figure~\ref{maps2} shows the full and low-resolution maps
(for radio contours and restoring beam see Table~\ref{contours}) of
the eight unidentified sources. Their radio source parameters are
listed in Table~\ref{tabsfr}.

For NGC~891, no components are detected in the full resolution map but
three components, which might be a triple source, are detected in the
low resolution map\footnote{SN~1986J, present in NGC~891 and detected
  in the radio band by VLA \citep{pereztorres02}, is located $\sim$1
  arcmin south of the optical centre, which is beyond the region
  considered in our analysis.}. The nominal optical centre of the
galaxy lies almost midway between two of the north components, being
about 2.5 arcsec from either one. In NGC~4013 and NGC~5273 radio
emission is detected in a circular shape (diameter of 2.5--3 arcsec)
around the optical centre, probably associated with a nuclear
star-forming ring. For these three cases, the identification of the
radio core is ambiguous, but the radio emission is likely associated
with the target galaxies. For the other 3 remaining unidentified
sources (NGC~5676, NGC~5907, and NGC~6015), radio components are
detected further than 20 arcsec from the optical nucleus ($\sim$
2kpc), pointing to a SF origin or background sources.

In addition, two radio sources appear in the field of NGC~2342 and
NGC~3198, respectively 4 arcsec and 6 arcsec from the detected central
cores, such that these off-nuclear sources are unidentified. In both
cases, the source components appear two-sided and elongated, similar
to radio jets, and hence these sources possibly belong to unknown
background AGN.

Using the same approach explained in Section~\ref{core-ident}, the
number of expected random radio detections within areas of radii 4 and
20 arcsec (the offsets measured above) are, respectively, 1.2 and
39.2. These values agree with the interpretation that the radio
sources found in the field of NGC~2342 and NGC~3198 and the
off-nuclear components in NGC~5676, NGC~5907, and NGC~6015 are
potentially background objects.

Although the core has not been ascertained in the unidentified
sources, these galaxies might still conceal active nuclei at the
galaxy centre, emitting below the detection limit or simply not
identified within a complex radio structure. HST and eMERLIN C-band
observations will provide better constraints on the nature of these
unidentified radio sources.

\begin{figure}
	\includegraphics[width=0.55\textwidth]{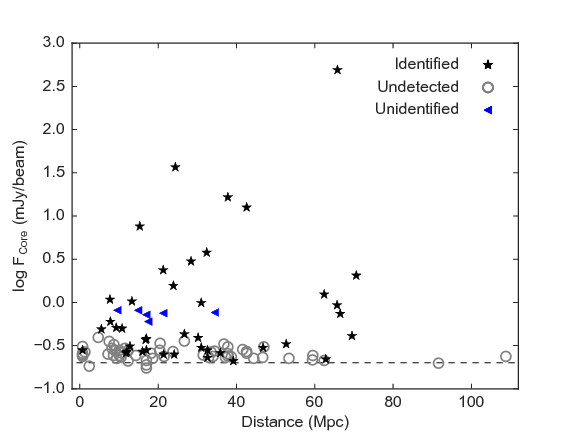}
        \includegraphics[width=0.55\textwidth]{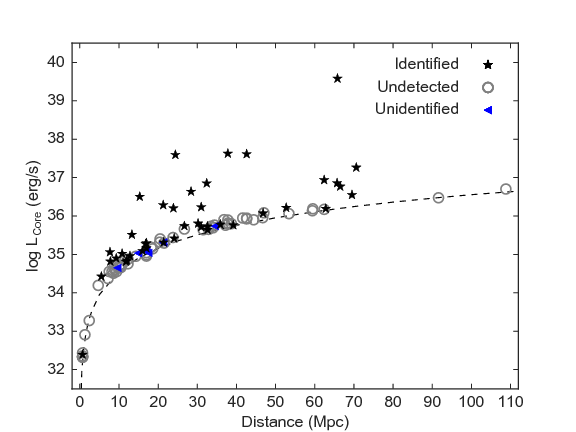}
        \caption{Radio core flux density (F$_{\rm core}$ in mJy beam$^{-1}$)
          (upper panel) and its luminosity (L$_{\rm core}$ in erg
          s$^{-1}$) (lower panel) as a function of the distance (Mpc)
          for the entire sample.  The dashed lines correspond to the
          flux density and luminosity curves at a 3$\sigma$ flux limit
          (0.2 mJy beam$^{-1}$). The different symbols correspond to
          identified sources, unidentified and
          undetected radio sources. For these last two types of
          sources, the values on the y-axis are meant as upper limits.}
    \label{distance}
\end{figure}

\subsection{Radio brightness}

The LeMMINGs survey detects for the first time faint radio cores ($<$
1mJy) on parsec scales for a large sample of nearby active and
quiescent galaxies. The 3$\sigma$ flux density limit of this survey is
$\sim$0.2 mJy (Fig.~\ref{distance}, upper panel), and we detect 
\onecolumn
\begin{center}
%\begin{longtable}{lC{1.85cm}C{2cm}C{0.3cm}ccC{1.8cm}C{1.7cm}C{0.8cm}cC{0.5cm}}
\begin{longtable}{lccccccccc}
\caption[Properties of the  sample.]{Optical-radio properties of the sample.} 
\label{optical} \\

%This is the header for the first page of the table...
\hline 
\hline
Name &  Hubble    & class  & class  & $\sigma$   & log(M$_{\rm BH}$) & L$_{\rm core}$  & L$_{\rm tot}$ & L$_{\rm [OIII]}$ &  log(L$_{\rm bol}$/L$_{\rm Edd}$) \\
     &   type     &  {\small Ho et al. (1997)} &  BPT   &  km s$^{-1}$ &  M$_{\odot}$  &  erg s$^{-1}$    & erg s$^{-1}$ & erg s$^{-1}$  & \\
 (1) & (2)& (3) & (4)  & (5) & (6)      & (7)& (8) & (9) & (10) \\    
\hline	
\endfirsthead

%This is the header for the remaining page(s) of the table...
\multicolumn{3}{c}{{\tablename} \thetable{} -- Continued} \\[0.5ex]
\hline
\hline
Name &  Hubble    & class  & class  & $\sigma$   & log(M$_{\rm BH}$) & log(L$_{\rm core}$)  & log(L$_{\rm tot}$) & log(L$_{\rm [OIII]}$) &  log(L$_{\rm bol}$/L$_{\rm Edd}$) \\
     &   type     &  {\small Ho et al. (1997)} &  BPT   &  km s$^{-1}$ &  M$_{\odot}$  & erg s$^{-1}$    & erg s$^{-1}$ & erg s$^{-1}$  & \\
 (1) & (2)& (3) & (4)  & (5) & (6)      & (7)& (8) & (9) & (10)  \\   
\hline 

\endhead

%This is the footer for all pages except the last page of the table...
\hline
  \multicolumn{10}{c}{{Continued on Next Page}} \\
\endfoot

%This is the footer for the last page of the table...
 \\[-1.8ex] 
\endlastfoot

%Now the data...
NGC~7817  & SAbc      & H      & H               &66.7  & 6.21  & $<$35.64 & $-$ &  39.29 & $-$1.51 \\
IC~10 	  & IBm?      & H      & H               &35.5  & 5.11  & $<$32.91 & $-$ &   37.13 & $-$2.57 \\
NGC~147   & dE5 pec   & ALG    & ALG$^{a}$          &22 & 4.28  & 32.39  & 32.92 &  $-$   &   $-$     \\ 
NGC~185   & dE3 pec   &S2      & L             &  19.9  & 4.10  &  $<$32.33 & $-$ & 34.63 & $-$4.06 \\
NGC~205   & dE5 pec   & ALG    & ALG$^{a}$   &23.3 & 4.34$^{*}$  & $<$32.36 & $-$ &  $-$   &    $-$   \\
NGC~221   & E2        & ALG    & ALG$^{a}$  &72.1  & 6.36$^{*}$  & $<$32.33 & $-$ &  $-$   &   $-$   \\
NGC~224   & SAb       & ALG    & ALG$^{a}$       &169.8 & 7.84  & $<$32.32 & $-$ &  $-$   &   $-$   \\
NGC~266   & SBab      &L1.9    & L              &229.6 & 8.37  & 36.94  &  37.02 & 39.43  &   $-$3.53  \\
NGC~278   & SABb      &H       & H              &47.6  & 5.62  & 34.81  &  35.83 & 37.47  &   $-$2.74   \\
NGC~315   & E+        &L1.9    & L        &303.7 & 8.92$^{*}$   & 39.58  &  39.61 & 39.43  &   $-$4.02  \\
NGC~404   & SA0       &L2      & L        &40    & 5.65$^{*}$   & $<$33.28 & $-$ & 37.16  &  $-$3.08  \\
NGC~410   & E+        &T2:     & L              &299.7 & 8.84   & 37.26  &  37.53 & $<$39.32  &   $<$-4.11  \\
NGC~507   & SA0       & ALG    & ALG            &307.7 & 8.88   & 36.86  &  37.07 &  $-$   &   $-$     \\ 
NGC~598   & SAcd      &H       & H$^{a}$    &21         & 4.20  &  $<$32.43 & $-$ & $<$34.63 & $<$-4.16 \\
IC~1727   & SBm       &T2/L2   & L              &136.8 & 7.47  &  $<$34.55 & $-$ & 37.34  & $-$4.72   \\
NGC~672   & SBcd      &H       & H         &$<$64.3 & $<$6.15  &  $<$34.55 & $-$ & 37.66  & $-$3.08   \\
NGC~697   & SABc      &H       & H	     &75       & 6.42  &  $<$35.95 & $-$ & 37.86  & $-$3.15  \\
NGC~777   & E1        &S2/L2:: & L$^{b}$  &       324.1 & 8.97  & 36.77  &  36.98 &  38.38 &    $-$5.18  \\
NGC~784   & SBdm      &H       & H$^{c}$         &35.5  & 5.11  & $<$34.19 & $-$ & 37.68  & $-$2.02   \\
NGC~812   & SAB0/a pec&H       & H               &120.9 & 7.25  & $<$36.70 & $-$ & 38.68 & $-$3.16  \\
NGC~818   & SABc      &H:      & H              &151.3 & 7.64   & $<$36.14 & $-$ & 38.46 & $-$3.77  \\
NGC~841   & SABab     &L1.9:   & L$^{d}$        &159.2 & 7.73   & $<$36.19 & $-$ & 38.74  & $-$3.58  \\
NGC~890   & SAB0      &ALG     & ALG            &210.9 & 8.22   & $<$36.06 & $-$ &  $-$   &   $-$   \\
NGC~891   & SAb?      &H       & H$^{c,e}$       &73.1  & 6.37   & $<$34.65 & $-$ &  36.29  & $-$4.66  \\
NGC~925   & SABd      &H       & H              &71.9  & 6.34   & $<$34.55 & $-$ & 37.21  & $-$3.72  \\
NGC~959   & SAdm      &H       & H 	        &43.6  & 5.47   & $<$34.65 & $-$ & 37.40  & $-$2.66   \\
NGC~972   & SAab      &H       & H              &102.8 & 6.97   & 35.31  &  36.06 &  38.64 &    $-$2.92  \\
NGC~2273  & SBa       &S2      & S              &148.9 & 7.61   & 36.64  &  37.98 &  40.43 &    $-$1.77  \\
NGC~2342  & S pec     &H       & H              &147.3 & 7.60   & 36.55  &  37.64 & 39.71  &   $-$2.48   \\
NGC~2268  & SABbc     &H       & H	        &143.3 & 7.55   & $<$35.76 & $-$ & 39.27   &   $-$2.87  \\
UGC~3828  & SABb      &H       & H              &73.9  & 6.39   & 36.07  &  36.63 &  38.84 &    $-$2.14  \\
NGC~2276  & SABc      &H       & H              &83.5  & 6.61   &  $<$35.90 & $-$ & 38.17  &  $-$3.03  \\
NGC~2300  & SA0       &ALG     & ALG            &261.1 & 8.60   & 36.23  &  36.41 &    $-$ &     $-$   \\
UGC~4028  & SABc?     &H       & H	        &80.5  & 6.54   & 36.22  &  36.70 & 38.88  &    $-$2.25  \\
NGC~2500  & SBd       &H       & H$^{c}$        &47.1  & 5.61   & $<$34.71 & $-$ &  36.55  &  $-$3.65 \\
NGC~2543  & SBb       &H       & H              &112.4 & 7.12   & $<$35.65 & $-$ & 38.55  & $-$3.16  \\
NGC~2537  & SBm pec   &H       & H              &63    & 6.11   & $<$34.60 & $-$ & 38.72  & $-$1.98 \\
NGC~2541  & SAcd      &T2/H:   & H$^{f}$	&53    & 5.81   & $<$34.72 & $-$ & 36.80  &  $-$3.60 \\
NGC~2639  & SAa?      &S1.9    & L$^{g}$        &179.3 & 7.94   & 37.61  &  38.50 &  39.60 &    $-$2.93  \\
NGC~2634  & E1        &ALG     & ALG            &181.1 & 7.96   & 35.81  &  35.87 &    $-$ &    $-$    \\  
NGC~2681  & SAB0/a    &L1.9    & L$^{h}$        &109.1 & 7.07   & 35.51  &  35.99 & 38.37  &   $-$3.29   \\
IC~520 	  & SABab?    &T2:     & L             &138.1 & 7.48    & $<$36.08 & $-$ & 39.03  &  $-$3.04  \\
NGC~2655  & SAB0/a    &S2      & L             &159.8 & 7.74    &  37.59  &  37.97 & 39.44  &   $-$2.89   \\
NGC~2715  & SABc      &H       & H             &84.6  & 6.63    &  $<$35.32 & $-$ & 37.79  &  $-$3.43   \\
NGC~2748  & SAbc      &H       & H       &83    & 7.65$^{*}$    & $<$35.44 & $-$ & 37.83   &  $-$4.41   \\
NGC~2841  & SAb       &L2      & L$^{c}$       &222   & 8.31    & 34.84  &  35.47 & 38.19  &   $-$4.63   \\
NGC~3184  & SABcd     &H       & H             &43.3  & 5.46    & $<$34.64 & $-$ & 37.31  &  $-$2.74    \\
NGC~3198  & SBc       &H       & H             &46.1  & 5.57    & 35.02  &  35.19 & 36.97  &   $-$3.19   \\
NGC~3294  & SAc       &H       & H	       &56.4  & 5.92    & $<$35.66 & $-$ & 38.33   &  $-$2.18    \\
NGC~3319  & SBcd      &H:      & L$^{i}$       &87.4  & 6.68    &  $<$34.84 & $-$ & 37.07   &  $-$4.20   \\
NGC~3414  & SA0 pec   &L2      & L$^{q}$ &236.8 & 8.40$^{*}$     & 36.20  &  36.33 & 39.06  &   $-$3.95   \\
NGC~3430  & SABc      &H       & H	      &50.4  & 5.72     & 35.74  &  36.19 & 37.74  &   $-$2.57  \\
NGC~3432  & SBm       &H       & H            &37    & 5.18     & 34.82  &  34.83 & 38.00  &     $-$1.77  \\
NGC~3583  & SBb       &H       & H            &131.7 & 7.40     & $<$35.69 & $-$ & 38.26   &  $-$3.73  \\
NGC~3600  & SAa?      &H       & H	      &49.8  & 5.70     & $<$34.66 & $-$ & 38.21   & $-$2.08  \\
NGC~3652  & SAcd?     &H       & H            &56.4  & 5.92     & $<$35.73 & $-$ & 38.51   & $-$2.00  \\
NGC~3665  & SA0       &H:      & H       &236.8 & 8.76$^{*}$    & 36.85  &  37.67 & 38.28  &   $-$4.73   \\
NGC~3675  & SAb       &T2      & L$^{e}$  &108   & 7.26$^{*}$   & 34.96  &  35.17 & 37.79  &   $-$3.85   \\
NGC~3726  & SBa pec   &H       & H             &41.5  & 5.38    & $<$34.99 & $-$ & 37.80   & $-$2.17  \\
NGC~3877  & SAc       &H       & H             &86.1  & 6.66    & $<$36.47 & $-$ & 37.86   & $-$3.39  \\
NGC~3893  & SABc      &H       & H             &85.3  & 6.64    & $<$34.96 & $-$ & 37.44   & $-$3.79  \\
NGC~3938  & SAc       &H::     & H$^{e,j}$      &29.1  & 4.76    & 35.16  &  35.49 & 37.61  &   $-$1.74   \\
NGC~3949  & SAbc      &H       & H             &82    & 6.57    & $<$35.05 & $-$ & 37.44   & $-$3.72   \\
NGC~4013  & SAb       &T2      & L$^{e,i}$      &86.5  & 6.67    & $<$35.10 & $-$ & 37.36   &  $-$3.90 \\
NGC~4051  & SABbc     &S1.2    & S$^{e,k}$ &89    & 6.10$^{*}$   & 35.29  &  36.92 & 40.17  &   $-$1.14   \\
NGC~4914  & E         &ALG     & ALG           &224.7 & 8.33    & $<$36.17 & $-$ &  $-$   &   $-$  \\
NGC~5005  & SABbc     &L1.9    & L        &172   & 8.27$^{*}$   & 36.29  &  37.54 & 39.41  &   $-$3.05   \\
NGC~5055  & SAbc      &T2      & L$^{l}$  &117 	& 8.92$^{*}$    & $<$34.37 & $-$ & 37.44  &  $-$6.07  \\
NGC~5112  & SBcd      &H       & H	  & $<$60.8& $<$6.05    & $<$35.40 & $-$ & 37.42   &  $-$3.22 \\
NGC~5194  & SAbc pec  &S2      & S          &96       & 6.85    & 35.06  &  36.14 & 38.91  &   $-$2.53   \\
NGC~5195  & IA0 pec   &L2:     & L$^{e}$      &124.8  & 7.31    & 34.90  &  35.59 & 37.84  &   $-$4.06   \\
NGC~5273  & SA0       &S1.5    & S  &71 	& 6.61$^{*}$    & $<$35.31 & $-$ & 39.82   &   $-$1.38 \\
NGC~5297  & SABc      &L2      & L$^{m}$      &61.3   & 6.07    & $<$35.89 & $-$ & 38.22   &  $-$2.44  \\
NGC~5353  & SA0       &L2/T2:  & L             &286.4 & 8.76    & 37.63  &  37.66 & 38.73  &   $-$4.62   \\
NGC~5371  & SABbc     &L2      & L             &179.8 & 7.94    & $<$35.79 & $-$ & 39.03   &   $-$3.50  \\
NGC~5377  & SBa       &L2      & L             &169.7 & 7.84    & 35.71  &  36.35 & 38.81  &   $-$3.62   \\
NGC~5383  & SBb pec   &H       & H	       &96.5  & 6.86    & $<$35.82 & $-$ & 38.07   &  $-$3.38   \\
NGC~5395  & SAb pec   &S2/L2   & L$^{i}$       &145.5 & 7.57    & $<$35.96 & $-$ & 38.66   &   $-$3.50   \\
NGC~5448  & SABa      &L2      & L             &124.5 & 7.30    &  35.73  &  36.46 & 38.55  &   $-$3.34   \\
NGC~5523  & SAcd      &H       & H             &30.1  & 4.82    & $<$35.29 & $-$ & 37.25   &  $-$2.16  \\
NGC~5557  & E1        &ALG     & ALG	       &295.3 & 8.81    & $<$35.94 & $-$ &  $-$   &    $-$   \\
NGC~5660  & SABc      &H       & H              &60.7 & 6.05    & $<$35.76 & $-$ & 38.10   &  $-$2.54   \\
NGC~5656  & SAab      &T2::    & L$^{i,n}$      &116.7 & 7.19    & $<$35.92 & $-$ & 37.99   &  $-$3.79   \\
NGC~5676  & SAbc      &H       & H	       &116.7 & 7.19    & $<$35.74 & $-$ &  37.96  &  $-$3.82  \\
NGC~5866  & SA0       &T2      & L             &169.1 & 7.84    & 36.50  &  36.76 & 37.50  &   $-$4.93   \\
NGC~5879  & SAbc      &T2/L2   & L       &73.9  & 6.62$^{*}$     & 35.27  &  35.39 & 37.89  &   $-$3.09   \\
NGC~5905  & SBb       &H       & H 	      & 174.6 & 7.89    & $<$35.90 & $-$ & 39.03    &  $-$3.45    \\
NGC~5907  & SAc       &H:      & H	       &120.2 & 7.24    &  $<$35.03 & $-$ & 36.88   &  $-$4.94  \\
NGC~5982  & E3        &L2::    & ALG           &239.4 & 8.44    & $<$35.80 & $-$ & $<$38.55  &  $<$-4.48  \\
NGC~5985  & SABb      &L2      & L              &157.6 & 7.71   &  35.76  &  36.30 & 38.76  &   $-$3.54   \\
NGC~6015  & SAcd      &H       & H              &43.5  & 5.47   &  $<$35.04 & $-$ & 37.18   &  $-$2.88  \\
NGC~6140  & SBcd pec  &H       & H	         &49.4  & 5.69  &  $<$35.15 & $-$ & 37.52  &  $-$2.76   \\
NGC~6702  & E         &L2::    & ALG	        &173.6 & 7.88   & 36.19  &  36.85 &   $-$  &    $-$    \\  
NGC~6703  & SA0       &L2::    & L              &179.9 & 7.95   & 35.78  &  35.91 & 38.46  &   $-$4.08   \\
NGC~6946  & SABcd     &H       & H              &55.8  & 5.90   & 34.43  &  35.73 & 37.03  &   $-$3.46   \\
NGC~6951  &SABbc     &S2       & L$^{o}$   &127.8 & 6.93$^{*}$   & 35.42  &  36.02 & 38.69  &   $-$3.25   \\
NGC~7217  & SAab      &L2      & L              &141.4 & 7.52   & 35.09  &  35.87 & 38.31  &   $-$3.80   \\
NGC~7331  & SAb       &T2      & L$^{p}$   &137.2 & 8.02$^{*}$   &  $<$34.95 & $-$ &  38.30 &  $-$3.76  \\
NGC~7332  & SA0 pec   &ALG     & ALG	   &124.1 & 7.08$^{*}$  &  $<$35.18 & $-$ &  $-$   &    $-$  \\
NGC~7457  & SA0?      &ALG     & ALG      &69.4	  & 6.95$^{*}$   & $<$34.88 & $-$ &  $-$   &     $-$  \\
NGC~7640  & SBc       &H       & H               &48.1 & 5.64   & $<$34.51 & $-$ & 36.84  &  $-$3.39 \\
NGC~7741  & SBcd      &H       & H               &29.4 & 4.78   & $<$34.7 & $-$ & 37.91   &   $-$1.46   \\
NGC~7798  & S         &H       & H               &75.1 & 6.42   & 35.64  &  36.54 & 38.64  &   $-$2.37   \\
%NGC~5548  & 14 17 59.513 &  +25 08 12.45 &09 &  J1419+2706 & & 67 	& SA0/a        &SEYFERT   &291 	& 8.78  \\
%NGC~2685  & 08 55 34.750 &  +58 44 03.87 &25 &  J0930+7420 &  {c} 	& 16.2 	& SB0 pec      &SEYFERT    &93.8 	& 6.81  \\
\hline
\hline
\end{longtable}
\end{center}
\vspace*{-1.1cm}\fontsize{8}{7}\selectfont Column description: (1)
source name; (2) morphological galaxy type given from RC3
\citep{devaucouleurs91}; (3) optical spectroscopic classification
based on \citet{ho97a}: H=H{\sc ii}, S=Seyfert, L=LINER, T=Transition
object, and ALG=Absorption line galaxy. The number attached to the
class letter designates the AGN type (between 1 and 2); quality
ratings are given by ':' and '::' for uncertain and highly uncertain
classification, respectively. Two classifiations are given for some
ambiguous cases, where the first entry corresponds to the adopted
choice; (4) optical spectroscopic classification based on BPT diagrams
and from the literature. See the notes for the classification based on
the literature; (5) stellar velocity dispersion (\kms) from
\citet{ho09}; (6) logarithm of the SMBH mass (M$_{\odot}$) determined
based on stellar velocity dispersions \citep{tremaine02} and the
galaxies marked with $^{*}$ have direct BH mass measurements from
\citet{vanderbosch16}. ; (7)-(8) radio core and total luminosities
(erg s$^{-1}$); (9) [O~III] luminosities from \citet{ho97a}; (10)
Eddington ratio, i.e. ratio between the bolometric and Eddington luminosity.
\\
Notes: letters correspond to following publications: $a$
\citet{lira07}; $b$ \citet{annibali10}; $c$ \citet{gavazzi13}; $d$
\citet{cidfernandes05}; $e$ \citet{goulding09}; $f$
\citet{moustakas06}; $g$ \citet{trippe10}; $h$ \citet{cappellari01};
$i$ SDSS; $j$ \citet{dale06}; $k$ \citet{balmaverde14}; $l$
\citet{balmaverde13}; $m$ \citet{rampazzo95}; $n$ \citet{leech89}; $o$
\citet{perez00}; $p$ \citet{keel83}, \citet{bower92}; $q$
\citet{falco99}, \citet{sarzi06}.  \twocolumn \normalfont\normalsize

\noindent
radio sources up to 70 Mpc; only two targets are more distant and are
not detected. Table~\ref{optical} collects the radio luminosities of
the sample. Fig.~\ref{distance}, lower panel, displays the core
luminosity-distance curve for our sample. The identified sources are
largely above the luminosity curve, while the undetected and
unidentified sources straddle the curve.

The core luminosities range between 10$^{32}$ to 10$^{40}$ erg
s$^{-1}$, with a median value of 5.8$\times$10$^{35}$ erg s$^{-1}$.
The sources in our sample extend to lower radio luminosities, on
average by a factor of 10 more than in previous surveys of the Palomar
sample, 10$^{35}$ erg s$^{-1}$ \citep{nagar02,filho06}, within
100-1000 times higher than the luminosity of Sgr~A*. The total radio
powers for our sample estimated from the low-resolution radio images
(median value of 2.2$\times$10$^{36}$ erg s$^{-1}$,
Table~\ref{optical}) cover the same range of core luminosities. The
radio core typically contributes $\sim$26 per cent of the total radio
emission.

%For Detected Galaxies, the medians are:
%Core Lum:                5.790000e+35
%Total Lum: 2.242640e+36
%For HII galaxies this is (core/total lum):
%3.225000e+35 and 1.343300e+36
%For (21 detected) LINERs (core/total lum):
%6.020000e+35 and 2.242640e+36
%For Seyferts (3):
%1.970000e+35 and 8.248750e+36 respectively for core and total lum
%For ALGs (5 detected):
%core lum: 1.560000e+36
%total lum: 2.587660e+36

We calculated the brightness temperatures for the detected targets
using the eMERLIN beam size and the peak flux densities. At the
typical observed unresolved scale of 0.2 arcsec, the brightness
temperatures of radio cores of $\sim$1mJy are of the order of 10$^{4}$
K, below the typical threshold of $\sim$10$^{6}$ K to discriminate
between an AGN and stellar origin \citep{falcke00}. However, the
calculated values do not preclude an active SMBH origin. Since the
brightness temperatures depend on the flux densities and the sizes, at
eMERLIN resolution and sensitivity only sources brighter than 5 mJy
beam$^{-1}$ can yield brightness temperatures greater than 10$^{6}$ K
and, thus, be identified with synchrotron emission from relativistic
jets. Only six sources exceed this core flux density limit and they
are associated with all types of radio morphologies observed.

%0.2 mJy 200mas -> 2708.33 K
%       180 mas  -> 3343.62
%100mJy            1.67181e+06 K
%100mJy  90 mas ->  6.68724e+06 K

\section{Optical data}
\label{optical-radio}

In this Section, we introduce the optical classifications for the AGN
and galaxies for our sample and we compare the different classes based
on their radio properties.

\begin{figure*}
	\includegraphics[width=0.4\textwidth,angle=90]{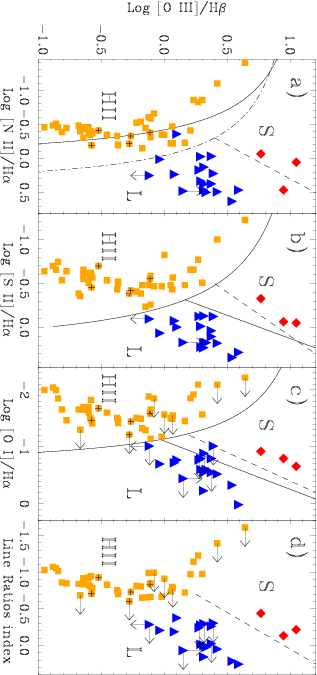}
        \caption{Diagnostic spectroscopic (BPT) diagrams for the 66
          galaxies classified in this paper based only on the data
          from \citet{ho97a} (the remaining sources are classified
          based on data from recent literature, see column 4 in
          Table~\ref{optical}): log([O~III]/H$\beta$) vs. a)
          log([N~II]/H$\alpha$), b) log([S~II]/H$\alpha$), c)
          vs. log([O~I]/H$\alpha$), and d) line ratios index (see
          definition in Section~\ref{optclass}).  In the first three
          panels, the solid lines separating star-forming galaxies,
          LINER, and Seyfert galaxies are from \citet{kewley06}. The
          dashed lines between Seyferts and LINERs in the four panels
          are introduced by \citet{buttiglione10}. In the diagram $a$,
          the sources between the solid and the dot-dashed lines
          (based on starburst models, \citep{kewley01,kauffmann03c}
          were classified as Transition galaxies by \citet{ho97a}, which we re-classify
          as LINER or H{\sc ii} galaxies based on the other diagrams.
          We mark LINERs as blue triangles, Seyferts as red diamonds,
          and H{\sc ii} galaxies as orange squares (with an additional
          plus for the jetted H{\sc ii}).}
    \label{bpt}
\end{figure*}

\subsection{Revised optical classification}
\label{optclass}

The optical nuclear spectra contain several narrow and broad emission
lines which diagnose the nature of the nuclear
emission. \citet{baldwin81} introduced a series of widely-used
diagnostic diagrams (BPT diagrams) to classify objects based on their
narrow emission line intensity ratios. In these diagrams, nebulae
photoionised by hot, young stars (H{\sc ii} regions) can be
distinguished from those photoionised by a harder radiation field,
such as that of an AGN. A suitable choice of emission line ratio
further delineates the excitation sequence in H{\sc ii} regions and
divides AGN into a high-excitation class (Seyferts) and a
low-excitation class (low-ionization nuclear emission-line regions,
LINERs, \citealt{heckman80}).

\citet{ho97a} detected emission lines for most of the Palomar sample,
which indicate the presence of active SMBHs and SF regions. They used
spectroscopic criteria similar to \citet{veilleux87} to separate
active nuclei from SF galaxies as they both used line ratios
insensitive to the reddening, [O~III]/H$\beta$, [N~II]/H$\alpha$,
[S~II]/H$\alpha$, [O~I]/H$\alpha$. We revise their classifications by
using the state-of-the-art spectroscopic diagnostic diagrams based on
the criteria introduced by \citet{kewley06} using the SDSS
emission-line galaxies (using BPT diagrams, mostly including
radio-quiet AGN) and by \citet{buttiglione10} using optical spectra
for the 3C radio galaxies (only radio-loud AGN) with the Telescopio
Nazionale Galileo.  The Kewley et al. and Buttiglione et al. schemes
marginally differ in the separation of LINERs and Seyferts from those
used by \citet{ho97a}: the low [O~III]/H$\beta$ Seyferts are now
reclassified as LINERs. Furthermore, we also estimate for all
narrow-lined objects the average of the low ionization lines ratios,
i.e.  1/3 (log([N~II]/H$\alpha$)+ log([S~II]/H$\alpha$)+
log([O~I]/H$\alpha$)) defined as the `line ratios index' by
\citet{buttiglione10}. This index appears to be more stable than the
other single line ratios as it is an averaged line ratio, and
represents a further diagnostic for a more robust separation between
LINERs and Seyferts.

For our sample of 103 sources, 66 are detected (i.e. having line
uncertainties smaller than 50 per cent in \citet{ho97a}) in at least
four emission lines, which should be sufficient for a secure
classification in the BPT diagrams (Table~\ref{optical}): these
galaxies are displayed in Fig.~\ref{bpt}. Of the remaining 37 sources,
23 are classified based on recent data from the literature (see notes
in Table~\ref{optical}). The classification of 12 galaxies, all ALGs,
remains unchanged from \citet{ho97a}, because we do not find further
spectra available in the literature with detectable emission lines. In
addition, we reclassify the LINERs, NGC~5982 and NGC~6702, as ALGs,
because they do not show any significant emission lines.

We choose to remove the class named `Transition galaxies' since it is
a composite class of H{\sc ii} and AGN objects, based only on one
diagnostic panel ([O~III]/H$\beta$ vs [N~II]/H$\alpha$ in the BPT
diagrams). Therefore the sources which fall in the `transition' region
in this panel (see Fig.~\ref{bpt}), are then classified as LINER or
H{\sc ii} depending on where they fall on the other diagnostic
diagrams. Finally, each object is classified as H{\sc ii}, LINER, or
Seyfert based on at least two diagnostic diagrams in case the third
one disagrees with the previous two.

The revised diagnostic diagrams mostly affect the number of
Seyferts: four Seyferts and two objects tentatively classified as
Seyfert/LINERs according to \citet{ho97a} are now classified as
LINERs (NGC~185, NGC~777, NGC~2639, NGC~2655, NGC~5395, and NGC~6951)
because of a low [O~III]/H$\beta$ as explained previously. The two
sources which show optical broad lines remain classified as type-1
Seyferts (NGC~4051 and NGC~5273). In addition, only one H{\sc ii} galaxy is
now considered a LINER (NGC~3319).

After the revised optical classification, the sample discussed in this
paper consists of 52 H{\sc ii} galaxies, 33 LINERs, 14 ALGs and 4
Seyferts. As we removed the Transition galaxies from the sample and
reclassified them as LINERs or H{\sc ii} regions, the comparison with the
full Palomar sample classified by \citet{ho97a} is not
straightforward.  LINERs and H{\sc ii} galaxies now account for
$\sim$45-50 per cent of the sample. Only Seyferts appear to be less
represented than in the original Palomar sample.

%In this paper we present the radio observation for 280 targets limited
%within 110 Mpc, randomly selected from the parental sample. 

When considering the galaxy morphological types, most of the sources
are late-type galaxies (from Sa to Sd, $\sim$75 per cent), with a
smaller fraction of elliptical and lenticulars (E and S0). Using the
stellar velocity dispersions measured from the optical spectra by
\citet{ho09} (see Table~\ref{optical}) and the empirical relation from
\citet{tremaine02}, we inferred their BH masses, which range between
10$^{4}$ and 10$^{9}$ M$_{\odot}$. The most massive BHs are, not
surprisingly, found in the elliptical galaxies, that host LINERs and
ALGs. We also derive the BH masses using the M$_{\rm BH}-\sigma$
relations from \citet{graham13} and \citet{kormendy13}. The different
relations lead to M$_{\rm BH}$ measurements\footnote{The median
  M$_{\rm BH}$ and standard deviations of the M$_{\rm BH}$
  distribution for our sample using the three M$_{\rm BH}-\sigma$
  relations are (in logarithmic scale) 6.85, 6.46, and 7.09
  M$_{\odot}$ and 1.18, 1.62, 1.29 M$_{\odot}$ for \citet{tremaine02},
  \citet{graham13} and \citet{kormendy13} respectively. The
  \citeauthor{tremaine02} correlation produces the smallest M$_{\rm
    BH}$ scatter for our sample.}, which agree with each other for
intermediate BH masses, $\sim$10$^{6}$-10$^{7}$ M$_{\odot}$, but
differ at high BH masses within 0.5 dex and at low BH masses within 1
dex. For 17 sources, we use the direct BH measurements (derived from
stellar and gas dynamics, mega-masers and reverberation mapping)
available from the BH mass compilation of \citet{vanderbosch16}. These
M$_{\rm BH}$ values are consistent with the values obtained using the
M$_{\rm BH}-\sigma$ relation from \citet{tremaine02}, which we then
choose to use for our work. However, the choice of the M$_{\rm
  BH}-\sigma$ relation used for our sample does not significantly
affect the scope of the work.

%The Palomar sample is magnitude limited and probes the local Universe
%up to a distance of 120 Mpc, with a median value of 20 Mpc. The
%galaxies we present in this work are randomly selected from the
%original sample and are similarly limited within a distance of 110 Mpc
%with 10 objects further than 60 Mpc and a median value of 20 Mpc. The
%sample presented here is representative of the entire Palomar sample
%and, then, of the population of nearby galaxies.

%sample: 74 LINER, 35 ALG, 28 Seyfert, 42 Trans, and
%101 H{\sc ii}. In this paper, we present the first release of the radio
%data for 103 galaxies: 33 LINERs, 14 ALG, 4 Seyfert and 52
%HII galaxies,which correspond to a range of 25-40\% of optical
% classes.

%LINER 45%
%Seyfert 14%
%HII 50%

%$2 Irregular
%14 Lenticular
%75 Spirals
%12 Ellipticals

%The Palomar sample is magnitude limited and probles the local Universe
%up to a distance of 120 Mpc, with a median value of 20 Mpc. The sample
%we present in this work is within 110 Mpc with 10 objects further than
%60 Mpc. 

\subsection{Radio vs optical classification}

Table~\ref{fraction} summarises the number of detected, unidentified,
and undetected sources dividing the sample by optical class.  

Although Seyferts have the highest radio detection rate (4/4), their
scarcity in our sample limits our conclusions. LINERs have the second
largest detection rate in the sample: 22 objects out of 34 ($\sim$65
per cent).  Absorption line galaxies have a smaller detection rate
than LINERs: 5 out of 14 galaxies ($\sim$36 per cent). The radio
detection fraction is yet lower for H{\sc ii} galaxies with 16 out of
51 ($\sim$31 per cent). Considering only those sources for which we
identify a radio core, results in the exclusion of six sources (one
LINER, one Seyfert, and four H{\sc ii} galaxies) such that the
detection fractions become: LINERs 21/34, Seyferts 3/4, ALGs 5/14 and
H{\sc ii} galaxies 12/51. 

By dividing the sample into the different radio morphological types, the
sources in the sample spread across all the radio categories
(Table~\ref{fraction}). The radio classes which imply a presence
of a jet (B, C and D) encompass all the optical classes.
LINERs show different radio structures, but mostly core/core--jet and
triple structures. ALGs are mostly core/core--jet, Seyferts show
edge-brightened radio morphologies, and H{\sc ii} galaxies are not
associated with large double-lobed radio morphologies. Although H{\sc ii} galaxies are
powered by SF according to the BPT diagrams, five sources show clear
jetted morphologies: two show one-sided jets, a possible sign of a
Doppler-boosted jet. In fact, one of these two is NGC~3665, which exhibits a
FR~I radio morphology extended over $\sim$3 kpc at the VLA scale
\citep{parma86}. Conversely, the remaining seven H{\sc ii} galaxies do not
show clear evidence of jets (single cores or complex morphologies).

Regarding relations with galaxy type, early-type galaxies (elliptical
and lenticular) are most frequently detected as radio sources
($\sim$50 per cent) and are typically associated with jetted radio
morphologies. Spiral galaxies, which are the most abundant host type
(73 per cent), are detected in radio in one third of cases, and are
associated with all types of radio structures. Only one irregular
galaxy is detected out of two in this sample.

%Detected by Hubble type:
%1/2 Irregular
%8/14 Lenticular
%6/12 Elliptical
%26/75 Spiral

%The detected sources shows radio structures on a typical scale
%$\sim$1\arcsec, which corresponds to a physical region of
%$\sim$100pc. The jet forms we detect range between $\sim$40pc to
%$\sim$500pc, with three cases (NGC~5005, NGC~5866, and NGC~6702) with
%longer jets. 

\begin{table}
	\centering
	\caption{Spectral--radio morphological classification breakdown}
	\label{fraction}
	\begin{tabular}{l|cccc|c} 
		\hline
                                   & \multicolumn{4}{c|}{optical class} \\
\hline
radio class                     &   LINER & ALG & Seyfert  &  H{\sc ii}  &  Tot  \\
		\hline

{\tiny core/core--jet (A)}         &    8     &  4  &    0    &  4     &  16  \\
{\tiny one-sided jet (B)}         &    0     &  0  &    0    &  2     &  2  \\
{\tiny triple (C)}                &    9     &  1  &    1    &  3     &  14 \\
{\tiny doubled-lobed (D)}         &    3     &  0  &    1    &  0     &  4  \\
{\tiny jet+complex (E)}           &    1     &   0 &    1    &  3     &  5 \\
\cmidrule(rl){2-6}
Tot                               &   21     &  5  &    3    &  12     &  41  \\
\hline	
unidentified                      &    1     &  0  &   1     &   4   & 6 \\
undetected                        &    12    &  9  &   0     &   35   &  56 \\      
	\hline
Tot                               &   34     &14  &    4     &  51    &  103  \\	
\hline
\end{tabular}
\begin{flushleft}
  Notes. The sample is divided in radio and optical classes based on
  their radio detection, core-identification or non-detection.
\end{flushleft}
\end{table}

%\subsection{Radio brightness}

\begin{figure}
	\includegraphics[width=0.5\textwidth]{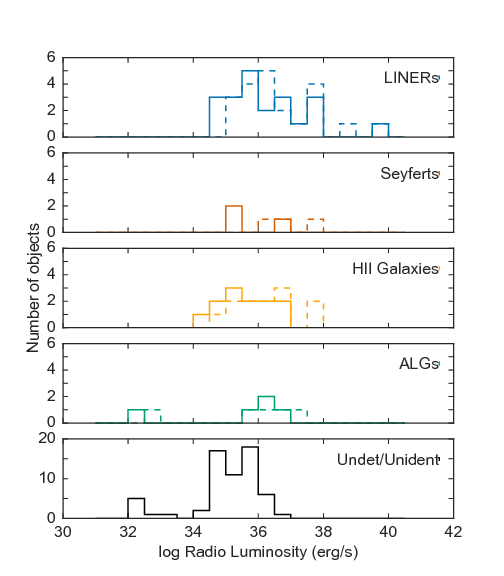}
	\includegraphics[width=0.5\textwidth]{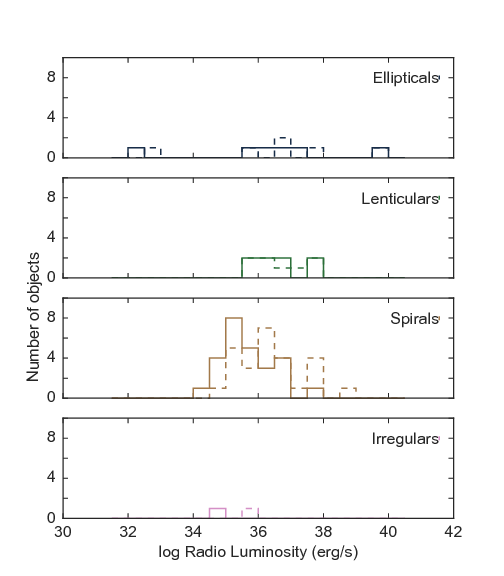}
        \caption{Histograms of the radio luminosity (erg s$^{-1}$) per
          optical class (upper plot) and host morphological type
          (lower plot). The solid-line histogram represents the radio
          core luminosity distribution and the dashed line corresponds
          to the total radio luminosity distribution. In the bottom
          panel of the upper plot, we also show the $3\,\sigma$ upper
          limit radio luminosity distribution for the undetected and
          unidentified sources.}
    \label{histo}
\end{figure}

\begin{figure*}
	\includegraphics[width=0.7\textwidth]{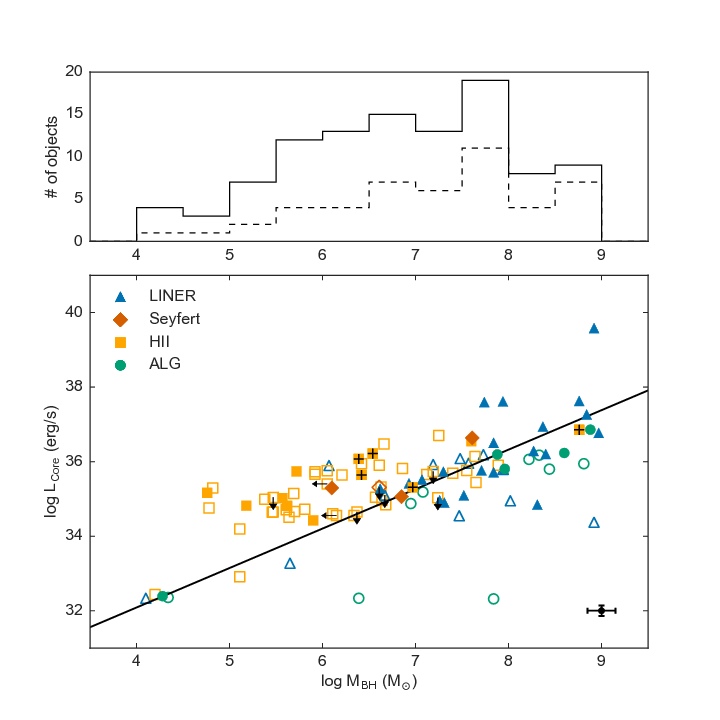}
        \caption{In the upper plot, we show the histograms of the
          entire sample (solid line) and of the detected/identified
          sources (dashed line) in bins of BH mass to estimate the
          detection fraction.  In the lower panel, we show the radio
          core luminosities (L$_{\rm core}$ in erg s$^{-1}$) as a
          function of the BH masses (M$_{\odot}$) for the sample,
          divided per optical class (symbol and color coded as in the
          legend). The jetted H{\sc ii} galaxies show an additional
          plus symbol. The filled symbols refer to the detected radio
          sources, while the empty symbols refer to undetected radio
          sources. The unidentified sources are the empty symbols with
          radio upper limits. The solid line represents the linear
          correlation found for all the jetted or active galaxies
          (corresponding to M$_{\rm BH}$ $\gtrsim$ 10$^{6.6}$
          M$_{\odot}$). In the bottom-right corner, we show the typical
          error bars associated with the points.}
\label{core_mbh}
\end{figure*}

\begin{figure}
        \includegraphics[width=0.489\textwidth]{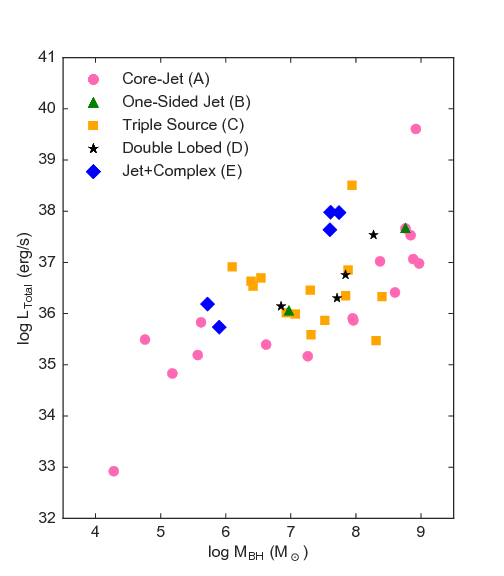}
        %\vspace{-0.7cm}
        \caption{The total radio luminosity (L$_{\rm tot}$ in erg
          s$^{-1}$) as a function of the BH masses
          (M$_{\odot}$) for the sample, divided per radio
          morphological class A, B, C, D, and E (symbol and color
          coded according to the legend).}
    \label{tot_mbh}
\end{figure}

%In terms of radio brightness, the active galaxies (LINERs and
%Seyferts) show the brightest radio cores, while most of the H{\sc ii}
%galaxies are on the edge of the detection limit curve. ALGs are
%associated with faint nuclei. 

Figure~\ref{histo} shows the luminosity distribution of the radio core
and total emission for our sample for each optical class. The LINERs
produce the largest radio core luminosities with a median value
of 6.0$\times$10$^{35}$ erg s$^{-1}$. ALGs show a slightly higher median
value (1.6$\times$10$^{36}$ erg s$^{-1}$), but are also among the
weakest radio sources. Seyferts and H{\sc ii} galaxies show the lowest
core power, 2.0$\times$10$^{35}$ and 3.2$\times$10$^{35}$ erg
s$^{-1}$, respectively. The undetected and unidentified galaxies have
radio luminosities ranging between 10$^{32}$ to 10$^{37}$ erg
s$^{-1}$, with a median radio luminosity of 1.7$\times$10$^{35}$ erg
s$^{-1}$.

The total radio luminosities of the sample generally cover the same
range of values as the core powers: LINERs and H{\sc ii} galaxies are
again at the two opposites of the luminosity range. The core
dominance, which represents the contribution of the radio core with
respect to the total flux density of the source, differs for each
optical class. The ALG are the most core dominated ($\sim$60 per
cent), followed by LINERs and H{\sc ii} galaxies with moderate core
dominance ($\sim$27 per cent), while the Seyferts have the smallest
core dominance ($\sim$2.4 per cent).

By dividing our sample by galaxy morphological type, early-type hosts
produce the highest and lowest radio luminosities, because they mostly
include LINERs and ALGs (Fig.~\ref{histo} lower plot). The most
numerous type of galaxies in our sample, spirals, spread across the
entire distribution of radio luminosities. The only irregular galaxy
detected in the radio band (NGC~5195) has low core and total
luminosities of 7.9$\times$10$^{34}$ and 3.9$\times$10$^{35}$ erg
s$^{-1}$, respectively.

%For Detected Galaxies, the medians are:
%Core Lum:                5.790000e+35
%Total Lum: 2.242640e+36
%For HII galaxies this is (core/total lum):
%3.225000e+35 and 1.343300e+36
%For (21 detected) LINERs (core/total lum):
%6.020000e+35 and 2.242640e+36
%For Seyferts (3):
%1.970000e+35 and 8.248750e+36 respectively for core and total lum
%For ALGs (5 detected):
%core lum: 1.560000e+36
%total lum: 2.587660e+36

For the brightness temperatures, the values mostly reflect the flux
density distribution. In fact, LINERs show the highest values, with
six sources exceeding 10$^{6}$ K, which can be thus associated with
synchrotron emission from relativistic jets. Seyferts, ALGs and H{\sc
  ii} have, respectively, lower brightness temperatures, down to
10$^{3}$ K. Therefore, we cannot use the brightness temperature as a
diagnostic of the presence of an AGN for the vast majority of the
sample.

%0.2 mJy 200mas -> 2708.33 K
%       180 mas  -> 3343.62
%100mJy            1.67181e+06 K
%100mJy  90 mas ->  6.68724e+06 K

\subsection{Radio properties vs BH mass}

As the BPT diagrams ascribe the ionization of the emission lines for
Seyferts and LINERs to their central AGN, their radio cores are
expected to be attributed to active SMBHs. Conversely, caution is
needed to confirm the AGN origin of the identified cores of H{\sc ii}
galaxies, since small star-forming clusters and weak AGN can both
account for the radio emission in regions of size 100 pc
\citep{condon82,varenius14,salak16}. Instead, ALGs might exhibit radio
activity from their SMBHs, while they are undetected in the optical
band.

Two possible diagnostics to identify active SMBHs are the BH mass,
M$_{\rm BH}$, and the radio luminosity (listed in
Table~\ref{optical}). The former can be used as an indicator of
nuclear activity, because active nuclei are preferentially associated
with massive BHs (e.g., \citealt{best05b,gallo10}). The latter roughly
assesses the probability of being radio-jet dominated: a sign of an
active SMBH. Furthermore, the two quantities are typically connected
in active nuclei: radio-emitting AGN tend to be radio louder as the BH
becomes more massive (e.g. \citealt{best05b}). The histogram in
Figure~\ref{core_mbh} (upper panel) describes the radio core detection
for our sample as a function of BH mass. It is evident that the
detection fraction increases with BH mass. For M$_{\rm BH} >$10$^{7}$
M$_{\odot}$, this fraction is $\sim$50 per cent reaching 85 per cent
in the last bin, $\sim$10$^{8}-$10$^{9}$ M$_{\odot}$, while it drops
to less than 40 per cent below 10$^{6}$ M$_{\odot}$.

Figure~\ref{core_mbh} (lower panel) also shows the distribution of
core luminosities as a function of BH mass. Two different situations
appear in this plot, roughly separated at 10$^{6}$-10$^{7}$
M$_{\odot}$: a positive radio-M$_{\rm BH}$ sequence appears for
M$_{\rm BH}$$\gtrsim$10$^{6.6}$ M$_{\odot}$, where all jetted or
active galaxies are included, while at lower BH masses a luminosity
plateau emerges. Despite the large scatter, we note that the M$_{\rm
  BH}$-L$_{\rm core}$ sequence includes LINERs and Seyferts, which
generally tend to have larger BH masses in local galaxies
\citep{kewley06,best05a} and clearly have an active nucleus as they
are detected in the emission line diagrams. Note that the ALGs, which
are not detected in emission lines, are also present along this
sequence, whilst H{\sc ii} galaxies are mostly at lower BH masses
where the flattening of the sequence is evident. Therefore, we fit the
data points present in this sequence with a linear relation and we
find a correlation in the form L$_{\rm core}$$\sim$M$_{\rm BH}$$^{1.0
  \pm 0.2}$ with a Pearson correlation coefficient (r-value) of 0.663
which indicates that the two quantities do not correlate with a
probability smaller than 6.5$\times$10$^{-5}$. Conversely, if we
search for a correlation including all the detected radio sources on
the entire range of BH masses, the correlation is drastically less
significant (r-value is 0.525 with a probability of
4.2$\times$10$^{-4}$ of being a fortuitous correlation) due to the
radio-M$_{\rm BH}$ break at $\sim$10$^{6.6}$ M$_{\odot}$.

In general, LINERs, Seyferts, and ALGs appear to be on the
correlation, suggesting that the radio core emission is probably
dominated by the AGN component. For H{\sc ii} galaxies, those which
display a jetted radio morphology, all with M$_{\rm BH}$>10$^{6}$
M$_{\odot}$, lie at the low end of the relation. Conversely, the
non-jetted H{\sc ii} galaxies fall on the flat part of the M$_{\rm
  BH}$-L$_{\rm core}$ relation. This different behaviour between the
jetted non-jetted H{\sc ii} galaxies suggests an AGN dominance for the
former and a SF dominance for the latter group.

Considering the radio morphological classes, all structures are
represented along the sequence (Fig~\ref{tot_mbh}). The
core/core--jet morphology spans the entire range of BH masses and
total radio luminosities, dominating the distribution below 10$^{6}$
M$_{\odot}$ and above 10$^{8}$ M$_{\odot}$. The remaining
morphological radio types are found mainly between 10$^{6}$ and
10$^{8}$ M$_{\odot}$. But, overall, there is no strong tendency for
specific radio morphologies to be associated with particular radio
luminosities.

%in realta sotto 10$^{39}$ non so se una liner e' un AGN

%Nagar02 vede la stessa cosa Mbh vs core

%galaxies masses?

\subsection{Radio properties and [O~III] luminosity}

[O~III] is a forbidden emission line, coming from the narrow-line
region extending from some hundreds of pc to kpc scale. Although it is
marginally dependent on orientation and obscuration
(e.g. \citealt{risaliti11,baldi13a,dicken14,bisogni17}), [O~III] line
luminosity, L$_{\rm [O~III]}$, is a good indicator of the bolometric
AGN luminosity (L$_{\rm bol}$=3500$\times$L$_{\rm [O~III]}$,
\citealt{heckman04}).

\begin{table}
	\centering
	\caption{FR~Is observed by MERLIN from literature.}
	\label{fr1}
	\begin{tabular}{l|cccc|c} 
		\hline\hline
name      &   z &  F$_{core}$ & L$_{core}$ & L$_{\rm [O~III]}$   & Ref\\
          &     & mJy         & erg s$^{-1}$ & erg s$^{-1}$ &  \\
		\hline
3C~66B  & 0.021258   & 180 &39.40 &  40.05  & FR97\\
3C~264  &  0.021718  & 170 &39.39 &   39.20 & BA97 \\
3C~78   &  0.028653  & 700 &40.25 &   39.41 & UN84\\
3C~338  &  0.030354  & 163 &39.67 &   39.57 & GI98\\
%3C~293  &   0.045034 & 1820& 41.07&   39.80  &  BE92\\
3C~274  &  0.004283  & 530 & 38.45&    38.99 & GO89\\
3C~189  &  0.042836  &  512&40.48 &    39.94 & BO00\\

\hline
\end{tabular}
\begin{flushleft}
  Column description. (1) name; (2) redshift; (3)-(4) radio core flux
  density and luminosity from MERLIN observations from literature data; (5) [O~III]
  luminosity from \citet{buttiglione10}; (6)  MERLIN radio references: FR97
  \citet{fraix97}; BA97 \citet{baum97}; UN84 \citet{unger84}; GI98
  \citet{giovannini98}; GO89 \citet{gonzalez89}; BO00 \citet{bondi00}.
\end{flushleft}
\end{table}

\begin{figure*}
	\includegraphics[width=0.645\textwidth,angle=90]{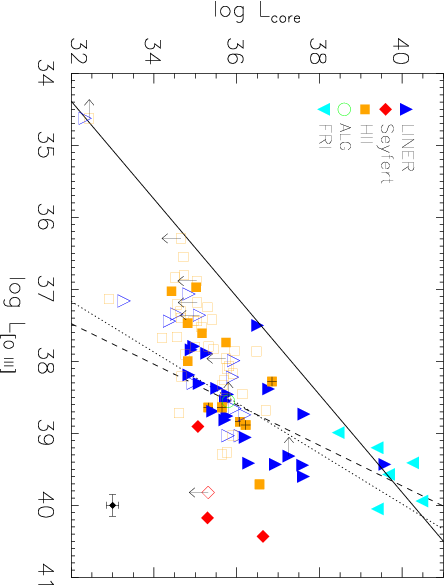}
        \caption{[O~III] luminosity ( L$_{\rm [O~III]}$ in erg
          s$^{-1}$) vs radio core luminosities (L$_{\rm core}$ in erg
          s$^{-1}$) for the LeMMINGs sample.  The different optical
          classes are coded (symbol and color) in the plot according
          to the legend. The jetted H{\sc ii} galaxies show an
          additional plus symbol. The filled symbols refer to the
          detected radio sources, while the empty symbols refer to
          undetected radio sources. The unidentified sources are the
          empty symbols with radio upper limits.  The FR~I radio
          galaxies are the filled cyan downward triangles. The dotted
          line represents the linear correlation by fitting only the
          LINERs, while the dashed line represents the fit by
          including the FR~I radio galaxies observed with MERLIN. The
          solid line correspond to the [O~III]--radio linear fit valid
          for FR~I radio galaxies using VLA radio data at 5 GHz from
          \citet{baldi15}. The bars in the bottom-right corner
          indicate the typical errors for the data points.}
    \label{coreo3}
\end{figure*}

\begin{figure}
	\includegraphics[width=0.8\columnwidth,angle=90]{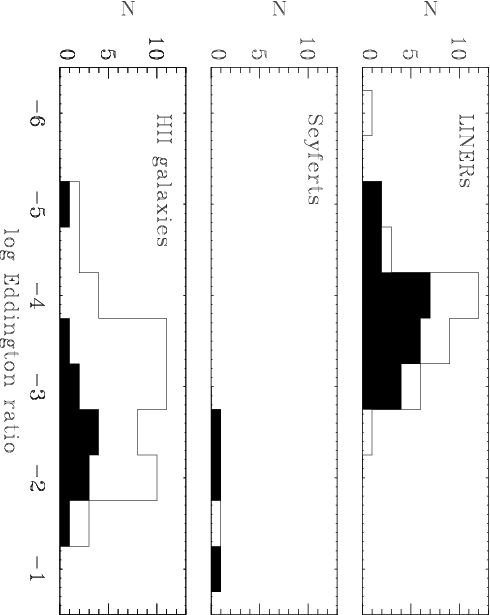}
        \caption{The Eddington ratio (ratio between the bolometric and
          Eddington luminosity) for the sources detected in [O~III]:
          LINERs, Seyferts and H{\sc ii} galaxies. The distribution of
          the identified sources is shown as a filled histogram.}
    \label{eddratio}
\end{figure}

Figure~\ref{coreo3} shows the radio core luminosity as a function of
the [O~III] luminosity (values taken from \citealt{ho97a}) for the
Palomar sample. The three optical classes generally cluster at
different L$_{\rm [O~III]}$. Seyferts have the highest [O~III]
luminosities, (mean value 6.9$\times$10$^{39}$ erg s$^{-1}$ with a
deviation standard of 0.80), while the H{\sc ii} regions lie on the
opposite side of the L$_{\rm [O~III]}$ range (mean value
9.1$\times$10$^{37}$ erg s$^{-1}$ and a deviation standard of
0.83). LINERs bridge the gap between the other two classes with
intermediate L$_{\rm [O~III]}$ (mean value 4.6$\times$10$^{38}$ erg
s$^{-1}$ and a standard deviation of 0.67). In addition, we note that
LINERs scatter along a specific [O~III]-radio trend. Therefore, we
test the hypothesis of a linear correlation for this class. We found a
tentative correlation in the form L$_{\rm [O~III]}$ $\propto$ L$_{\rm
  core}$$^{0.35\pm0.20}$.  The corresponding Pearson correlation
coefficient (r-value) is 0.672 which indicates that the two
luminosities for LINERs do not correlate with a probability of
8$\times$10$^{-4}$. The galaxies with a core/core--jet morphology or a
triple structure tend to be on the correlation, since they are mostly
classified as LINERs. Furthermore, most of the elliptical and
lenticular galaxies (unbarred galaxies in general) follow the
correlation.

We test the possibility of a linear correlation which might encompass
the LINERs in our sample with low-luminosity radio-loud AGN with a
LINER optical spectrum, FR~I radio galaxies. This hypothesis is
supported by the idea of a common central engine, an ADAF disc coupled
with a jet, typically attributed to LINERs
\citep{falcke04,nemmen14}. Hence, we collect all the 1.5-GHz MERLIN
observations of FR~I radio galaxies from the literature at z$<$0.05
which have an angular resolution similar to our eMERLIN
observations. We find six sources, which we list in Table~\ref{fr1}.
Figure~\ref{coreo3} shows that FR~I radio galaxies extend the LINER
correlation at higher luminosities. Although FR~Is are more powerful
than the LINERs in our sample by a factor 10$-$10$^{3}$ in radio, they
are still in the low-luminosity regime (L$_{\rm H\alpha}<$10$^{40}$
erg s$^{-1}$). The linear correlation (L$_{\rm [O~III]}$ $\propto$
L$_{\rm core}$$^{0.29\pm0.15}$) has a r-value of 0.766 with a
probability of 3$\times$10$^{-6}$ to be fortuitous. This correlation
is consistent within the errors with the one found only for the LINERs
of the sample.

Seyfert galaxies typically lie below the correlation found for LINERs
and are up to a factor of 100 more luminous in [O~III]. Conversely,
the jetted H{\sc ii} galaxies lie on the correlation, while the
non-jetted ones (single core component or with a complex radio
morphology) fall above: less luminous in [O~III] by a factor of 30 and
more luminous in radio by some orders of magnitudes because of the
steepness of the [O~III]-radio relation. The only exception is an H{\sc
  ii} galaxy with a complex radio morphology, NGC~2342, which lies
above the correlation found for LINERs. Furthermore, we include in
Figure~\ref{coreo3}, NGC~5982, which is the only ALG of the sample, for
which an upper limit on the [O~III] luminosity is available from
\citet{ho97a}.

Using the bolometric and Eddington luminosities, we can also calculate
the Eddington ratio, L$_{\rm bol}$/L$_{\rm Edd}$, an indicator of the
accretion rate onto the SMBH (listed in Table~\ref{optical}).  We
derive the values for the detected radio sources of our sample and
present them in a histogram in Fig.~\ref{eddratio}. Assuming that line
emission is powered by the AGN for H{\sc ii} galaxies, we note that,
although less luminous in line emission, they have intermediate
Eddington ratios between LINERs and Seyferts.  The Eddington ratios of
LINERs are below 10$^{-3}$, while the values of Seyferts are above it,
e.g. the typical threshold used to separate between low- and
high-accretion onto the SMBHs \citep{best12}. The H{\sc ii} galaxy
with the lowest Eddington rate is the FR~I, NGC~3665, consistent with
the other LINERs. We do not find any correlation between the Eddington
ratio and the radio core luminosity for our sample.

\begin{figure*}
        \includegraphics[width=0.63\textwidth,angle=90]{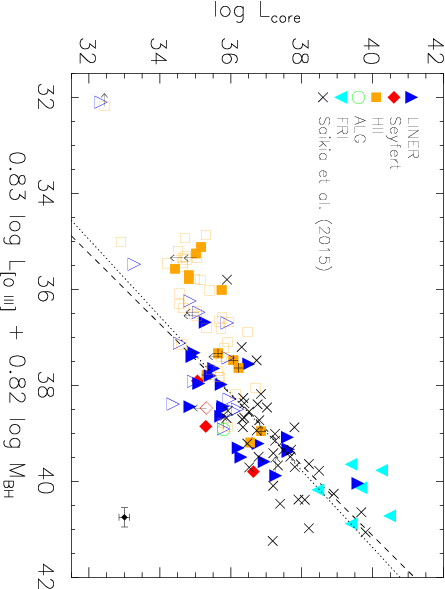}
        \caption{Projection of the fundamental plane of BH activity
          (FPBHA) in the optical band for the LeMMINGs sample,
          i.e. core luminosities vs 0.83$\times$log L$_{\rm [O~III]}$
          + 0.82$\times$log M$_{\rm BH}$ with the luminosities in erg
          s$^{-1}$ and BH masses in M$_{\odot}$. The different classes
          are coded (color and symbol) according to the legend. The
          jetted H{\sc ii} galaxies show an additional plus
          symbol. The filled symbols refer to the detected radio
          sources, while the empty symbols refer to undetected radio
          sources. The unidentified sources are the empty symbols with
          radio upper limits. The FR~I radio galaxies are the filled
          cyan downward triangles. The black crosses are the Palomar
          galaxies used by \citet{saikia15} in the optical FPBHA.  The
          dotted correlation is found for the entire sample, but
          excluding the non-jetted H{\sc ii} galaxies and the dashed
          correlation is obtained for LINERs and FR~Is. The
          correlations are parameterised as expressed by
          \citet{saikia15}. The bars in the bottom-right corner
          indicate the typical uncertainties for the data points.}
    \label{fbhpo3}
\end{figure*}

\subsection{The Fundamental Plane of Black Hole Activity in the
  Optical Band}

An attempt at unifying all active BHs comes from the empirical
fundamental plane of BH activity (FPBHA) relation
\citep{merloni03,falcke04,gultekin09}, which suggests a scale
invariance in accretion and jet production. This plane relates radio
and X-ray luminosities and BH mass for a sample of compact objects:
from galactic BHs to AGN. By using a similar approach,
\citet{saikia15} introduced a new FPBHA using the [O~III] luminosity
as a tracer of the accretion, instead of the X-ray luminosity and
found an analogous correlation for active BHs.

Since BH masses and emission line luminosities are available for our
sample, we plot our data by using the same parameterisation expressed
by \citet{saikia15} for the entire sample (Fig.~\ref{fbhpo3}). The
plane identified by \citet{saikia15} used radio cores, detected at 15
GHz with the VLA by \citet{nagar05} at a resolution ($\sim$0.13 arcsec)
comparable with our observations. This difference in frequency yields
a difference in radio powers (in $\nu$L$_{\nu}$). In fact five sources
in common with our sample have eMERLIN core luminosities, on average,
lower than the VLA counterparts by a factor $\sim$4. This corresponds
to a spectral index between 1.5 and 15 GHz of $\alpha$ $\sim -0.4$
($F_{\nu}$ $\sim\nu^{\alpha}$).

When considering our sample in the optical FPBHA, LINERs and Seyferts
appear to follow the same trend, which extends to FR~I radio
galaxies. The jetted and non-jetted H{\sc ii} galaxies appear in two
different areas of the plot: the former sources cluster where LINERs
lie, while the latter are offset by a factor $\sim$100 because of
their smaller BH masses and lower [O~III] luminosities. Therefore, by
excluding the non-jetted H{\sc ii} galaxies we find another
correlation which extends over four order of magnitudes, by using the
same parameterisation of \citet{saikia15}:

\vspace{-0.5cm}
%$$\log L_{\rm core} = (0.83 \,\ \log L_{\rm [O~III]} + 0.82 \,\, \log M_{\rm BH})(1.2\pm0.2) + (-11.6\pm5.9) $$
$$\log L_{\rm core} = (0.83 \,\ \log L_{\rm [O~III]} + 0.82 \,\, \log M_{\rm BH}) m + q $$

\noindent
where $m$ = 1.25$\pm$0.15 and $q$ = $-$11.6$\pm$5.9, while
\citeauthor{saikia15} found $m$ = 1 and $q$ = -3.08. The units are erg
s$^{-1}$ for the luminosities and M$_{\odot}$ for the BH masses. 
The corresponding Pearson correlation coefficient (r-value) is 0.815 which
indicates that the two quantities do not correlate with a probability
of $<$10$^{-7}$. If we consider only LINERs and FR~Is, we find that
$m$=1.37$\pm$0.17 and $q$= $-$16.6$\pm$6.7. The corresponding r-value
is 0.847 with a probability of $<$10$^{-7}$ of being a fortuitous
correlation. This correlation is still consistent with that found for
the entire sample (excluding the non-jetted H{\sc ii} galaxies) within
the errors.  Furthermore, most of the elliptical and lenticular galaxies
lie on the correlation.

Although radio core luminosity, BH mass and [O~III] luminosity are
found to correlate for LINERs, the scatter in the optical FPBHA is
larger ($\sim$ 1 dex) than those found in the other correlations with
L$_{\rm core}$ (0.6--0.8 dex). However, the FPBHA is not driven by any
stronger dependence of one of the three quantities which establish the
fundamental plane. This confirms the validity of the FPBHA at least
for the LINERs of our sample. The small number of Seyferts and jetted
H{\sc ii} galaxies in our sample limits an analogous analysis
performed for these classes.  For the non-jetted H{\sc ii} galaxies,
which induce the break in the optical FPBHA at low radio powers, we
have verified that assigning a possible core luminosities via the
radio--M$_{\rm BH}$ relation found at larger BH masses, these sources
nicely line up on the fundamental plane. This test further
corroborates our suggestion that the radio emission in the non-jetted
H{\sc ii} galaxies is mostly powered by SF, because otherwise we would
not expect these galaxies to necessarily depart from the optical
FPBHA.

In general, in the optical FPBHA we also note that the three optical
classes slightly scatter in radio luminosities along the correlation,
by decreasing L$_{core}$ in the order: H{\sc ii} galaxies, Seyferts,
and LINERs.

\vspace{-0.6cm}
\section{Discussion}
\label{discussion}

This eMERLIN radio legacy survey represents the deepest radio study of
the Palomar sample to date, reaching rms of
$\sim$70$\mu$Jy. Furthermore, the LeMMINGs survey differs from
previous studies of the Palomar sample, as we have deliberately
observed all the optical classes including the galaxies which are not
classified as AGN based on the emission line ratios, i.e. H{\sc ii}
galaxies and ALGs.  Here we discuss the results of 103 radio sources,
randomly selected from the full LeMMINGs `shallow' sample (280
sources).

%Focusing only in the central galaxy regions (at least on an area of
%0.73 arcmin$^{2}$), the high-resolution ($\sim$150 mas) radio maps of
%the LeMMINGs survey reveal that nearly half of the sample (47/103)
%emit radio emission above 0.2 mJy, which corresponds to a luminosity
%of 10$^{35}$ erg s$^{-1}$ at a median distance of 20 Mpc, on a typical
%scale of $\sim$100 pc. More than half of the sources,
%classified as active based on their emission-line ratios show a
%compact radio source or/and jet structures. Half of the sources which
%are not classified as powered by an AGN based on the BPT diagrams,
%show jetted radio appearances, a possible sign of active SMBHs.

%As we reclassify the galaxies listed as `transition' by \citet{ho97a}
%into either LINERs or H{\sc ii} galaxies based on the BPT diagrams, the
%comparison with previous radio studies is non-trivial. 

The total detection rate (47/103, $\sim$46 per cent) is consistent
with previous VLA/VLBA observations of the Palomar sample (see
\citealt{ho08} for a review), which mostly focused on LINERs and
Seyferts which are typically the brightest radio sources. For 40 per
cent (41/103) of the sample the radio emission can be associated with
the optical galaxy centre, suggesting a relation with the active SMBH.

%Half of the detected sources show clear jetted morphologies (triple
%sources, twin jets, double lobed, one-sided jets, 20/41) and five
%sources show a complex structure which might hide a possible jet and
%diffuse SF. The most frequent radio class observed in our sample is
%core/core--jet morphology (16/41). 

More than half of the sources, classified as active based on their
emission-line ratios show a compact radio source and/or jet
structures. Half of the sources which are not classified as powered by
an AGN based on the BPT diagrams, show jetted radio appearances, a
possible sign of active SMBHs.  Half of the detected sources show
clear jetted morphologies (triple sources, twin jets, double lobed,
one-sided jets, 20/41) and five sources show a complex structure which
might hide a possible jet and diffuse SF. 

The most frequent radio class observed in our sample is the
core/core--jet morphology (16/41). The nature of these unresolved
radio components, whether an unresolved radio jet base or a SF nuclear
region on the scale of $<$100 pc, can be revealed by using further
diagnostics such as [O~III] luminosities and BH masses in absence of
clear jet-like structures.  We will discuss the core origin and the
results of each optical class in the following sub-sections.

\subsection{LINERs}

LINERs show the highest number of radio detections in the Palomar
sample and are hosted by elliptical, lenticular and spiral galaxies.
The detection rate (22/34, 64 per cent) is slightly higher than a
previous radio survey of the Palomar sample ($\sim$44 per cent,
\citealt{ho08}). The eMERLIN observations reveal that LINERs are
associated with single radio core components (8/21), triple radio
sources which turn into twin jets at lower resolutions (9/21),
double-lobed structures (3/21) and complex morphologies (1/21).  Only
one object lacks a core identification. Their radio luminosities are
on average higher than Seyferts and H{\sc ii} galaxies by a factor
2-3.

Although the typical emission line ratios of LINERs can be reproduced
either by an AGN, shocks or post-AGB stars
\citep{allen08,sarzi10,capetti11b,singh13}, several factors point to
an AGN origin of their radio emission:

\begin{itemize}
\item They are mostly associated with symmetric jets extended on
  $\sim$0.05-1.7 kpc-scales. It is unlikely that SF can reproduce such
  radio morphologies and radio luminosities, since SF typically
  dominates over the AGN emission below 10$^{34}$ erg s$^{-1}$ in the
  sub-mJy regime \citep{bonzini13,padovani16}.

\item The radio jets are brightened closer to the cores, similar to
  the edge-darkened morphologies of FR~Is. These
  characteristics suggest that the jets are probably launched
  collimated and relativistically, and their bulk speed decreases
  along the jet propagation axis, similar to what is seen in nearby
  low-luminosity radio galaxies
  \citep{parma87,morganti87,falcke00,giovannini05}.

\item Radio core luminosities correlate with the BH masses,
  highlighting the role of the SMBHs in the production of radio
  emission in this class.

\item The highest radio core flux densities are associated with the
  LINERs, pointing to high brightness temperatures: at eMERLIN
  sensitivity and resolution, six LINERs have bright radio cores ($>$
  5 mJy) with brightness temperatures $>$10$^{6}$ K, indicative of
  synchrotron emission from relativistic jets. High brightness
  temperatures ($>$10$^{8}$ K) of radio cores detected with the VLBA
  for LINERs \citep{falcke00} point to the same interpretation, which
  is analogous to luminous radio-loud AGN.
\end{itemize}

LINERs are associated with different types of host galaxy, but with
the largest BH masses ($>$10$^{7}$ M$_{\odot}$) and typically accrete at
low rates ($<$10$^{-3}$ Eddington rate). The most accepted scenario
for this class is a radiatively inefficient disc model, usually as an
advection-dominated, geometrically thick, optically thin accretion
disc (ADAF, see review from \citealt{narayan98,narayan08}). Since
LINERs tend to be radio louder than other optical classes
\citep{capetti06,kharb12}, this effect can be ascribed to an ADAF disc
which is efficient at producing jets, as suggested by theoretical
studies including analytical work (e.g.,
\citealt{narayan91,meier01,nemmen07,begelman12}) and numerical simulations
(e.g., \citealt{tchekhovskoy11,mckinney12,yuan12}).

%The non-thermal
%synchrotron emission from core and a jet dominates the entire spectral
%output of the AGN at different wavelengths
%\citep{koerding08}. Forthcoming observations in C band will reveal the
%spectral properties of the cores whether they are flat-spectrum as
%expected by synchrotron self-absorption.

The combination of a jet and an ADAF disc (JDAF, \citealt{falcke04}) is
also attributed to FR~I radio galaxies.  We find that LINERs and FR~Is
follow similar correlations in the [O~III]-radio plane and in the
optical FPBHA. Hence, LINERs share the same central engine as FR~Is
and appear to be scaled-down versions of FR~Is. This result is
analogous to what has been found for low-luminosity radio-loud LINERs
in early-type galaxies (e.g., \citealt{balmaverde06b}).

%This affinity
%suggests that that LINERs and FR~Is can be possibly unified in a
%single picture of low-power accreting jetted AGN with a LINER
%spectrum, as suggested by \citet{falcke04}.

\subsection{Seyferts}

Only four Seyferts are present in this sample, based on the revised
BPT diagrams. Seyferts are mostly found in spiral galaxies and are all
detected at 1.5 GHz. One source, (NGC~5273, a lenticular galaxy) does
not show a radio core, but a ring of radio emission probably
associated with SF. The radio morphologies observed for the three
detected and identified Seyferts are: a triple source, a double-lobed
source and a complex morphology with a possible jet, all suggestive of
a jetted structure on a scale of $\sim$50-400 pc.  Our detection rate
(4/4) is higher than previous radio surveys of the Palomar sample for
this class of sources ($\sim$47 per cent, \citealt{ho08}). Although
the small number of Seyferts in our sample limits the interpretation
of our results about this class, we can, however, highlight their
properties as compared to the LINERs.

The high emission line ratios typical of Seyferts require
photo-ionization from an AGN \citep{kewley06}. However, the question
remains as to whether these active SMBHs or SF can account for the
observed radio properties of Seyferts.  In our survey, the jets of
Seyferts appear to be more edge-brightened than LINERs and similar to
the radio morphologies of local Seyferts (e.g.
\citealt{kukula93,kukula95,gallimore96,morganti99,wrobel00,kharb06}). For
the three detected Seyferts, diffuse radio structures appear in low
resolution images resembling radio lobes, which are observed to be
more common in Seyfert galaxies than LINERs
\citep{baum93,gallimore06}. Lobe structures indicate that the jets are
not necessarily relativistic nearer to the core and they eventually
terminate in a bow shock which plows into the surrounding interstellar
medium, resulting in a bubble-like structure. An AGN-driven scenario
can explain pc-kpc scale lobes, but we cannot rule out the possibility
that starburst super-winds cause the observed morphologies
\citep{pedlar85,heckman93,colbert96}.

The [O~III] line excess observed for our Seyferts with respect to the
[O~III]-radio correlation established for LINERs (Fig.~\ref{coreo3})
invokes the presence of a brighter ionizing source at the centre of
these galaxies than an ADAF disc. The high [O~III] luminosities, a
proxy of the AGN bolometric power, lead to high accretion rates
($>$10$^{-3}$ Eddington rate), providing further evidence for a
different central engine from that present in LINERs. A thin disc
(standard disc, \citealt{shakura73}) is commonly used to account for
the multi-band properties of Seyferts. These disc systems are less
efficient in launching relativistic radio jets than ADAF discs, but
are still able to produce jets, though less collimated and slower than
those in LINERs (\citealt{yuan14} and references therein). In fact,
the Seyferts in our sample roughly follow the optical FPBHA traced by
the LINERs, possibly suggesting a comparable jet-disc symbiosis. The
large fraction of Seyferts associated with jets in our sample and in
previous studies of local Seyferts suggests that low-power Seyferts
exist in an intermediate regime between the jet-ADAF dominated LINERs
and the near-Eddington luminous QSOs, which are typically radio-quiet
\citep{ho02,ho08,kauffmann08,sikora08}. This intermediate stage could be
a consequence of an evolutionary transition of disc physics from an
ADAF to a standard disc by increasing the accretion rate and/or disc
radiative efficiency \citep{trippe14}. However, a substantial paucity
of radio studies of QSOs, which are typically at higher redshifts, in
literature, could also affect this observational evidence. In fact,
generally, the nature of the radio emission in radio-quiet AGN,
whether a scaled-down version of the radio-loud jet
\citep{barvainis96,gallimore06}, coronal emission from magnetic
activity above the accretion disc as an outflow
\citep{field93,laor08,behar15}, or thermal free--free
emission/absorption \citep{gallimore04} still remains to be fully
understood.
% This
% intermediate stage is probably a consequence of both a evolutionary
% transition between low and high luminosities and a specific disc
% configuration of a thin standard disc with a hot corona, where the
% jet is anchored (at the inner disc radius). The jet collimates
% within a narrow nozzle near the SMBH and then it fans out to form a
% diffuse radio morphology
% \citep{donea96,donea02,markoff05,king11}. By increasing the
% accretion rate, such a jet launching mechanism becomes less
% efficient \citep{sikora07} and the physical origin of the
% parsec-scale radio emission in high-accreting QSOs, whether a
% scaled-down version of the radio-loud jet
% \citep{barvainis96,gallimore06}, coronal emission from magnetic
% activity above the accretion disc as an outflow
% \citep{field93,laor08}, or thermal free--free emission/absorption
% \citep{gallimore04} remains to be resolved.

\subsection{Absorption line galaxies}

As discussed in the Introduction, although ALGs lack evidence of
BH activity, they may still hide an active BH. In our sample we
detected radio emission associated with the innermost region of the
galaxy for $\sim$36 per cent (5/14) of the ALGs. They appear as a
single radio core for four cases and a triple source for one
galaxy. The ALGs in our survey are elliptical or lenticular
galaxies, which harbour SMBHs with masses larger than 10$^{8}$
M$_{\odot}$, typical of M$_{\rm BH}$ limits set for radio-loud AGN
\citep{chiaberge11}. The radio luminosities for these sources are
similar to LINERs. Only NGC~147 has an exceptionally low BH mass of
1.9$\times$10$^{4}$ M$_{\odot}$ in a dwarf elliptical galaxy
associated with the lowest radio luminosity ($\sim$10$^{32}$ erg
s$^{-1}$) of the sample.

The lack of availability of detected emission lines for our ALGs
prevents further analysis on either AGN bolometric luminosities or SF
rates. Therefore a complementary multi-band study of these galaxies is
needed first to study whether they hide active SMBHs.  A
cross-matching of optical SDSS \citep{stoughton02} and radio FIRST
\citep{becker95} data of local radio galaxies reveals that no-emission
line radio sources (slightly higher in radio luminosities,
10$^{39}$-10$^{41}$ erg s$^{-1}$, than our sample) mostly show host
and nuclear properties similar to LINERs \citep{baldi10b} and show
extended radio structures for 27 per cent of the sample. In summary, our
high-resolution radio observations and previous studies argue for the
presence of active SMBHs associated with radio jets in, at least, one
third of the ALGs.

The lack of evidence of an active SMBH in a large fraction of ALGs
might be interpreted within a nuclear recurrence scenario due to an
intermittent accretion phenomenon \citep{reynolds97,czerny09}. To
account for the large fraction of the undetected and the five detected
ALGs, the duration of the active phase must cover a wide range of
values with short active periods favoured over the longer
ones. Assuming a jet bulk speed of $\sim$0.02$c$ typical for low-power
radio galaxies \citep{massaglia16}, the most extended radio jet
detected for an ALG ($\sim$700pc for NGC~6702) in our sample sets a
limit on the age of $\sim$1.3$\times$10$^{5}$ years. Therefore, this
value infers an upper limit on the radio activity of ALGs, which is
consistent with the typical duty cycles of radio AGN,
i.e. 10$^{4}$-10$^{8}$ years estimated at different wavelengths (see
\citealt{morganti17} for a review). This recurrent nuclear activity
might account for their high core dominance, possibly interpreted as a
sign of a young radio age.

%Since ALGs host the most massive BHs in elliptical galaxies, they are
%expected to be accreting due to their large gravitational potential
%and so active in optical and possibly in radio.  However, the problem
%of starving massive BHs have been discussed much in the last decades
%(e.g \citealt{fabian88,kormendy01}) and the inactivity of ALGs falls
%in this investigation. 

%The angular resolution of our
%observations set the limit on the age ...

\subsection{H{\sc ii} galaxies}

H{\sc ii} galaxies are SF-dominated nuclei, based on the
emission line ratios, but their classification does not preclude the
possibility of the presence of a weak AGN at their galaxy centre. The
relatively few available radio observations of H{\sc ii} galaxies and
single-source studies generally lead to low detection rates ($\sim$7.5
per cent, \citealt{ulvestad02} and references therein).  We detected
radio emission for 31 per cent (16/51) of the H{\sc ii} sample.
For 12 H{\sc ii} galaxies, five sources show jetted structures
(one-sided and twin jets), four show single core detections and three
show complex morphologies. However for four sources a clear radio core
component is missing from the eMERLIN maps. The radio luminosities
of H{\sc ii} galaxies are, on average, lower than those of LINERs by a
factor $\sim$2. The radio structure sizes are of order 50-500 pc.

%Only \citet{ulvestad02} have completed a survey
%of a well-defined sub-sample of 40 H{\sc ii} type nuclei in the Palomar
%sample, but they found that none of them has a compact radio nucleus
%at the flux levels of those in LLAGN in the sample. 

Two different types of H{\sc ii} galaxies emerge from our eMERLIN
survey: jetted and non-jetted. The H{\sc ii} galaxies with jetted
morphologies (NGC~972, NGC~3665, UGC~3828, NGC~7798, and UGC~4028),
possibly indicating the presence of an AGN, are found associated with
BH masses larger than 10$^{6}$ M$_{\odot}$. The emission line ratios
of these sources do not differ from the other H{\sc ii} galaxies in the BPT
diagrams (see Fig.\ref{bpt}). As further evidence of a possible AGN,
these five sources lie on the [O~III]--radio correlation found for
LINERs and on the optical FPBHA. One source, NGC~3665, shows an FR~I
radio morphology at VLA resolution \citep{parma86}. Two sources
show one-sided jets, which might indicate the Doppler boosting effect
of relativistic jets. This group of H{\sc ii} galaxies probably host a
weak AGN whose optical signature has simply been overpowered by the
dominant optical signal from nuclear SF. Their hidden LLAGN appears
more consistent with a sub-Eddington LINER rather than a more luminous
Seyfert. In fact the radio jets of this group are morphologically
similar (e.g., core-brightened) to the LINERs in our sample and
roughly follow the relations traced by the LINERs.

Conversely, the situation appears different for the remaining seven
H{\sc ii} galaxies (NGC~278, NGC~3198, NGC~3938, NGC~3432, NGC~6946,
NGC~2342, and NGC~3430), which do not show clear jet structures. The
BH masses of these sources are below 10$^{6}$ M$_{\odot}$ and they
show clear [O~III] line deficit of a factor $\sim$30 and a large radio
excess with respect to the [O~III]--radio correlation found for
LINERs. This result seems to follow the idea that the low-radiation
field generated by SF can account for both their observed radio and
emission line properties.  The only exception to this picture is
NGC~2342, which has a complex radio morphology, has a BH mass of
4$\times$10$^{7}$ M$_{\odot}$, lies on the optical FPBHA and has the
highest [O~III] luminosity of all detected H{\sc ii} galaxies.

For non-jetted H{\sc ii} galaxies, we consider the case of non-thermal
SF to be the primary origin of their radio counterparts (an assumption
which is justified at 20 cm where most of the radio emission is of a
non-thermal nature). We calculate the expected nuclear supernova rates
$\nu_{SN}$ assuming that the total radio emission is dominated by
cosmic rays accelerated in supernova remnants (SNR), using the formula
from \citet{condon92}, L$_{radio}$/(1.4$\times$10$^{38}$ erg s$^{-1}$)
= 11 $\times$($\nu_{SN}$/yr$^{-1}$). The calculated supernova rates
correspond to 2$\times$10$^{-5}$$-$6$\times$10$^{-4}$ yr$^{-1}$ in the
central core region of $\sim$50 pc. These values are roughly similar
to the supernova rates expected in M82 (15-30 yr$^{-1}$ for the entire
galaxy, \citealt{muxlow94,fenech08}) extrapolated to a region 100
times smaller, or 3-6$\times$10$^{-4}$ yr$^{-1}$.  Furthermore,
assuming that a fraction of radio emission comes from thermal stellar
processes, the predicted free--free radio emission expected from the
H$\alpha$ luminosities (10$^{37.7-38.8}$ erg s$^{-1}$,
\citealt{ho97a}) due to thermal stellar emission
\citep{ulvestad81,filho02} would be conservatively higher within a
factor $\sim$10 than the measured radio luminosities of AGN-dominated
sources, i.e. Seyferts and LINERs. This would correspond to SF rates
of 4$\times$10$^{-4}$-5$\times$10$^{-3}$ M$_{\odot}$ yr$^{-1}$ in the
central radio core, derived from the H$\alpha$ luminosities (assuming
a Salpeter mass function for 0.1-100 M$_{\odot}$ and solar
metallicity, \citealt{kennicutt98}). In conclusion all these
calculations favour a SF origin of the radio emission, possibly from a
pc-scale nuclear starburst, for the non-jetted H{\sc ii} galaxies.

\subsection{Any clear star-forming galaxies?}
\label{anysfg}

Considering the identified galaxies of our sample which do not show
clear jetted morphologies, we do not find any evidence for radio
sources which generally resemble diffuse star-forming galaxies
\citep{muxlow10,murphy17,herrero17}, based only on the radio
properties. However, single radio components might hide unresolved
nuclear star-bursts and complex radio morphologies might also be due to
the presence of star-forming regions, possibly induced by radio jets
(for example, see NGC~2273 and NGC~2655). The only cases where the
radio-optical results support for a possible star-forming scenario,
are the non-jetted H~II galaxies.

For the unidentified galaxies, the nature of the detected radio
emission is unclear. However, two objects (NGC~4013 and NGC~5273)
clearly show circum-nuclear radio-emitting rings, which were not seen
before at lower resolution. These two sources are the most evident
cases of SF in our sample, based on their radio morphology.

For the objects further than 4 Mpc, which are the majority of the
sample, the spatial frequencies covered by eMERLIN are suited to
detect compact SNR \citep{westcott17}. However, the snapshot imaging
technique used in this survey is not appropriate for detecting
diffuse, low-brightness radio emission ($<$10$^{5}$ K), typical of old
SNR. Conversely, young SNR and H{\sc ii} regions could
be detected since they are more compact and brighter. At distances
less than $\sim$4 Mpc, VLA data are required to study old SNR ($>$
400--500 yrs), which would be resolved out by the eMERLIN
baselines. Within this distance, only one galaxy (NGC~147) has been
detected in our survey, showing a single radio component. The addition
of shorter spacing VLA data to this project will enhance the ability
to detect more diffuse lower surface brightness emission from SF
products.

\vspace{-0.5cm}
\subsection{Host galaxies}

In this sample, we find that half of the early-type galaxies
(ellipticals and lenticulars) host a radio-emitting active SMBH, while
this occurs only in one third of the late-type galaxies (spirals and
irregulars). More precisely, our radio survey detects $\sim$60 per
cent of galaxies earlier than Sb. The detection rate of radio AGN
dramatically drops toward later Hubble types: we detect radio cores
for 22 per cent of Sc-Sd galaxies only. Our results agree with
previous radio studies on the Palomar sample
\citep{ho01a,ulvestad01a,nagar05,ho08}, which found flat-spectrum
radio cores predominantly in massive early-type galaxies
\citep{sadler89,wrobel91b,capetti09,miller09,nyland16}.

The LeMMINGs survey detects nuclear activity, predominantly in the
form of jet morphologies, for SMBHs more massive than 10$^{6}$
M$_{\odot}$ but the detection fraction drastically decreases below
10$^{7}$ M$_{\odot}$.  We also find that the core luminosities of the
jetted galaxies correlate with the BH mass. These results confirm past
radio continuum studies that showed that radio brightness in galaxies
is a strong function of the host mass or BH mass, proportional to
M$_{\rm BH}$$^{\alpha}$ (with a broad range of $\alpha$ values,
typically $1-2$,
\citealt{laor00,best05b,nagar02,mauch07}). Consequently, this
dependence also explains the different detection rates between
early-type and late-type galaxies, \citep{decarli07,gallo08}, due to
their different BH masses.

The association between radio detection, host type and BH mass is an
indirect consequence of a more fundamental relationship between radio
luminosity and optical bulge luminosity (or mass)
\citep{ho02,nagar02}.  Massive elliptical galaxies, which typically
host large SMBHs, are more efficient in producing radio emission than
less massive spiral galaxies and, thus, turn out to be radio-loud for
M$_{\rm BH} >$10$^{8}$ M$_{\odot}$ \citep{chiaberge11}.

%Systematic spectroscopic observations are
%more efficient than radio surveys to detect active BHs by searching for
%broad-lines AGN below 10$^{6}$ M$_{\odot}$ (see \citealt{green07}).

%22% 9/41

\section{Summary and conclusions}
\label{overview}

This paper presents the first release of radio data and results from
the eMERLIN legacy survey, LeMMINGs, aimed at studying a sample of
nearby (active and quiescent) galaxies. We observed 103 sources from
the Palomar sample \citep{ho97a} with $\delta>$ 20$^{\circ}$ (one
third of the entire legacy program) at 1.5 GHz with the eMERLIN array,
reaching a sensitivity of $\sim$70$\mu$Jy and an angular resolution of
$\sim$150 mas. The remaining radio images of this survey and the
multiband study of the sample will be released and addressed in
forthcoming papers.

First, we have updated the optical spectroscopic classification of the
sample based on the state-of-the-art diagrams. We used diagnostic
schemes presented in \citet{kewley06} and \citet{buttiglione10} to
classify the sources as LINERs, Seyferts, H{\sc ii} galaxies, and ALGs
(34:4:51:14).

Our radio survey detected significant radio emission in the innermost
region of the galaxies (0.73 arcmin$^{2}$) for 47 objects out of 103
($\sim$46 per cent) with flux densities $\gtrsim$0.2 mJy. For 41
sources we identified the radio core within the structure, associated
with the optical galaxy centre. We resolved parsec-scale radio
structures with a broad variety of morphologies: core/core--jet,
one-sided jet, triple sources, twin jets, double lobed, and complex
shapes with extents of 40--1700 pc. It is important to stress that
such detected structures mostly appeared as compact in previous VLA
observations. The jetted radio sources (25/41) are more common than
single cores (16/41).  While the jetted morphologies, which are
associated with galaxies with M$_{\rm BH}>$ 10$^{6}$ M$_{\odot}$, are
interpreted as a possible sign of an active SMBH, more caution is
needed to interpret the single unresolved components associated with
no jets, whether they are a simple unresolved jet base or star-forming
cores.

%Considering the different optical classifications, the most common detected radio
%galaxies are LINERs and early-type galaxies, typically showing single
%cores or jetted morphologies. The other optical classes, which
%correspond to the remaining half of the detected sources, cover all
%radio morphologies. Seyferts show radio morphologies that are more
%edge-brightened than those in LINERs.

%The radio core luminosities we observe (10$^{32}$-10$^{40}$ erg
%s$^{-1}$) are lower than previous radio surveys of the Palomar sample,
%10$^{35}$ erg s$^{-1}$ \citep{nagar05,filho06}, due to higher
%sensitivity (by a factor $\sim$4) and higher angular resolution (by a
%factor 2-10 depending on frequency), thanks to the capabilities of
%eMERLIN. LINERs emit the highest radio powers, whilst H{\sc ii}
%nuclei lie on the lower side of the radio luminosity range. ALGs and
%Seyferts reside in an intermediate luminosity regime. The radio cores
%typically contribute to $\sim$26 per cent of the total detected
%emission. Concerning the host type, half of the early-type galaxies
%and one third of the late-type galaxies are detected in our survey.

The radio core powers (10$^{32}$-10$^{40}$ erg s$^{-1}$) are lower
than those measured from previous radio surveys of the Palomar sample,
10$^{35}$ erg s$^{-1}$ \citep{nagar05,filho06}, due to higher
sensitivity (by a factor $\sim$4) and higher angular resolution (by a
factor 2-10 depending on frequency) of eMERLIN.  The most common
detected radio galaxies are LINERs and early-type galaxies, typically
showing single cores or jetted morphologies.  Seyferts show more
edge-brightened radio morphologies than those of LINERs and have
intermediate luminosities.  ALGs mostly show single cores and are
highly core-dominated.  H{\sc ii} galaxies lie on the lower side of
the radio luminosity range and show both jetted and non-jetted radio
morphologies for the detected H{\sc ii} galaxies.  The radio cores
typically contribute to $\sim$26 per cent of the total detected
emission. Concerning the host type, half of the early-type galaxies
and one third of the late-type galaxies are detected in our survey.

To investigate the nature of the radio emission, we look for empirical
correlations between core luminosities and BH masses and [O~III] line
luminosity, all good diagnostics of the SMBH activity. We find that the
core power, indicative of the jet energetics, correlates with M$_{\rm
  BH}$ for masses $\gtrsim$10$^{7}$ M$_{\odot}$, but a clear
break emerges at lower masses.  Ellipticals, lenticulars and radio
jetted galaxies lie on the correlation. This result is possibly a
consequence of a fundamental link between SMBH activity and radio jet
production. By assuming a hierarchical galaxy evolution, massive
galaxies (typically early-type) nurse their SMBHs through mergers,
which grow in mass and evolve to low-Eddington ratios
\citep{volonteri13}.  Massive SMBHs provide the adequate conditions
for supporting the launch of a jet \citep{falcke04}, which accounts
for the empirical radio--M$_{\rm BH}$ relation. Conversely, for SMBHs
with masses $\lesssim$10$^{6}$ M$_{\odot}$, which are generally
found in less-massive late-type galaxies, the radio detection rate
decreases in our sample. It is likely that a larger contribution from
SF in the radio band might account for the flattening of the
radio-M$_{\rm BH}$ relation.

[O~III] line luminosity is a good indicator of AGN bolometric luminosity,
and is found to broadly correlate with radio core power. However, the
radio emission efficiency, i.e. the fraction of the radio emission
produced with respect to the bolometric luminosity of the AGN, is
different for each optical class. This result is in agreement with
previous multi-band studies for different classes of LLAGN and extends
to more powerful radio galaxies (e.g.,
\citealt{chiaberge99,nagar05,panessa07,balmaverde06a,balmaverde06b,hardcastle09,baldi10,degasperin11,panessa13,asmus15,mingo16}).
The radio properties and the location in the radio--[O~III] plane
attribute different nuclear characteristics, resulting from
fundamentally different BH physics, which we discuss further below:

\begin{itemize}
\item {\bf LINERs}. Our eMERLIN observations show a prevalence of
  jetted structures associated with LINERs (13/21). Furthermore, we
  find evidence of an affinity between the LINERs in our sample and
  FR~I radio galaxies, as they follow similar correlations in the
  [O~III]--radio plane and in the optical FPBHA. This result points to
  a common origin of the nuclear emission: the non-thermal synchrotron
  emission from a jet, which dominates the spectral output of the AGN
  at different wavelengths \citep{koerding08}. The low Eddington rate
  of LINERs and their similarities with FR~Is suggest that a
  radiatively inefficient disc, probably an ADAF disc, which also aids
  in the jet production due to its funnel-like structure and poloidal
  magnetic field, is the most viable accretion mode at the centre of
  LINERs. These results are in agreement with previous multi-band
  studies which proposed a jet+ADAF model (JDAF, \citealt{falcke04})
  \citep{balmaverde06b,balmaverde06a,panessa07,baldi09} for LINER-like
  LLAGN, which also accounts for the inverse correlation between their
  Eddington rate and radio-loudness \citep{ho02,ho08}.

 %In fact, this
 % work adds further evidence that LINERs consist of a homogeneous
 % population of radio sources typically associated with massive
 % (early-type) galaxies and massive SMBHs (typically $>$10$^{7}$
 % M$_{\odot}$), whereby the low accretion rate, set by hot gas, scales
 % with jet kinetic power
 % (e.g. \citealt{falcke04,allen06,balmaverde08}). Such a heterogeneity
 % is likely a consequence of a hierarchical evolution of early-type
 % galaxies (e.g. \citealt{schawinski07,schawinski10,heckman14}).

\item {\bf Seyferts}. Although this class is limited in number in our
  sample, we find that Seyferts are associated with radio jets (3/4),
  which appear more edge-brightened than those observed in
  LINERs. This type of morphology points to jet bulk speeds lower than
  those inferred from the core-brightened jets in LINERs. The high
  emission line ratios, large [O~III] luminosities and the large
  [O~III] excess with respect to the [O~III]-radio core correlation
  found for the synchrotron-dominated LINERs indicate a different
  central engine for Seyferts: a radiatively efficient disc, accreting
  at higher rates than in LINERs. Our results emphasise the increasing
  observational evidence that, kpc-scale jets are more commonly
  observed in local Seyferts than in more luminous Seyferts and
  QSOs. As also supported by the shared optical FPBHA with LINERs,
  low-power Seyferts are in an intermediate position between the
  jetted LINERs and the radio-quiet near-Eddington QSOs
  \citep{balick82,ho02,ho08,bush14}, suggesting a continuum of radio
  properties of LLAGN \citep{kharb14}. This intermediate stage is
  probably a consequence of a luminosity/accretion dependence on the
  jet efficiency \citep{trippe14} from a thick to a thin accretion
  disc \citep{donea96,donea02,markoff05,king11}.

  %In addition, Seyferts are found to roughly follow the LINERs in the
  %optical FPBHA. Therefore, the moderate radio luminosities and BH
  %masses of low-power Seyferts, as observed in this survey and
  %previous studies, emphasise the increasing evidence that
  %low-luminosity Seyferts are in an intermediate position, which
  %bridges the gap between the jetted LINERs and the radio-quiet
  %near-Eddington QSOs \citep{balick82,ho02,ho08,bush14}. This
  %intermediate stage is probably a consequence of both a evolutionary
  %transition between low and high luminosities and a specific disc
  %configuration of a thin standard disc with a hot corona, where the
  %jet is anchored (at the inner disc radius). The jet collimates
  %within a narrow nozzle near the SMBH and then it fans out to form a
  %diffuse radio morphology
  %\citep{donea96,donea02,markoff05,king11}. By increasing the
  %accretion rate, such a jet launching mechanism becomes less
  %efficient \citep{sikora07} and the physical origin of the
  %parsec-scale radio emission in high-accreting QSOs, whether a
  %scaled-down version of the radio-loud jet
  %\citep{barvainis96,gallimore06}, coronal emission from magnetic
  %activity above the accretion disc as an outflow
  %\citep{field93,laor08}, or thermal free--free emission/absorption
  %\citep{gallimore04} remains to be resolved.

%intermediate
%https://www.aanda.org/articles/aa/full_html/2014/01/aa22486-13/aa22486-13.html
%http://adsabs.harvard.edu/cgi-bin/bib_query?arXiv:1309.6921
%https://www.aanda.org/articles/aa/abs/2007/29/aa7578-07/aa7578-07.html
%http://ned.ipac.caltech.edu/level5/Sept01/Balick/Balick3_1.html

%#########

\item {\bf H{\sc ii} galaxies} and {\bf ALGs} are a mixed population
  of galaxies which either host weakly active or silent SMBHs, and may
  be overwhelmed by SF. On one hand, eMERLIN observations of H{\sc ii}
  galaxies reveal a sub-population of jetted sources (5/51), which has
  M$_{\rm BH}$ $>$ 10$^6$ M$_{\odot}$, follows the LINERs in the
  optical FPBHA: these sources are possibly powered by weakly active
  SMBHs, similar to LINERs. The remaining non-jetted H{\sc ii}
  galaxies better reconcile with a SF scenario. On the other hand, the
  detected ALGs (5/14) are associated with core--jet/triple radio
  morphologies and are commonly hosted in massive ellipticals, similar
  to LINERs. Furthermore, for ALGs a possible radio nuclear recurrence
  scenario might reconcile with the occasional lack of activity
  observed in massive early-type galaxies, due to general duty cycles
  of nuclear activity \citep{morganti17}.

\end{itemize}

In conclusion, to a first approximation, AGN with different BH masses
and hosted in different galaxies, can produce jets with similar
power, size, and morphology \citep{gendre13,baldi18}. This result
suggests that (currently) directly unobservable quantities, such as BH
spin, magnetic field strength, and accretion rotation, might play an
important role in the jet launching mechanism
\citep{garofalo10,tchekhovskoy12,garofalo13}, regardless of the AGN
and host properties. In addition, in our survey, independent of their
optical classes, the active galaxies and the jetted H{\sc ii} galaxies
appear to broadly follow a similar correlation in the optical FPBHA,
stretching their luminosities up to FR~I radio galaxies. Therefore, in
the low-luminosity regime, the physical process which regulates the
conversion of the accretion flow into radiative and kinetic jet energy
could be universal across the entire SMBH mass scale and for different
optical classes. Such a process does not necessarily require the same
disc--jet coupling since the classes are differently powered. Instead,
it points to a common scaling relation in terms of BH properties
(mass, spin, disc-BH alignment and co-rotation), accretion, and jet
production, which are distinct for each AGN class, but become similar
when all are properly combined in the FPBHA, as validated by
magnetic-hydrodynamic simulations \citep{heinz03}. However, we should
point out that a slight separation of the different optical classes is
still evident across the optical FPBHA (with the Seyferts at lower
radio luminosities than LINERs), which was not observed in previous
works at lower VLA resolution. Resolving the parsec-scale jet base,
thanks to eMERLIN, has helped bring this effect in the fundamental
plane to light. The origin of this stratification in the optical FPBHA
might be still reminiscent of the different accretion modes for each
optical class.

%However, all jetted AGN follow
%the optical fundamental plane of BH activity, suggesting a scale
%invariance between the accretion and the jet mechanism.

%H{\sc ii} nuclei are generally found below the correlation
%being less luminous in line, while Seyferts show a line excess. Once
%the BH mass is included as parameter in the correlation to establish
%the optical fundamental plane of BH activity \citep{saikia15}, the
%situation appears clearer: LINER, Seyferts and the jetted H{\sc ii}
%galaxies share the same line-radio correlation, following the FR~I
%radio galaxies. These matches are indicative of a common underlying
%physics which associate the accretion properties with the jet
%launching mechanism.

%. However their radio emission efficiency is
%$\sim$10$^{-6}$, smaller than H{\sc ii} galaxies which, conversely, are
%dramatically less efficient to launch jet but their radio output
%contributes more in the total bolometric luminosity. This can be
%ascribed to a substantial addition of thermal emission from
%SF. Oppositely, low-luminosity Seyferts, based also on previous radio
%studies, show similar radio detections, suggesting a similar jet
%production to LINERs, but significant disparities indicates
%fundamental different AGN physics. Seyferts show evidences of higher
%Eddington ratios ($>$10$^{-3}$), more luminous optical and X-ray
%nuclear counterparts than LINERs. They are interpreted as
%characteristics of optically thick geometrically thin standard
%disc. The radio emission efficiency is, however, lower than LINERs,
%suggesting a different jet-disc performance.

The LeMMINGs project has uncovered new radio active SMBHs in the local
Universe, which were not detected and identified before. This
discovery is fundamental for conducting a fair census of the local BH
population and to provide robust constraints on cosmological galaxy
evolution models \citep{shankar09}. The local BH demographics and the
study of the BH activity is currently still limited to a few
single-band observations, circumscribed to small samples. The LeMMINGs
legacy survey will address these topics with a multi-band approach
applied to a complete sample, pushing the low end of the radio
luminosity function. In fact, further eMERLIN observations at C band
(5 GHz) will, for the first time, extend the radio luminosity function
of the local Universe down to 10 times the luminosity of Sgr~A* and
help to discriminate SF from genuine AGN activity. Furthermore,
complementary data, which include optical (HST), X-ray (Chandra), and
infrared (Spitzer) band, will unveil the origin of the parsec-scale
radio emission and properties of the central engines of the local
LLAGN population.

%The unprecedented
%sensitivity and spatial resolution of the new-generation Square
%Kilometre Array will eventually clarify the puzzling physics of the
%jet-disc connection in LLAGN.

\vspace{-0.5cm}
\section*{Acknowledgements}

The authors thank the referee for a quick publication and the helpful
comments from A. Laor and A. Capetti for the interpretation of the
results.  RDB and IMcH acknowledge the support of STFC under grant
[ST/M001326/1] and IMcH thanks the Royal Society for the award of a
Royal Society Leverhulme Trust Senior Research Fellowship.  We
acknowledge funding from the University of Southampton for a Mayflower
studentship afforded to DW.  EB and JW acknowledge support from the
UK's Science and Technology Facilities Council [grant number
ST/M503514/1] and [grant number ST/M001008/1], respectively. CGM
acknowledges financial support from STFC. JHK acknowledges financial
support from the European Union's Horizon 2020 research and innovation
programme under Marie Sk\l{}odowska-Curie grant agreement No 721463 to
the SUNDIAL ITN network, and from the Spanish Ministry of Economy and
Competitiveness (MINECO) under grant number AYA2016-76219-P. DMF
wishes to acknowledge funding from an STFC Q10 consolidated grant
[ST/M001334/1].  BTD acknowledges support from a Spanish postdoctoral
fellowship `Ayudas para la atracci\'on del talento
investigador. Modalidad 2: j\'ovenes investigadores, financiadas por
la Comunidad de Madrid' under grant number 2016-T2/TIC-2039.  FP has
received funding from the European Union's Horizon 2020 Programme
under the AHEAD project (grant agreement No 654215). We also
acknowledge the Jodrell Bank Centre for Astrophysics, which is funded
by the STFC. eMERLIN and formerly, MERLIN, is a National Facility
operated by the University of Manchester at Jodrell Bank Observatory
on behalf of STFC. This publication has received funding from the
European Union's Horizon 2020 research and innovation programme under
grant agreement No 730562 [RadioNet].

%%%%%%%%%%%%%%%%%%%%%%%%%%%%%%%%%%%%%%%%%%%%%%%%%%



\bibliographystyle{mn2e}
\bibliography{my} % if your bibtex file is called example.bib

%%%%%%%%%%%%%%%%%%%%%%%%%%%%%%%%%%%%%%%%%%%%%%%%%%

%%%%%%%%%%%%%%%%% APPENDICES %%%%%%%%%%%%%%%%%%%%%

\appendix

\section{Radio data}
\label{app}

\begin{figure*}
	\includegraphics[width=0.93\textwidth]{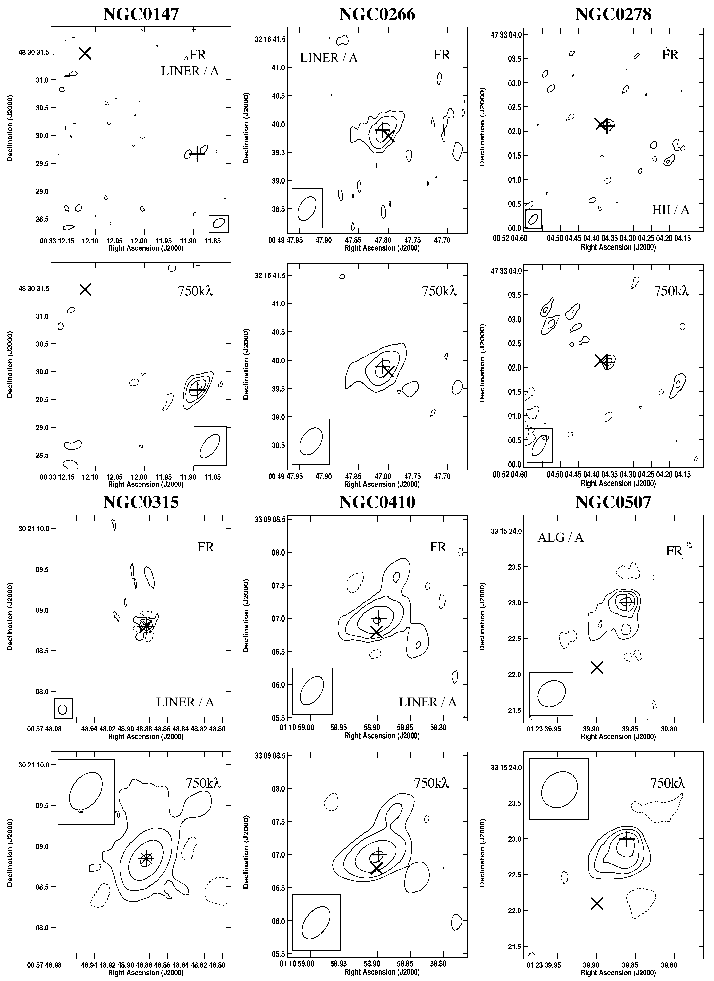}
        \caption{eMERLIN 1.5-GHz images of the identified Palomar galaxies. See last page of this figure for details.}
    \label{maps1}
\end{figure*}

In the Appendix \ref{app} we present the radio images of the 47
detected objects of the Palomar sample studied here (Fig.~\ref{maps1}
and \ref{maps2}). Table~\ref{tabdet} and \ref{tabsfr} list the source
parameters of the radio components detected in the images for the
identified and unidentified sources,
respectively. Table~\ref{contours} provides the radio contours and the
properties of the restoring beams of the radio maps.

\addtocounter{figure}{-1}
\begin{figure*}
%\ContinuedFloat
	\includegraphics[width=0.93\textwidth]{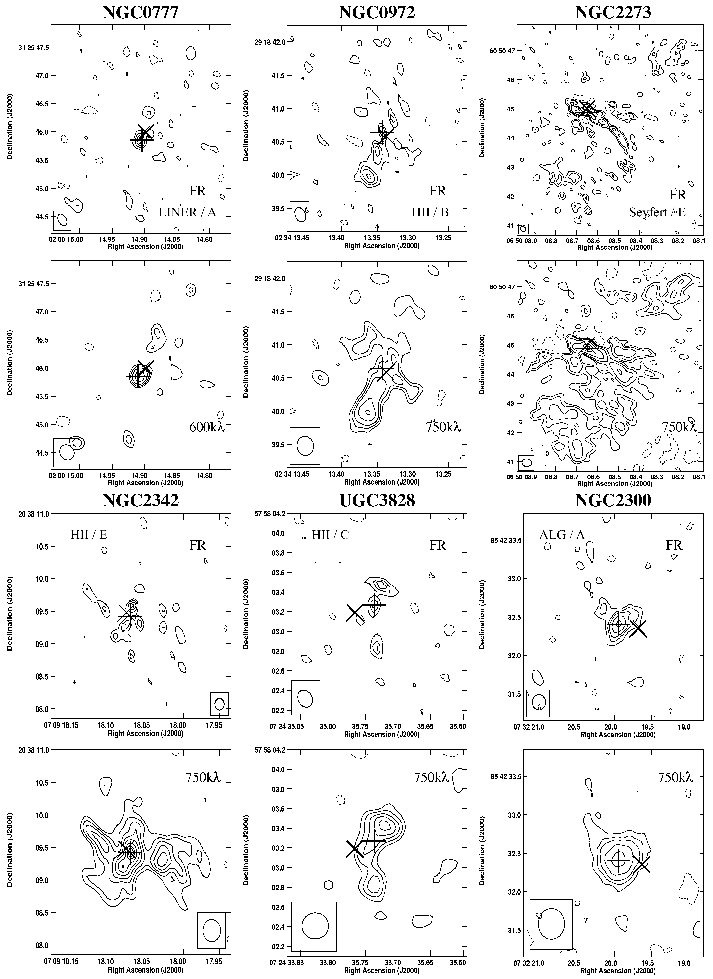}
    \caption{eMERLIN 1.5-GHz images of the identified Palomar galaxies. See last page of this figure for details.}
\end{figure*}

\addtocounter{figure}{-1}
\begin{figure*}
%\ContinuedFloat
	\includegraphics[width=0.93\textwidth]{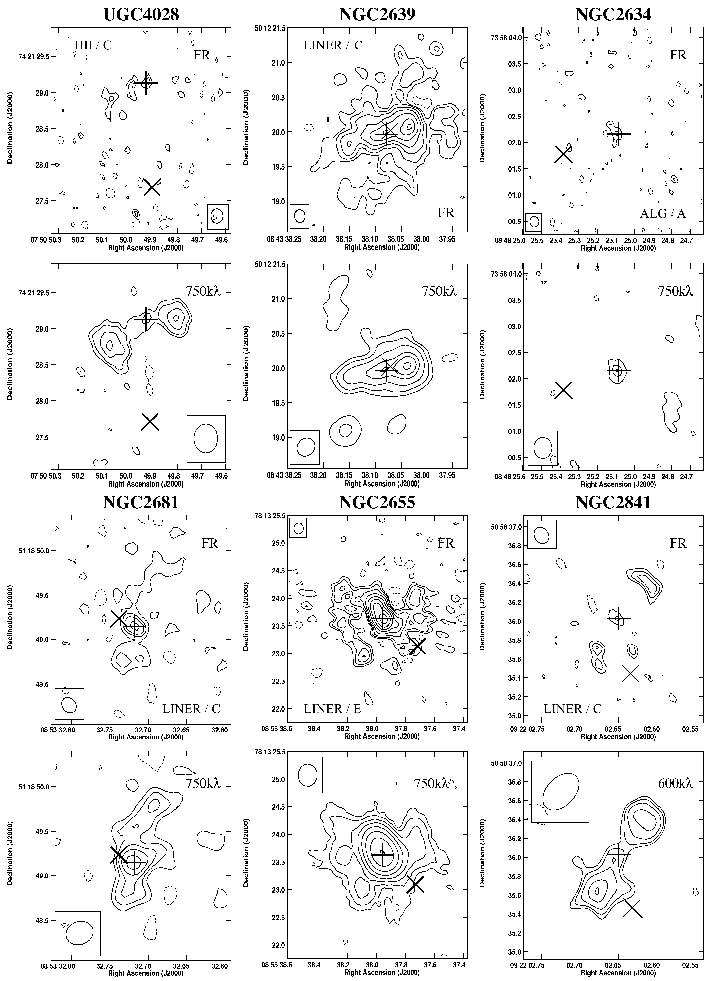}
    \caption{eMERLIN 1.5-GHz images of the identified Palomar galaxies. See last page of this figure for details.}
\end{figure*}

\addtocounter{figure}{-1}
\begin{figure*}
%\ContinuedFloat
	\includegraphics[width=0.93\textwidth]{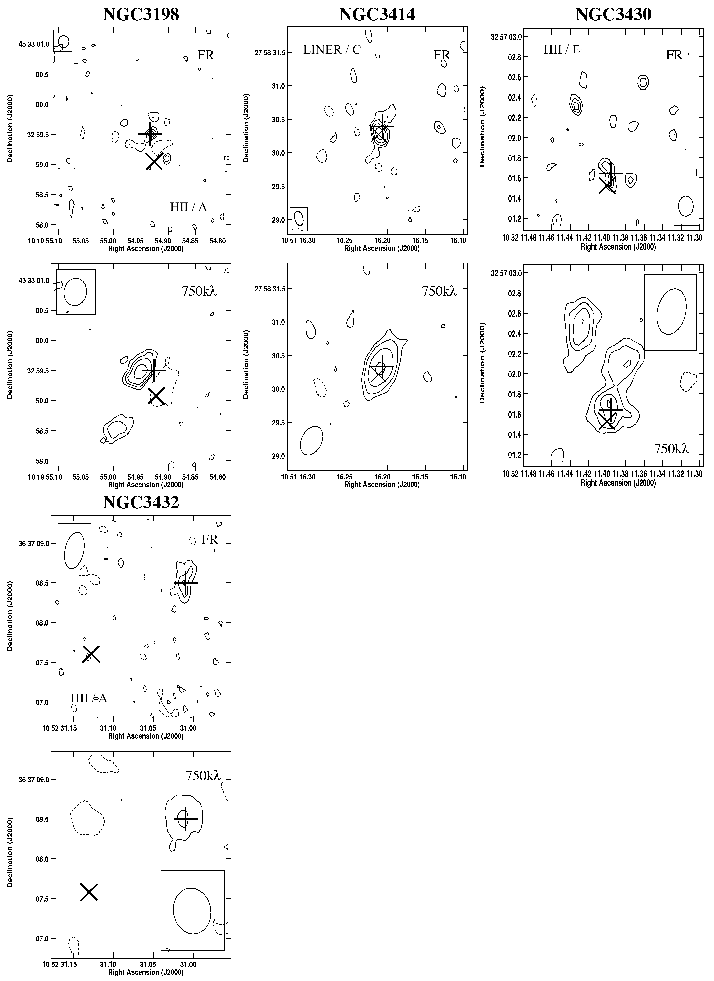}
    \caption{eMERLIN 1.5-GHz images of the identified Palomar galaxies. See last page of this figure for details.}
\end{figure*}

\addtocounter{figure}{-1}
\begin{figure*}
%\ContinuedFloat
	\includegraphics[width=0.93\textwidth]{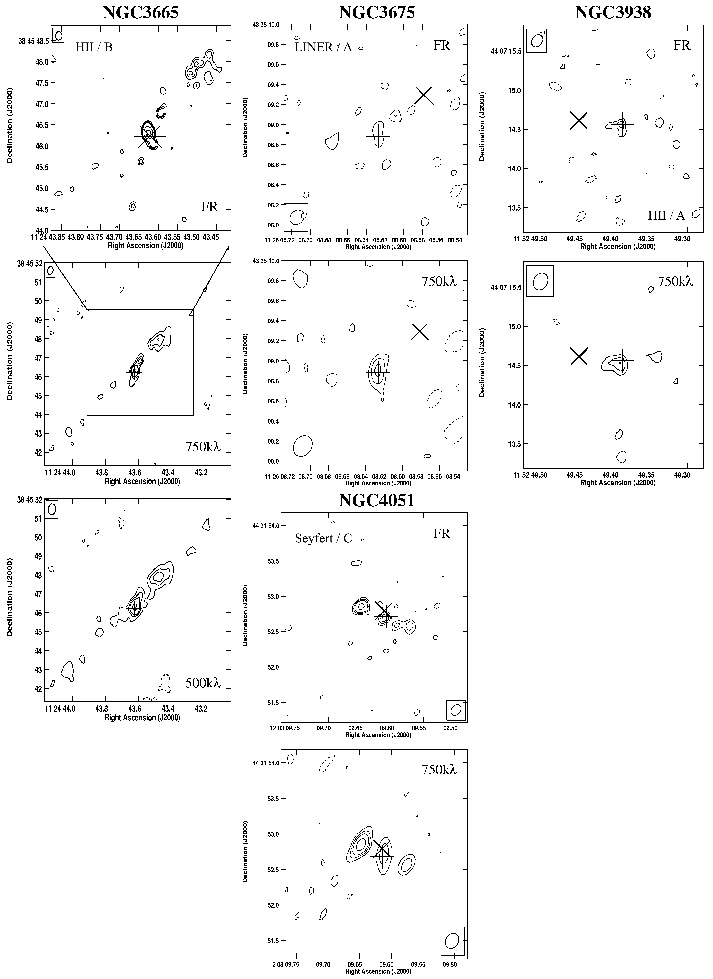}
    \caption{eMERLIN 1.5-GHz images of the identified Palomar galaxies. See last page of this figure for details.}
\end{figure*}

\addtocounter{figure}{-1}
\begin{figure*}
%\ContinuedFloat
	\includegraphics[width=0.93\textwidth]{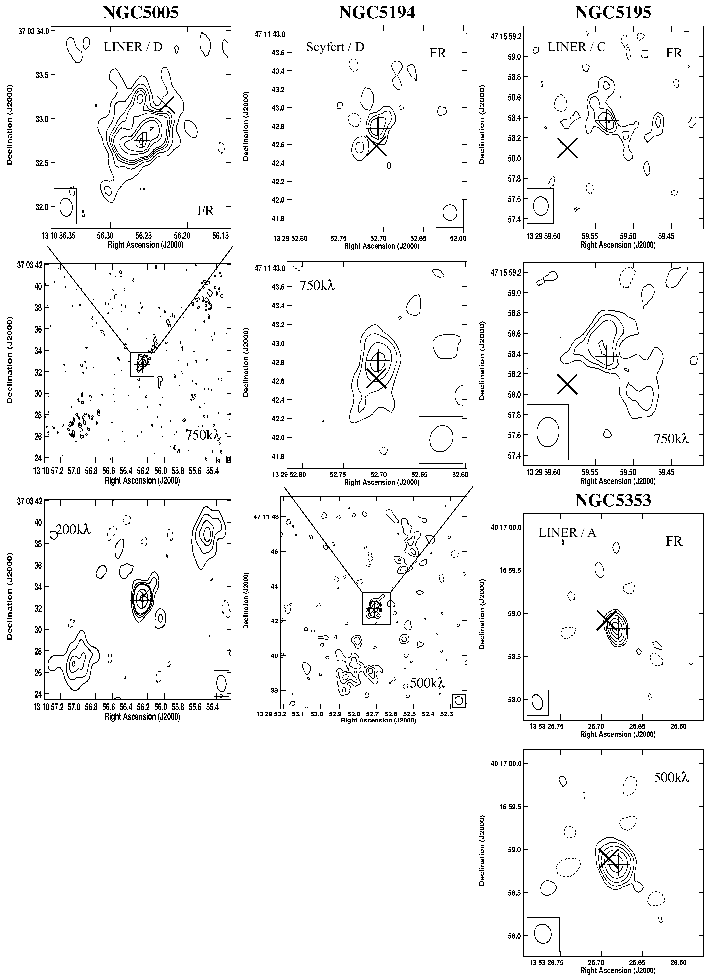}
    \caption{eMERLIN 1.5-GHz images of the identified Palomar galaxies. See last page of this figure for details.}
\end{figure*}

\addtocounter{figure}{-1}
\begin{figure*}
%\ContinuedFloat
	\includegraphics[width=0.93\textwidth]{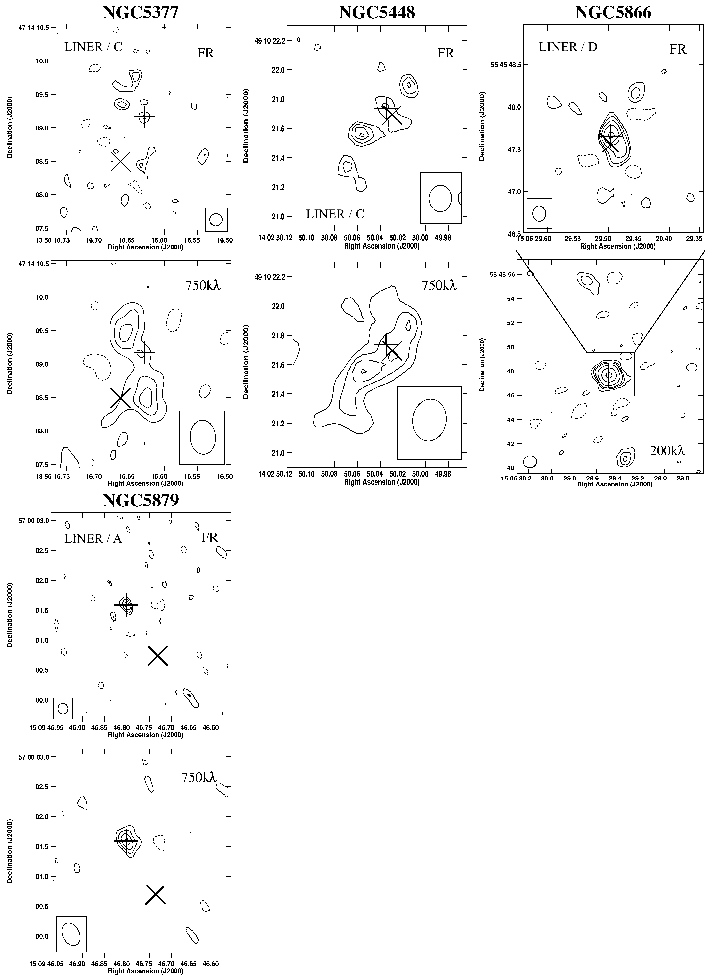}
    \caption{eMERLIN 1.5-GHz images of the identified Palomar galaxies. See last page of this figure for details.}
\end{figure*}

\addtocounter{figure}{-1}
\begin{figure*}
%\ContinuedFloat
	\includegraphics[width=0.93\textwidth]{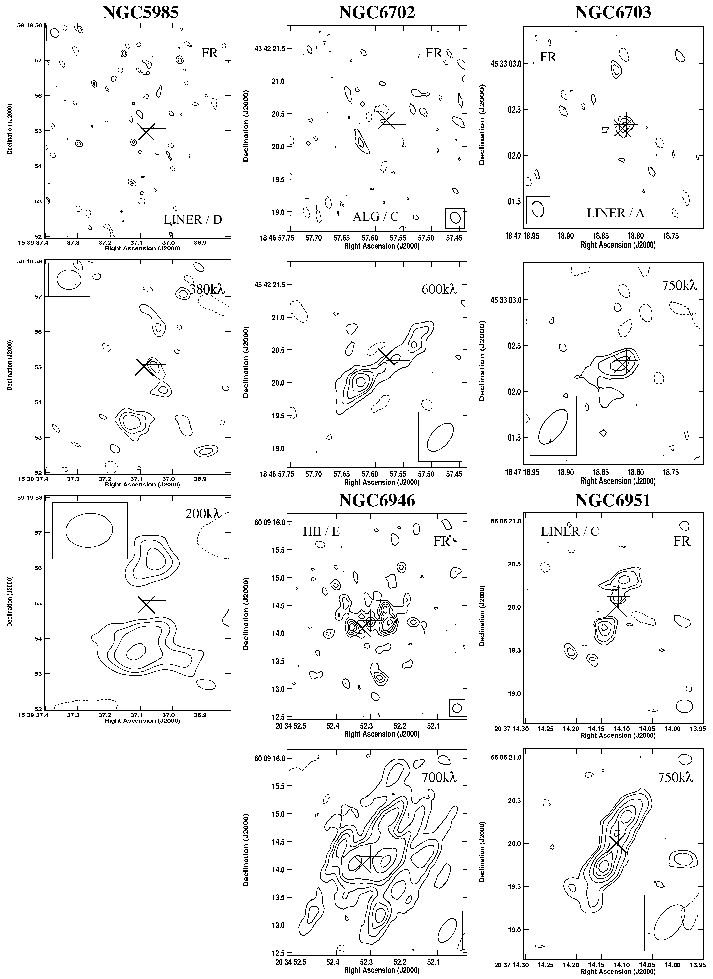}
    \caption{eMERLIN 1.5-GHz images of the identified Palomar galaxies. See last page of this figure for details.}
\end{figure*}

\addtocounter{figure}{-1}
\begin{figure*}
%\ContinuedFloat
	\includegraphics[width=0.65\textwidth]{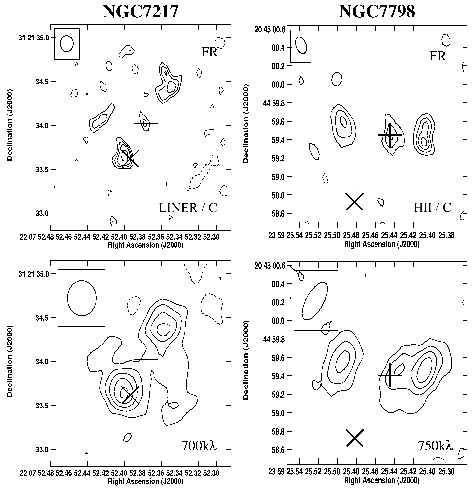}
        \caption{(the quality of the images has been reduced due to
          size limit for ArXiv submission) eMERLIN 1.5-GHz radio
          images of the galaxies with an identified radio core. For
          each galaxy two panels are shown. The upper panel shows the
          full-resolution map, while the lower panel shows the
          low-resolution map obtained with a uv-tapered scale written
          in the panel (in k$\lambda$). For four galaxies (NGC~3665,
          NGC~5005, NGC~5194, and NGC~5985) a third radio map is
          presented corresponding to a lower resolution map (see the
          scale and map parameters in Tab~\ref{contours}). The maps
          for the each source are on the same physical scale, unless
          there is a zoom-in map, marked with a box around the
          selected source. The restoring beam is presented as an
          ellipse on one of the corners of each of the maps.  The
          contour levels of the images are presented in
          Table~\ref{contours}. The $\times$ mark indicates the
          optical galaxy centre taken from NED, while the $+$ symbol
          marks the radio core position, if identified. In the upper
          panels, the optical (LINER, Seyfert, H{\sc ii}, and ALG) and
          radio (A, B, C, D, E, see Section~\ref{core-ident} for
          description) classifications of the sources are reported.}
\end{figure*}

\begin{figure*}
	\includegraphics[width=0.93\textwidth]{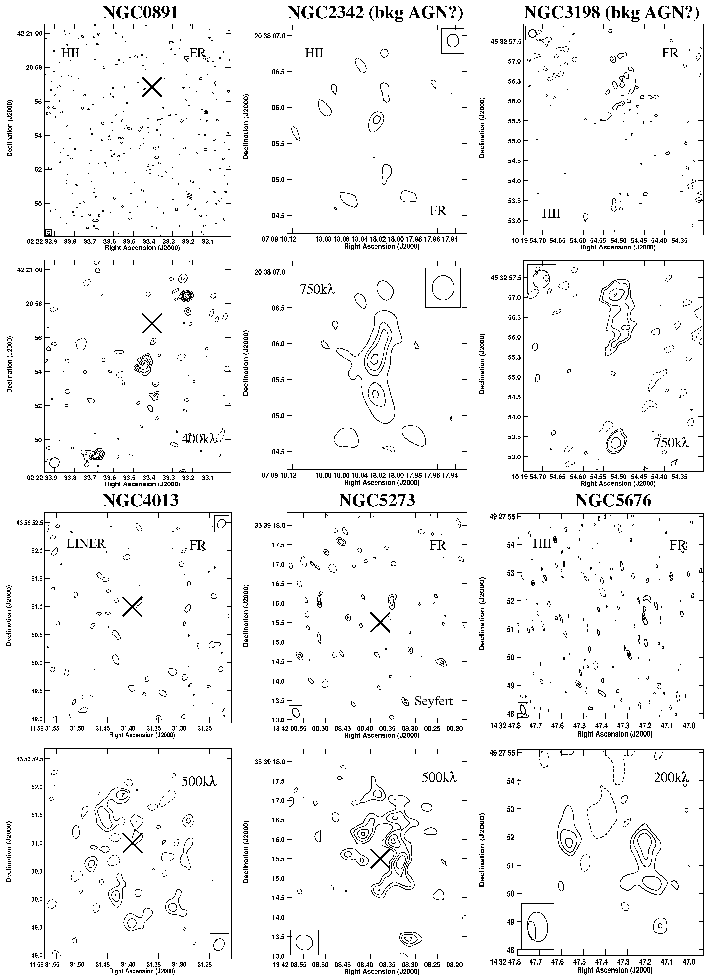}
        \caption{eMERLIN 1.5-GHz images of the unidentified Palomar galaxies. See last page of this figure for details.}
  \label{maps2}
\end{figure*}

\addtocounter{figure}{-1}
\begin{figure*}
%\ContinuedFloat
	\includegraphics[width=0.65\textwidth]{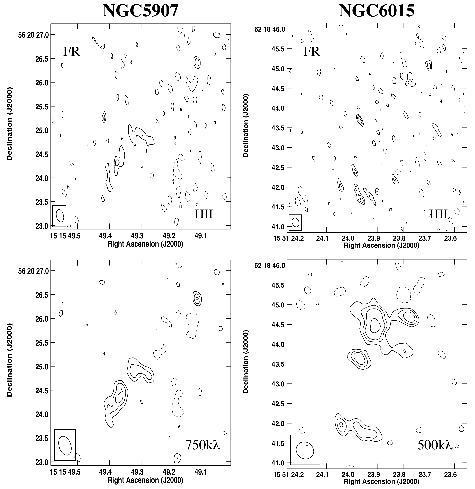}
        \caption{(the quality of the images has been reduced due to
          size limit for ArXiv submission) eMERLIN 1.5-GHz radio
          images of the galaxies, which are detected but no radio core
          is identified. For each galaxy two panels are shown. The
          upper panel shows the full-resolution (FR) map, while the
          lower panel shows the low-resolution map obtained with a
          uv-tapered scale written in the panel (in k$\lambda$). The
          restoring beam is presented as an ellipse on one of the
          corners of each of the maps.  The contour levels of the maps
          are presented in Table~\ref{contours}. The $\times$ mark
          indicates the optical galaxy centre taken from NED. The maps
          for the each source are on the same physical scale. The
          images presented for NGC~2342 and NGC~3198 show jet-like
          features of possible background AGN, while NGC~5676,
          NGC~5907, and NGC~6015 show off-nuclear regions, possibly
          attributed to SF (see Section~\ref{unidentified} for
          details).  In the upper panels, the optical (LINER, Seyfert,
          H{\sc ii}, and ALG) classification of the sources is
          reported.}
\end{figure*}

\onecolumn
\fontsize{7}{10}\selectfont
\begin{landscape}
\begin{center}
\begin{longtable}{C{1.2cm}|C{0.4cm}ccccccc|C{0.05cm}ccccc|C{1.8cm}}

\caption[Properties of the sample.]{Properties of the detected sources.} 
\label{tabdet} \\

%This is the header for the first page of the table...
\hline \hline

%This is the header for the first page of the table...
\hline \hline

     &   \multicolumn{8}{c}{Full resolution}                                                   &  \multicolumn{7}{|c}{Low resolution} \\
name & comp & $\theta_{\rm M}\times\theta_{\rm m}$  & PA$_{\rm d}$ & rms  &  $\alpha$(J2000) & $\delta$(J2000) &  F$_{\rm peak}$ & F$_{\rm tot}$ & &  $u{-}v$     &  $\theta_{\rm M}\times\theta_{\rm m}$ &  PA$_{\rm d}$ & rms &  F$_{\rm peak}$ &morph/size \\
     &        &arcsec & deg  & mJB &                 &                &  mJB       &  mJy      & &  k$\lambda$ & arcsec & deg   &  mJB &   mJB      & \\

\hline	
\endfirsthead

%This is the header for the remaining page(s) of the table...
\multicolumn{3}{c}{{\tablename} \thetable{} -- Continued} \\[0.5ex]
\hline \hline 

   &   \multicolumn{8}{c}{Full resolution}                                                   &  \multicolumn{7}{|c}{Low resolution} \\
name & comp & $\theta_{\rm M}\times\theta_{\rm m}$  & PA$_{d}$ & rms  &  $\alpha$(J2000) & $\delta$(J2000) &  F$_{\rm peak}$ & F$_{\rm tot}$ &  & $u{-}v$    &  $\theta_{\rm M}\times\theta_{\rm m}$ &  PA$_{d}$ & rms &  F$_{\rm peak}$  & morph/size \\
     &        &arcsec & deg  & mJB &                 &                &  mJB       &  mJy      &  & k$\lambda$ & arcsec & deg   &  mJB &   mJB       & \\

\hline
\endhead

%This is the footer for all pages except the last page of the table...
\hline
  \multicolumn{16}{c}{{Continued on Next Page}} \\
\endfoot

%This is the footer for the last page of the table...
  \\[-1.8ex] 
\endlastfoot

%Now the data...&  &      &  0.24$\times$0.17 & 37.2  &        & 
NGC~147  &      &  0.13$\times$$<$0.198  &  28.40    &   0.073       &   00  33 11.894 &  48 30 29.69  &    $<$0.28  &  0.95   &  &750  &  0.39$\times$0.07 & 142.8  &   0.070     &  0.5$\pm$0.09  & core (A)  \\
\cmidrule(rl){15-15}
          & Tot  &                     &          &        &               &              &                &      &     &                      &        &        &  & 0.95$\pm$0.10 &            \\
\hline
NGC~266  &  core  &  0.16$\times$$<$0.26 &   174.1  &  0.095  & 00 49 47.810 &  32 16 39.83  &  1.24$\pm$0.10   &  1.31  &  &750  &   0.20$\times$0.12 &  159.0  &  0.10    &  1.28$\pm$0.10    & core-jet (A) 1.5$\arcsec\to$480pc  \\
\cmidrule(rl){15-15}
  & Tot  &                     &          &        &               &              &                &      &     &                      &        &   &       &  1.5$\pm$0.1           &  \\
\hline
NGC~278  &      &   0.19$\times$$<$0.34   & 109.0  &   0.050  &    00 52 04.203 &  47 33  01.38  &  0.26$\pm$0.06  & 0.29  &  &750   &  0.48$\times$0.05 &  157.3  &  0.054     &   2.1$\pm$0.05     & \multirow{3}{1.8cm}{\centering  core (A) + components} \\  
             &  core   &  0.17$\times$$<$0.25 & 62.4  &  0.050  &   00 52 04.368  &  47 33 02.13 &  0.26$\pm$0.06 &  0.34   &  &750    & 0.16$\times$$<$0.30 &  146.6 &    0.054     &   0.24$\pm$0.05  &   \\
          &             &  0.04$\times$$<$0.28 &  74.7  &  0.050  &  00 52 0.4527  &  47 33 02.899 &   0.18$\pm$0.06  & 0.20  &   &750   &   0.17$\times$$<$0.45 &  159.8  &  0.054  &  0.24$\pm$0.05  &    \\
\cmidrule(rl){15-15}
      & Tot  &                     &          &        &               &              &                &      &     &               &       &        &          &  2.7$\pm$0.2 &            \\
\hline
NGC~315 & core  &  0.01$\times$0.08 &   67.3  &  5.5     &    00 57 48.883 &   30 21 08.82  &    490$\pm$30  & 503  &   & 750  &  0.10$\times$$<$0.21  &   71.0   &  4.0   &   331$\pm$26  & core-jet (A) \\
\cmidrule(rl){15-15}
    & Tot  &                     &          &        &               &              &                &      &     &      &                &        &          &  520$\pm$50     &            \\
\hline
NGC~410 & core  &  0.53$\times$0.11 &  89.6  &  0.14    &     01 10 58.901 & 33 09 06.98 &  2.05$\pm$0.14  & 3.62 &  &750  &  0.55$\times$0.10  &    93.6 &  0.16   &  2.21$\pm$0.16  &   core-jet (A)\\
\cmidrule(rl){15-15}
    & Tot  &                     &          &        &               &              &                &      &    & &                      &        &          &  3.8$\pm$0.2     &            \\
\hline
NGC~507 & core &   0.24$\times$$<$0.41 &  68.4 &  0.18    &  01 23 39.863  &   33 15 23.00  &   0.93$\pm$0.18 & 1.02 & &750 & 0.42$\times$0.24 &  16.7  &   0.26   &  1.01$\pm$0.26  &   core-jet (A) \\
\cmidrule(rl){15-15}
    & Tot  &                     &          &        &               &              &                &      &     &     &                 &        &          &  1.5$\pm$0.1     &            \\
\hline
NGC~777 &     & 0.17$\times$$<$0.23 & 152.6  & 0.077  &  02 00 14.894 & 31 25 46.33  &  0.36$\pm$0.09 & 0.59 & &600 &  0.88$\times$$<$0.22 &  163.8 & 0.081 & 0.32$\pm$0.12  & \multirow{3}{1.8cm}{\centering core (A) + components} \\
      & core & 0.17$\times$$<$0.14 & 171.7    & 0.077  &  02 00 14.908 & 31 25 45.84  &  0.74$\pm$0.10 & 0.84 & &600 &  0.16$\times$$<$0.25 &  108.9 &  0.081 & 0.84$\pm$0.12  & \\
\cmidrule(rl){15-15}
       & Tot  &                     &          &        &               &              &                &      &     &    &                  &        &        &  1.2$\pm$0.2 &            \\
    \hline
NGC~972  & core & 0.06$\times$$<$0.06    & 164.2 & 0.069  &   02 34 13.34 & 29 18 40.65 &  0.25$\pm$0.08 & 0.38  & &750 &  0.42$\times$0.26 &  157.6  & 0.069  & 0.26$\pm$0.09   &\multirow{3}{1.8cm}{\centering one-sided jet (B) 1.7$\arcsec\to$185pc}\\
        &     & $<$0.24$\times$$<$0.17 &  162.2 & 0.069  &   02 34 13.35 & 29 18 40.40 &  0.30$\pm$0.08 & 0.33  & &750 &  0.45$\times$0.29 &   148.4 & 0.069  & 0.35$\pm$0.09 &\\
        &     &   0.40$\times$0.27     & 152.5 & 0.069  &   02 34 13.56 & 29 18 34.99 &  0.38$\pm$0.08 & 1.44  & &750 &  0.39$\times$0.26 &   145.8 & 0.069  & 0.55$\pm$0.09 &  \\
\cmidrule(rl){15-15}
        & Tot  &                     &          &        &               &              &                &      & &    &                      &        &          &  1.4$\pm$0.2 &             \\
\hline
NGC~2273  &      & 0.19$\times$0.06 &  72.1  & 0.930 & 06 50 08.577   & 60 50 44.55 &  1.64$\pm$0.10 &  2.26 & &750 &       &    & 0.12   &  $<$3.25       &    \multirow{4}{1.8cm}{\centering twin jet + extended emission (E) 2.2$\arcsec\to$280pc}\\
        & core & 0.25$\times$0.16 &  169.4 & 0.930 & 06 50 08.647   & 60 50 44.93 &  2.99$\pm$0.10 &  7.12 & &750 & 0.29$\times$0.19 &  61.9 & 0.12   &  4.87$\pm$0.13 & \\
        &      & 0.30$\times$0.11 &  48.3 & 0.930 & 06 50 08.701   & 60 50 44.94 &  1.68$\pm$0.09 &  4.17 & &750 &             &  & 0.12   &  $<$4.82       &   \\
\cmidrule(rl){15-15}
         & Tot  &                     &          &        &               &              &                &  &    &     &                      &        &          & 66.1$\pm$3.0 &            \\
\hline
NGC~2342&         &                      &        &  0.065  &  07 09 18.029 & 20 38 09.26   &  $<$0.41  &     & &750  & 0.53$\times$0.34 &  164.3 & 0.068  & 0.51$\pm$0.09 &   \multirow{3}{1.8cm}{\centering complex (E) 2.5$\arcsec\to$865pc}\\  % su ogg sotto
        & core   &  0.17$\times$0.08      &   151.6 &  0.065  &  07 09 18.069 & 20 38 09.59 &  0.41$\pm$0.08& 0.54 & &750  & 0.53$\times$0.28 & 172.4 & 0.068  & $<$5.8  &    \\  
         &       &  0.26$\times$0.03    &   133.5 &  0.065  &  07 09 18.080 & 20 38 09.28 &  0.30$\pm$0.08& 0.52 & &750  & 0.63$\times$0.34 &  135.2 & 0.068  & 0.59$\pm$0.09 &  \\
\cmidrule(rl){15-15}
         & Tot  &                     &          &        &               &              &                &      &     &  &                    &        &        &  5.0$\pm$0.2 &            \\
\hline
UGC~3828    & lobeN     & 0.22$\times$$<$0.12 & 73.2  & 0.071  &  07 24 35.726 & 57 58 03.47 & 0.32$\pm$0.10& 0.51   & &750  &  0.32$\times$0.19&  93.3 & 0.078 & 0.49$\pm$0.11 &  \multirow{3}{1.8cm}{\centering double jet (C) 1$\arcsec\to$238pc}\\
          & lobeS     & 0.16$\times$$<$0.29 & 1.4  & 0.071  &  07 24 35.733 &  57 58 02.83& 0.29$\pm$0.10& 0.35   & &750  &  0.23$\times$$<$0.38&  178.1 & 0.078 & 0.38$\pm$0.12 &   \\ 
           & core & 0.13$\times$0.23 & 156.69  & 0.071  &  07 24 35.7372& 57 58 03.261& 0.30$\pm$0.10& 0.31   & &750  &                         &        & 0.078 & $<$4.8    &    \\
 \cmidrule(rl){15-15}
      & Tot  &                     &          &        &               &              &                &      &  &   &                      &        &          &  1.09$\pm$0.10 &            \\
\hline
NGC~2300  & core  & 0.18$\times$0.06  &  138.2  &  0.060       & 07 32 19.945 &  85 42 32.41  &   0.99$\pm$0.08 & 1.25 & & 750  & 0.22$\times$0.09  &  116.3 & 0.065     &  1.21$\pm$0.08  &   core-jet (A)  0.75$\arcsec\to$103pc \\
 \cmidrule(rl){15-15}
    & Tot  &                     &          &        &               &              &                &      &     &    &                  &        &          &  1.5$\pm$0.1 &             \\
\hline
UGC~4028 &  lobeW &            &              &    0.10     &  07 50 49.810  &   74 21 29.15   &   $<$0.30        &    &  &750  &  0.21$\times$0.26 & 53.2  &  0.072       &  0.44$\pm$0.08 &  \multirow{2}{1.8cm}{\centering twin jets (C) 1.5$\arcsec\to$406pc}\\
       & core   &  $>$0.18$\times$$>$14  &    155   &  0.10       & 07 50 49.931   &  74 21 29.13    &  0.33$\pm$0.10     &  0.33  &  &750 & 0.12$\times$$<$0.26  &  127.8 &   0.072   &   0.23$\pm$0.08 &   \\
       &  lobeW &            &              &   0.10      &  07 50 50.0749  &   74 21 28.77   &   $<$0.30       &      &  &750 &  0.46$\times$0.24 & 55.96   & 0.072     &  0.43$\pm$0.08 &   \\
 \cmidrule(rl){15-15}
     & Tot  &                     &          &        &               &              &                &      &     &          &            &        &          &  1.0$\pm$0.1      &           \\
\hline
NGC~2639 &      &  0.32$\times$0.25 & 121.1 & 0.13  &  08 43 38.034  &  50 12 20.07& 17.5$\pm$0.1  &  25.2 &  &750 & 0.38$\times$0.26 & 102.1 & 0.29   & 24.5$\pm$0.3  & \multirow{3}{1.8cm}{\centering twin jets (C)  1.5$\arcsec\to$340pc}\\
        & core &  0.42$\times$0.08 &108.4 & 0.13  &  08 43 38.080  & 50 12 19.99 & 12.6$\pm$0.1  &  17.8 &  &750 & 0.42$\times$0.18 & 105.0 & 0.29   &17.6$\pm$0.3   & \\
        &      &  0.29$\times$0.20 &141.8 & 0.13  &  08 43 38.151  & 50 12 19.82 & 3.40$\pm$0.1   & 8.3   &  &750 &                 &  & 0.29  &$<$11.4       &   \\
 \cmidrule(rl){15-15}
     & Tot  &                     &          &        &               &              &                &      &     &                      &        &    &      &  98$\pm$10 &            \\
\hline
NGC~2634 &  core  & 0.25$\times$$<$0.3 &  38.8 &  0.082       &  08 48 25.095 & 73 58 02.15 &      0.39$\pm$0.11 &  0.59 &  &750   & 0.10$\times$$<$0.4 & 31.4 &  0.081  & 0.41$\pm$0.12   &       core (A) \\
\cmidrule(rl){15-15}
     & Tot  &                     &          &        &               &              &                &      &     &       &               &        &          &  0.45$\pm$0.1 &            \\
\hline

NGC~2681     & jetN  &                &       &  0.92 &   08 53 32.693 &   51 18 49.78&  $<$3.7       &       & &750 &  0.68$\times$0.18 & 141.1 & 0.099 &0.57$\pm$0.12  & \multirow{3}{1.8cm}{\centering twin jets (C) 1.5$\arcsec\to$70pc}  \\
          & core & 0.17$\times$0.07 &  56.7 &  0.92 &   08 53 32.718 & 51 18 49.15 & 1.03$\pm$0.10  &  1.49 & &750 &  0.20$\times$0.16 & 175.2 & 0.099 &1.22$\pm$0.12 &  \\
          &      & 0.20$\times$0.16 &  169.5 &  0.92 &   08 53 32.733 &  51 18 48.77& 0.473$\pm$0.10 &  0.90 & &750 &  0.41$\times$0.28 &  161.6 & 0.099 &0.54$\pm$0.12 &   \\
\cmidrule(rl){15-15} 
     & Tot  &                     &          &        &               &              &                &      & &    &                      &        &          &  3.1$\pm$0.2 &            \\
\hline
NGC~2655 &  core & 0.17$\times$0.07  & 40.8  &  0.12    &    08 55 37.950 &   78 13 23.63  &  36.81$\pm$0.13  & 51.5 &  &750  &  0.37$\times$0.10  & 38.2  & 0.16 &  53.04$\pm$0.17       & \multirow{2}{1.8cm}{\centering \tiny{S-shaped jet + components (E) 2$\arcsec\to$204pc}}\\
      &        & 0.17$\times$0.10  &  148.0 &  0.12   &    08 55 38.09  &  78 13 22.96   &   1.62$\pm$0.13  & 2.35 &  &750  & 0.45$\times$0.31  & 170.1  &  0.16 & 2.09$\pm$0.17   &        \\
   & Tot  &                     &          &        &               &              &                &      &     &     &                 &        &          &  88$\pm$2     &            \\
\hline
NGC~2841 
         & lobeN &  0.45$\times$$<$0.12 & 50.4  & 0.074 &  09 22 02.611 & 50 58 36.39 & 0.40$\pm$0.09 &  0.51   & &600 &  0.36$\times$$<$0.40 &  46.9 & 0.081 & 0.51$\pm$0.11 &  \multirow{3}{1.8cm}{\centering twin jets (C) 1.5$\arcsec\to$68pc}\\
         & core & 0.19$\times$$<$0.12 & 52.7   & 0.074 &  09 22 02.652 & 50 58 36.03 & 0.27$\pm$0.09 &  0.31   & &600 &  0.36$\times$$<$0.45 &   166.9 & 0.081 & 0.31$\pm$0.11 &  \\
         & lobeS &  0.02$\times$$<$0.19 & 52.7  & 0.074 &   09 22 02.676&  50 58 35.65& 0.31$\pm$0.09 &  0.63   & &600 &  0.23$\times$$<$0.51 & 114.2  & 0.081 & 0.52$\pm$0.11 &  \\
\cmidrule(rl){15-15}
      & Tot  &                     &          &        &               &              &                &      &   &  &                      &        &          &  1.15$\pm$0.15 &           \\
\hline
NGC~3198 & core       &   0.15$\times$$<$0.08 &  125.9 & 0.114     &  10 19 54.931 & 45 32 59.50  &  0.50$\pm$0.12 & 0.52  & & 750  &   0.45$\times$$<$0.05  &  137.7  & 0.81     & 0.49$\pm$0.09 & \multirow{2}{1.8cm}{\centering core+ component (A)}   \\
       &        &                       &       &  0.114     &  10 19 54.994 &  45 32 58.53 &    $<$0.33      &       & & 750  &   0.47$\times$$<$0.19  &  120.2  &  0.81    &  0.30$\pm$0.09 &    \\
\cmidrule(rl){15-15}
      & Tot  &                     &          &        &               &              &                &      &     &    &                  &        &          &  0.74$\pm$0.15 &            \\
\hline
NGC~3414 & core  & 0.07$\times$$<$0.12 &  162.4 & 0.060     &   10 51 16.211 &   27 58 30.28  &   1.56$\pm$0.10  & 1.67 &  &750 &   0.32$\times$0.07 &  159.7  & 0.11     &   1.89$\pm$0.13  & twin jet (C) 1$\arcsec\to$95pc \\
\cmidrule(rl){15-15}
      & Tot  &                     &          &        &               &              &                &      &     &       &               &        &          &  2.1$\pm$0.1 &           \\
\hline
NGC~3430 &      &            &                   &    0.10       &    10 52 11.384  &    32 57 02.11 &   $<$0.30      &         &  & 750  & 0.29$\times$$>$0.28  &  113.5 & 0.069       &   0.33$\pm$0.08 &  \multirow{4}{1.8cm}{\centering double-lobed/complex (E) 2.2$\arcsec\to$230pc}   \\
      & core   &             &                   &    0.10       &   10 52 11.403  &    32 57 01.63 &   $<$0.30      &         &  &750   & 0.42$\times$$>$0.29 &  102.6 & 0.069      &   0.43$\pm$0.08 &    \\  
     &         &    0.17$\times$$<$0.13   &    167.4 &  0.10       & 10 52 11.433  &    32 57 02.32 &   0.41$\pm$0.11 & 0.42  &   & 750   & 0.64 $\times$0.59 &    55.3 & 0.069      &   0.42$\pm$0.11 &   \\  
 \cmidrule(rl){15-15}
     & Tot  &                     &          &        &               &              &                &      &     &              &        &        &          &  1.2$\pm$0.2      &            \\
\hline  
NGC~3432 &  core    &    $>$0.38$\times$$>$0.21 & 173.4 &    0.11      &  10 52 31.012   & 36 37 08.50  & 0.60$\pm$0.14   &   0.61  &  &750  &  $>$0.53$\times$$>$0.44  &  142.7  &  0.13     &  0.47$\pm$0.14  & core (A)  \\ 
\cmidrule(rl){15-15}
      & Tot  &                     &          &        &               &              &                &      &     &         &             &        &          &  0.62$\pm$0.15      &            \\
\hline
NGC~3665 &   \multirow{2}{*}{blobN} &            &         &    \multirow{2}{*}{0.061}       &   \multirow{2}{*}{11 24 43.482}  &   \multirow{2}{*}{38 45 47.89}  &    \multirow{2}{*}{$<$5.3}  &  &  \rdelim\{{2}{20pt}& 750 & 1.60$\times$0.74   &     109.7  &   0.35& 1.71$\pm$0.35 &   \multirow{4}{1.8cm}{\centering extended one-sided jet (B) 4.2$\arcsec\to$580pc}   \\
       &       &           &          &              &                              &                               &                          &                   &  & 500  &  1.78$\times$0.82 & 107.8      &  0.45     & 2.27$\pm$0.45  &    \\
      &    \multirow{2}{*}{jetN}  &          &         &    \multirow{2}{*}{0.061}       &   \multirow{2}{*}{11 24 43.592}   &  \multirow{2}{*}{38 45 46.80} &   \multirow{2}{*}{0.71$\pm$0.08} & \multirow{2}{*}{1.13}   & \rdelim\{{2}{20pt} & 750  &0.64$\times$0.25  &   143.5 & 0.35 & 2.00$\times$0.35  &    \\ 
      &       &           &          &              &                              &                               &                          &                     & &500  &           &        &         &   $<$3.5  &   \\
      & \multirow{2}{*}{core} &  \multirow{2}{*}{0.10$\times$0.01} &  \multirow{2}{*}{20.5}  &  \multirow{2}{*}{0.061}    &   \multirow{2}{*}{11 24 43.626}  & \multirow{2}{*}{8 45 46.31}  &  \multirow{2}{*}{3.78$\pm$0.08} &  \multirow{2}{*}{4.25}  &  \rdelim\{{2}{20pt} &750  &  0.34$\times$0.22  &  0.47 &  0.35   &    4.09$\pm$0.35   &   \\ 
     &                        &                                    &                         &                       &                                 &                                &                                 &                        &  &500   &  0.44$\times$0.32   &  167.4 &  0.45   &   4.71$\pm$0.44   &    \\
\cmidrule(rl){15-15}
     & Tot  &                     &          &        &               &              &                &      &     &      &                &        &          &  25$\pm$3 &            \\
\hline
NGC~3675  &  core  &             &          &      0.080     &     11 26 08.627  & 43  35 08.91  &  $<$0.53   &       & &  750 &  0.31 $\times$$<$0.15  & 174.5  & 0.077   &   0.31$\pm$0.08  &   core + comp (A) \\
\cmidrule(rl){15-15}
       & Tot  &                     &          &        &               &              &                &      &     &     &                 &        &          &  0.5$\pm$0.1 &            \\
\hline
NGC~3938 & core   &   $<$0.15$\times$$<$0.07 &   144.6  &  0.074      &     11 52 49.393 & 44 07 14.57 &  0.28$\pm$0.07 & 0.33 & &750 &  0.41$\times$0.08 & 90.4 &  0.083  & 0.32$\pm$0.08  & core (A) \\
\cmidrule(rl){15-15}
          & Tot  &                     &          &        &               &              &                &      &     &      &                &        &        &  0.6$\pm$0.1 &           \\
\hline
NGC~4051  & lobeW   &  0.28$\times$0.07  &   132.2 & 0.071  & 12 03 09.576 & 44 31 52.572 & 0.35$\pm$0.07 & 0.57  & & 750 &  0.28$\times$0.07  & 132.2 &  0.20    &   8.41$\pm$2.1    & \multirow{3}{1.8cm}{\centering triple source  (C) 1$\arcsec\to$50pc}\\
        & core   & 0.15$\times$0.11  &  145.2  &   0.071  & 12 03 09.612 & 44 31 52.69 &  0.38$\pm$0.07 & 0.63  &  &750 & 0.47$\times$0.14  & 175.3 &  0.20    &   7.96$\pm$2.1   &  \\
        & lobeE  & 0.10$\times$0.06  &  78.3   &  0.071   & 12 03 09.648 &  44 31 52.85&  0.86$\pm$0.07 &  1.08 & &750  & 0.29$\times$0.08  & 144.4 &  0.20    &  15.9$\pm$2.1    &   \\ 
\cmidrule(rl){15-15}
     & Tot  &                     &          &        &               &              &                &      &     &                      &        &          &  64.5$\pm$7.0 &          \\
\hline
NGC~5005 &  lobeN &      &       & 0.093 &  13 10 55.512&  37 03 38.96 &    $<$0.24      &     & &200 & 2.41$\times$2.32 & 153.7  & 0.34   &  2.8$\pm$0.3  & \multirow{5}{1.8cm}{\centering double-lobed jets (D) 26$\arcsec\to$1.7kpc}\\
       &       & 0.54$\times$0.28 &  106.3  & 0.093 &  13 10 56.247&  37 03 32.87 & 2.08$\pm$0.09  &  10.6& & 750 &     &      & 0.090  &  $<$8.4       & \\
       &  \multirow{2}{*}{core}  &   \multirow{2}{*}{0.34$\times$0.23} &    \multirow{2}{*}{138.3} &  \multirow{2}{*}{0.093} &   \multirow{2}{*}{13 10 56.260} &   \multirow{2}{*}{37 03 32.74} &  \multirow{2}{*}{2.37$\pm$0.09}  &  \multirow{2}{*}{7.49} & \rdelim\{{2}{20pt} & 750 & 0.56$\times$0.28 &  127.8 & 0.090  &  4.44$\pm$0.10 &   \\
         &       &                   &        &       &               &              &                &      & & 200&  0.88$\times$0.41 &  128.7 & 0.34   & 15.2$\pm$0.3    & \\
       &lobeS &   &         & 0.093 &  13 10 57.009& 37 03 26.81  &    $<$0.29      &      & & 200  & 2.20$\times$2.10 & 123.1  & 0.34   &  2.5$\pm$0.3  &\\
\cmidrule(rl){15-15}
    & Tot  &                     &          &        &               &              &         &       &      &     &                      &        &          & 42.3$\pm$3.0 &            \\ 
\hline
NGC~5194  &lobeN  &           &       &   0.101  &   13 29 52.510     &   47 11 45.97  &   $<$0.34              &       &  & 500 &      &      & 0.078     &  4.7$\pm$0.2$^{tot}$ &  \\
        & \multirow{2}{*}{core}  &  \multirow{2}{*}{0.18$\times$0.09} &  \multirow{2}{*}{145.3}  &  \multirow{2}{*}{0.091} &  \multirow{2}{*}{13 29 52.708} &  \multirow{2}{*}{47 11 42.80} &   \multirow{2}{*}{1.08$\pm$0.09}  & \multirow{2}{*}{1.74} & \rdelim\{{2}{20pt}  & 750 & 0.23$\times$0.10 & 159.8 &  0.076  &  1.4$\pm$0.1 &  \multirow{3}{1.8cm}{\centering double-lobed (D) on 10$\arcsec\to$370pc} \\
         &       &                  &         &       &               &             &                 &       & & 500 & 0.50$\times$0.10 & 170.3 & 0.078     & 1.7$\pm$0.1  &   \\
        &lobeS  &              &         &   0.111    &   13 29 52.819 &  47 11 38.92  &  $<$0.35             &       &  &500 &     &          & 0.078     &  5.8$\pm$0.2$^{tot}$ &    \\
\cmidrule(rl){15-15}
        & Tot  &                     &          &        &               &              &                &      &     &     &                 &        &          & 13.1$\pm$1.5 &            \\ 
\hline
NGC~5195 & jetW &                 &     &  0.080 &    13 29 59.493  &    47 15 58.07 &   $<$0.24       &       &  & 750 &    0.93$\times$0.59 &   35.9 &  0.11 & 0.33$\pm$0.1   & \multirow{3}{1.8cm}{\centering twin jet (C) 1$\arcsec\to$37pc}\\
         & core & 0.21$\times$0.02 & 55.0 & 0.080 &  13 29 59.535 &  47 15 58.37 & 0.51$\pm$0.09 & 0.77 & &750 &    &     & 0.11 & $<$6.3   &  \\
         & jetE &                 &     &  0.080 &   13 29 59.539 &    47 15 58.50 &   $<$0.25       &        &  &750 & 0.54$\times$0.34  &   158.5  &  0.11 & 0.61$\pm$0.1  & \\
        & Tot  &                     &          &        &               &              &                &      & &    &                      &        &          & 2.5$\pm$0.2 &          \\ 
\hline
NGC~5353 &  core & 0.12$\times$0.04 & 8.4  &   0.18 &  13 53 26.682 & 40 16 58.83 & 16.5$\pm$0.2 &   21.0 & &500 & 0.12$\times$0.07 & 21.1 & 0.14  & 17.6$\pm$0.2 &  core (A) \\
 \cmidrule(rl){15-15}
& Tot  &                     &          &        &               &              &                &      & &    &                      &        &          & 17.9$\pm$0.1 &           \\ 
\hline
NGC~5377   & jetS &      &         & 0.091 &   13 56 16.627   &    47 14 08.52    & $<$5.1         &      & &500 & 0.60$\times$0.21 & 4.0 & 0.11     &  0.53$\pm$0.12 &  \multirow{3}{1.8cm}{\centering twin jets (C)   1.8$\arcsec\to$227pc}\\
         & core & 0.11$\times$0.04 & 168.2 & 0.091 &  13 56 16.631 &  47 14 09.16 &  0.30$\pm$0.09 & 0.36 & &500 & 0.19$\times$$<$0.33 & 57.8 & 0.11     &  0.38$\pm$0.13 &  \\
         & jetN &              &    & 0.091 &    13 56 16.658   &   47 14 09.49    & $<$1.0         &      & &500 & 0.50$\times$0.27 & 156.1 & 0.11     &  0.53$\pm$0.12 & \\
\cmidrule(rl){15-15}
         & Tot  &                     &          &        &               &              &                &   &   &     &                      &        &          & 1.3$\pm$0.2 &            \\
\hline
NGC~5448 &      &  0.10$\times$$<$0.21 &  28.0 &  0.086 & 14 02 50.015  & 49 10 21.89 &  0.35$\pm$0.10 & 0.50 & &750 & 0.19$\times$0.14 & 21.9 & 0.082 & 0.37$\pm$0.11 &  \multirow{3}{1.8cm}{\centering twin jet (C) 1$\arcsec\to$141pc} \\
        & core &  0.22$\times$0.05 &   59.1  &  0.086 & 14 02 50.035  & 49 10 21.75 &  0.28$\pm$0.09 & 0.45 & &750 & 0.59$\times$0.34 & 137.8 & 0.082 & 0.36$\pm$0.11 &   \\
        &      &  0.18$\times$$<$0.12 &  86.1 &  0.086 & 14 02 50.057  & 49 10 21.56 &  0.39$\pm$0.10 & 0.53 & &750 & 0.43$\times$$<$0.21 & 98.6 & 0.082 & 0.45$\pm$0.11 &\\
         &      &  0.19$\times$0.10 &  53.5 &  0.086 & 14 02 50.069  & 49 10 21.34 &  0.30$\pm$0.10 & 0.40 & & 750 & 0.69$\times$0.24 & 172.6 & 0.082 & 0.32$\pm$0.10 & \\ 
        & Tot  &                     &          &        &               &              &                &     & &     &                      &        &          & 1.5$\pm$0.2 &           \\
\hline
NGC~5866 & lobeN    &      &       &  0.097 & 15 06 29.344 & 55 45 40.77 & $<$0.32         &     & & 200  & 1.01$\times$$<$0.66 & 56.0  &  0.133    & 0.95$\pm$0.19 &   \multirow{3}{1.8cm}{\centering core+jet, two lobes (D)  16$\arcsec\to$960pc}\\
       &  core  & 0.12$\times$0.07  & 20.8   &  0.097 & 15 06 29.492 & 55 45 47.66 & 7.6$\pm$0.1     & 9.3 & & 200  & 0.44$\times$$<$0.22 & 12.3  &  0.133    & 10.2$\pm$0.2 &  \\
       &  lobeS   &     &    &  0.097 & 15 06 29.686 & 55 45 55.49 & $<$0.37         &     &  &200  & 1.11$\times$$<$0.75 & 163.8  &  0.133    & 0.83$\pm$0.19 &   \\
\cmidrule(rl){15-15}
     & Tot  &                     &          &        &               &              &         &       &      &     &                      &        &          &  13.6$\pm$0.5 &            \\ 
\hline
NGC~5879 & core &  0.19$\times$$<$0.14 & 27.2 &  0.071  & 15 09 46.801 & 57 00 01.58 & 0.37$\pm$0.10   & 0.51 & & 750 & 0.21$\times$0.16 &  142.3 &  0.068  & 0.37$\pm$0.09 &  core (A) \\
\cmidrule(rl){15-15}
         & Tot  &                     &          &        &               &              &                &     &  &     &                      &        &          & 0.49$\pm$0.12 &            \\ 
\hline
NGC~5985 &  \multirow{2}{*}{lobeN}    & \multirow{2}{*}{} & \multirow{2}{*}{} &  \multirow{2}{*}{0.057}  & \multirow{2}{*}{15 39 37.031} & \multirow{2}{*}{59 19 55.95} & \multirow{2}{*}{$<$0.31}   &   \multirow{2}{*}{}  & \rdelim\{{2}{20pt} & 380 & 1.94$\times$$<$0.21 & 22.9  &  0.066   & 0.22$\pm$0.09 &     \multirow{3}{1.8cm}{\centering double-lobed (D) 4$\arcsec\to$530pc}\\
        &       &                  &        &       &               &              &                &      & &200  &  1.06$\times$$<$1.91 & 2.5   & 0.34   & 0.25$\pm$0.10   &   \\   
        &  core &                 &         &  0.057  & 15 39 37.062 & 59 19 55.09 & $<$0.17      &        & &380 & 0.36$\times$0.32 & 15.7  &  0.066   & 0.21$\pm$0.09 &   \\
        & \multirow{2}{*}{lobeS}    &  \multirow{2}{*}{} & \multirow{2}{*}{} &  \multirow{2}{*}{0.057}  & \multirow{2}{*}{15 39 37.130} & \multirow{2}{*}{59 19 53.45} & $<$0.32      &        & \rdelim\{{2}{20pt} & 380 & 0.45$\times$0.25 & 40.8  &  0.066  &  0.30$\pm$0.09 &  \\
        &       &                  &        &       &               &              &                &      & & 200 &  1.35$\times$0.66 & 145.1  &   0.34 & 0.35$\pm$0.10   &   \\
\cmidrule(rl){15-15}
        & Tot  &                     &          &        &               &              &                &      & &    &                      &        &          & 0.73$\pm$0.15 &             \\ 
\hline
NGC~6702  &      &  0.08$\times$$<$0.24 & 19.6   & 0.074   &   18 46 57.534 & 45 42 20.57 & 0.20$\pm$0.08  & 0.23 & &600 & 0.20$\times$$<$0.57 &  133.1 & 0.066  & 0.27$\pm$0.09 &  \multirow{3}{1.8cm}{\centering twin jet (C) 2$\arcsec\to$656pc}\\
        & core &                     &         & 0.074   &   18 46 57.564 & 45 42 20.36 &  $<$0.22       &      & &600 & 0.36$\times$0.10 & 96.6 & 0.066  & 0.22$\pm$0.09 &   \\
        &      & 0.27$\times$0.10    & 28.5   & 0.074   &   18 46 57.626  & 45 42 20.04 & 0.30$\pm$0.08  & 0.34 & &600 & 0.29$\times$0.56 & 137.5 & 0.066  & 0.42$\pm$0.09 &   \\
 \cmidrule(rl){15-15}
    & Tot  &                     &          &        &               &              &                &      &     & &                     &        &          &  1.0$\pm$0.1 &            \\
\hline
NGC~6703 & core &  0.18$\times$0.04 & 128.8  & 0.066    &   18 47 18.823 & 45 33 02.34 & 0.26$\pm$0.08  & 0.42 & &750 & 0.30$\times$$<$0.24 & 72.7 & 0.063  & 0.32$\pm$0.08 &  core+jet (A) \\
\cmidrule(rl){15-15} 
      & Tot  &                     &          &        &               &              &                &      & &    &                      &        &          &  0.35$\pm$0.09 &            \\
\hline
NGC~6946 &      &  0.04$\times$$<$0.16 &  66.4 & 0.057 & 20 34 52.172  &  60 09 14.20 & 0.37$\pm$0.07  & 0.37 & &700 & 0.77$\times$0.39 & 114.1 &  0.067  &  0.48$\pm$0.07 &\multirow{3}{1.8cm}{\centering jet+complex (E) 2.3$\arcsec\to$42pc}\\
       &      &  0.38$\times$0.16   & 133.8 & 0.057 & 20 34 52.242  &  60 09 14.20 & 0.83$\pm$0.08  & 1.61 & &700 & 0.55$\times$0.28 & 123.1 &  0.067  & 1.59$\pm$0.08 &   \\
       &      &  0.24$\times$0.10 &   0.33  & 0.057 & 20 34 52.255  &  60 09 14.42 & 0.61$\pm$0.07  & 1.31 & &700 &                  &       &  0.067  &  $<$2.57      &  \\
       &      &  0.21$\times$0.03 & 56.0 & 0.057 & 20 34 52.273  &  60 09 13.20 & 0.48$\pm$0.07  & 0.72 & &700 & 0.34$\times$$<$0.50 & 124.5 &  0.067  &  0.63$\pm$0.08 &  \\
       & core &  0.09$\times$0.04 & 12.5   & 0.057 & 20 34 52.299  &  60 09 14.20 & 0.49$\pm$0.07  & 0.57 & &700 &                    &       &  0.067  &    $<$0.35    &  \\
       &       &  0.17$\times$0.10 & 12.7 & 0.057 & 20 34 52.320  &  60 09 14.10 & 0.67$\pm$0.07  & 1.11 & &700 & 0.60$\times$0.32 & 116.0 &  0.067  & 1.28$\pm$0.08 &    \\
\cmidrule(rl){15-15}
      & Tot  &                     &          &        &               &              &                &    &  &     &                      &        &          &  10.0$\pm$0.3 &           \\ 
\hline
NGC~6951 &      &  0.23$\times$0.11 & 97.6  & 0.055 & 20 37 14.106  &  66 06 20.31 & 0.40$\pm$0.08  & 0.66 & &750 & 0.05$\times$$<$0.21 &  142.6 &  0.065  &  0.63$\pm$0.09  & \multirow{3}{1.8cm}{\centering twin jet (C) 1.6$\arcsec\to$175pc}\\
       & core &  0.06$\times$$<$0.23 & 83.3  & 0.055 & 20 37 14.118  &  66 06 20.10 & 0.25$\pm$0.08  & 0.24 & &750 & 0.16$\times$$<$0.26 & 136.2 &  0.065  &  0.56$\pm$0.09  &   \\
       &      &  0.14$\times$0.06 &172.9  & 0.055 & 20 37 14.144  &  66 06 19.76 & 0.51$\pm$0.08  & 0.73 & &750 & 0.09$\times$$<$0.26 & 130.6 &  0.065  &  0.73$\pm$0.09 &    \\
 \cmidrule(rl){15-15}
      & Tot  &                     &          &        &               &              &                &   &   &     &                      &        &          &  1.0$\pm$0.1 &            \\ 
\hline
NGC~7217 &      &  0.24$\times$0.11 &   138.2 &  0.075 &  22 07 52.350&  31 21 34.44 & 0.37$\pm$0.09  & 0.72 & &700 &  0.19$\times$$<$0.17 &  99.0  & 0.063      & 0.67$\pm$0.08  & \multirow{3}{1.8cm}{\centering  twin lobes (C) 1.5$\arcsec\to$120pc}\\
          & core &  0.14$\times$$<$0.11 &  27.5  &  0.075 & 22 07 52.378 & 31 21 34.03 &  0.27$\pm$0.09  & 0.28 & &700 &                  &        & 0.063      & $<$1.00        &  \\
         &      &  0.19$\times$0.12 &  4.2  &  0.075 & 22 07 52.400 & 31 21 33.64 &  0.55$\pm$0.09  & 0.79 & &700 &  0.21$\times$$<$0.21 &  8.9  & 0.063      & 0.79$\pm$0.08  &   \\
 \cmidrule(rl){15-15}
     & Tot  &                     &          &        &               &              &                &      & &    &                      &        &          &  1.6$\pm$0.2 &            \\ 
\hline
NGC~7798  &          &    0.14$\times$0.12  &  22.1   &   0.031   &  23 59 25.410  &  20 44 59.38  &  0.33$\pm$0.04  &  0.53   &  & 750  &  0.29$\times$0.21 &  2.0      &  0.058    &      0.51$\pm$0.07  &   \multirow{3}{1.8cm}{\centering  triple source (C) 1.4$\arcsec\to$240pc} \\
              &  core &    0.22$\times$0.06   &  162.2   &  0.031 &    23 59 25.444 & 20 44 59.43  & 0.23$\pm$0.04  &  0.43    &  & 750   &                               &               &   0.058   &   $<$0.34          &  \\
             &           &   0.35$\times$0.20   &   28.1    &  0.031  & 23 59  25.494   &  20  44 59.57 & 0.23$\pm$0.04  &  0.82  &  & 750  &  0.33$\times$0.27 &   173.7  &  0.058   &  0.36$\pm$0.07  &    \\
\cmidrule(rl){15-15}
            & Tot  &                     &          &        &               &              &                &      &     &                      &        &          &  1.8$\pm$0.1 &            \\
\hline
%NGC 777 & 1 &      & 0.23$\times$0.15 &  36.4   & 0.077  &  02 00 14.91 & 31 25 45.84  &  0.74$\pm$0.10 & 0.84 & 600 &  0.23$\times$0.15 &  36.4   & 0.077  &  0.9 & \rdelim\}{3}{20pt}[twin jets] \\
%        & 1 & core & 0.23$\times$0.15 &  36.4   & 0.077  &  02 00 14.91 & 31 25 45.84  &  0.74$\pm$0.10 & 0.84 & 600 &  0.23$\times$0.15 &  36.4   & 0.077  &  0.9 & \\
%        & 1 &      & 0.23$\times$0.15 &  36.4   & 0.077  &  02 00 14.91 & 31 25 45.84  &  0.74$\pm$0.10 & 0.84 & 600 &  0.23$\times$0.15 &  36.4   & 0.077  &  0.9 &\\
\hline
\end{longtable}
\end{center}
\vspace*{-1cm} \fontsize{6}{7}\selectfont Column description: (1)
galaxy name; (2) radio component: core, jet, lobe, blob or
unidentified component if not labeled (N or S stand for North or
South); (3) deconvolved FWHM dimensions (major $\times$ minor axes,
$\theta_{\rm M}\times\theta_{\rm m}$) of the fitted component, determined from
an ellitpical Gaussian fit from the full-resolution radio map; (4) PA
of the deconvolved component, PA$_{\rm d}$ from the full-resolution radio
map (degree); (5) rms of the radio map close to the specific component from the
full-resolution radio map (mJB, mJy beam$^{-1}$); (6)-(7) radio position in epoch J2000; (8)
peak flux density in mJy beam$^{-1}$, F$_{\rm peak}$ from the full-resolution
radio map: this represents the radio core flux density; (9) integrated flux
density, $F_{tot}$ in mJy, from the full-resolution radio map;
(10) $uv$-taper scale of the low-resolution radio map in k$\lambda$;
(11) deconvolved FWHM dimensions (major $\times$ minor axes,
$\theta_{\rm M}\times\theta_{\rm m}$) of the fitted component, determined from
an ellitpical Gaussian fit from the low-resolution radio map; (12) PA
of the deconvolved component, PA$_{\rm d}$, from the low-resolution radio
map (degree); (13) rms of the radio map close to the specific component from
the low-resolution radio map (mJy beam$^{-1}$); (14) peak flux density in mJy beam$^{-1}$,
F$_{\rm peak}$ from the low-resolution radio map. For NGC~5194 we give the
total integrated flux densities of the radio lobes instead of the peak
flux densities.  At the bottom of each target the total flux density of the
radio source associated with the galaxy is given in mJy, measured from
the low-resolution map; (15) radio morphology (A, B, C, D, E) and size
in arcsec and pc (see Section~\ref{core-ident}).  \twocolumn
\end{landscape}
\normalfont\normalsize 
\onecolumn
\fontsize{7}{10}\selectfont
\begin{landscape}
\begin{center}
\begin{longtable}{C{1.2cm}|C{0.4cm}ccccccc|C{0.05cm}ccccc|C{1.9cm}}

\caption[Properties of the sample.]{Properties of the unidentified sources.} 
\label{tabsfr} \\

%This is the header for the first page of the table...
\hline \hline

%This is the header for the first page of the table...
\hline \hline

     &   \multicolumn{8}{c}{Full resolution}                                                   &  \multicolumn{7}{|c}{Low resolution} \\
name & comp & $\theta_{\rm M}\times\theta_{\rm m}$  & PA$_{\rm d}$ & rms  &  $\alpha$(J2000) & $\delta$(J2000) &  F$_{\rm peak}$ & F$_{\rm tot}$ & &  $u{-}v$    &  $\theta_{\rm M}\times\theta_{\rm m}$ &  PA$_{\rm d}$ & rms &  F$_{\rm peak}$ &morph/size \\
     &        &arcsec & deg  & mJB &                 &                &  mJB       &  mJy      & &  k$\lambda$ & arcsec & deg   &  mJB &   mJB      & \\

\hline	
\endfirsthead

%This is the header for the remaining page(s) of the table...
\multicolumn{3}{c}{{\tablename} \thetable{} -- Continued} \\[0.5ex]
\hline \hline 

   &   \multicolumn{8}{c}{Full resolution}                                                   &  \multicolumn{7}{|c}{Low resolution} \\
name & comp & $\theta_{M}\times\theta_{m}$  & PA$_{d}$ & rms  &  $\alpha$(J2000) & $\delta$(J2000) &  F$_{\rm peak}$ & F$_{\rm tot}$ &  & $u{-}v$    &  $\theta_{M}\times\theta_{m}$ &  PA$_{d}$ & rms &  F$_{\rm peak}$  & morph/size \\
     &        &arcsec & deg  & mJB &                 &                &  mJB       &  mJy      &  & k$\lambda$ & arcsec & deg   &  mJB &   mJB       & \\

\hline
\endhead

%This is the footer for all pages except the last page of the table...
\hline
  \multicolumn{16}{c}{{Continued on Next Page}} \\
\endfoot

%This is the footer for the last page of the table...
  \\[-1.8ex] 
\endlastfoot

%Now the data...&  &      &  0.24$\times$0.17 & 37.2  &        & 
NGC~891  &      &  0.23$\times$0.10 &  10.6  & 0.063  &   02 22 33.230 & 42 20 58.56 &  0.27$\pm$0.07 & 0.40  & &400 &  0.22$\times$0.45 & 74.6 & 0.060  & 0.40$\pm$0.08   &\multirow{4}{1.8cm}{\centering multi-components} \\
       &      &                   &       & 0.063  &   02 22 33.441 & 42 20 54.70 &  $<$0.25       &       & &400 &  0.50$\times$0.30 & 12.9 & 0.060  & 0.30$\pm$0.08   &\\
       &      &                   &       & 0.063  &   02 22 33.451 & 42 20 54.27 &  $<$0.20       &       & &400 &  1.0$\times$$<$0.50 & 120.0 & 0.060  & 0.30$\pm$0.08   &\\
       &      &  0.24$\times$0.17 & 65.6  & 0.063  &   02 22 33.705 & 42 20 48.80 &  0.25$\pm$0.08 & 0.50  & &400 &  0.76$\times$0.41&  139.854 & 0.060  & 0.34$\pm$0.08  & \\
 \cmidrule(rl){15-15}
      & Tot  &                     &          &        &               &              &                &      &  &   &                      &        &        &  1.5$\pm$0.2           &  \\
 \hline
NGC~2342  &     &  0.47$\times$$<$0.13 &  11.2 &  0.065  &  07 09 18.018 & 20 38 06.25 &  0.22$\pm$0.08& 0.36 & &750  & 0.64$\times$<$$0.22 & 163.2 & 0.068  & 0.46$\pm$0.09 &  \multirow{3}{1.8cm}{\centering twin-jets?  2$\arcsec\to$692pc, 4$\arcsec$ offset}\\  % su ogg sotto
 (field)   &     &                      &       &  0.065  &  07 09 18.027 & 20 38 05.29 &   $<$0.40     &      & &750  & 0.44$\times$0.16 &  32.1 & 0.068  & 0.42$\pm$0.09 &   \\  %giu ogg sotto
         & core?    &  0.13$\times$0.11    &    136.8 &  0.065  &  07 09 18.027 & 20 38 05.83 &  0.39$\pm$0.08& 0.48 & &750  & 0.34$\times$0.11 & 163.5 & 0.068  & 0.55$\pm$0.09 &  \\ %core ogg sotto
\cmidrule(rl){15-15}
      & Tot  &                     &          &        &               &              &                &      &     &                      &        &         &  &  1.7$\pm$0.2      &            \\
\hline
NGC~3198  &     &  $>$0.19$\times$$>$0.08 &  166.6  & 0.114      &  10 19 54.519   &  45 32 57.07 &  0.35$\pm$0.12  &   0.40 &  &750 &  0.37$\times$$>$0.25 & 105.7  & 0.81     &   0.61$\pm$0.09 & \multirow{4}{1.8cm}{\centering elongated jet? 4.5$\arcsec\to$202pc, 6$\arcsec$ offset}  \\
(field)   &      &                          &         & 0.114      &   10 19 54.520  &  45 32 53.35 &  $<$0.34        &        &  &750 &  $>$0.29$\times$$>$0.32 & 171.3 &  0.81  &    0.62$\pm$0.09 &     \\
\cmidrule(rl){15-15}
          & Tot  &                     &          &        &               &              &                &      &     &                      &        &         &  &  2.1$\pm$0.2      &            \\
%\vspace{0.1mm}
\hline 
NGC~4013 &    &          &    &    0.070   &   11 58 31.321  &  43 56 49.86  &    $<$ 0.24    &    & & 500  &   0.20$\times$0.02  & 58.4   &  0.064   & 0.23$\pm$0.06  &    \multirow{6}{1.8cm}{\centering star-foring ring? 1.5$\arcsec\to$90pc}\\
             &     &       &   &  0.070        &   11 58 31.402 & 43 56 49.57  &  $<$0.24      &     &  &500  &  0.18$\times$0.07  &  64.0   &   0.064   & 0.25$\pm$0.06  &  \\
              &     &      &    &  0.070       &   11 58 31.422 &  43 56 51.85 &   $<$0.24   &      &  &500  &  0.20$\times$$<$0.15 &  75.9  &  0.064  &0.28$\pm$0.06 &  \\
             &     &       &   &   0.070     &    11 58 31.433  &  43 56 50.05 &  $<$0.24   &     &   & 500  & 0.18$\times$$<$0.24  & 25.2  &   0.064  &  0.27$\pm$0.06 &  \\
             &     &       &   &   0.070     &     11 58 31.455 &  43 56 50.46  & $<$0.24  &     &   &500    &   0.47$\times$0.15  &  29.0    &   0.064   &  0.27$\pm$0.06 &   \\
           &     &        &  &    0.070     &       11  58  31.481 &  43 56 50.63  & $<$0.21 &    &   &500    &  0.08$\times$$<$0.22   &  118.7  & 0.064 &  0.24$\pm$0.06 &  \\
\cmidrule(rl){15-15}
     & Tot  &                     &          &        &               &              &                &   &   &     &                      &        &        &  3.8$\pm$0.3 &            \\ 
\hline
NGC~5273 &   &     &    & 0.057 &  13 42 08.312  &  35 39 13.45   &  $<$0.19  &                      &  &500  &  0.35$\times$$<$ 0.28 &   92.5  &  0.049 & 0.25$\pm$0.05 &    \multirow{6}{1.8cm}{\centering star-forming ring? 3$\arcsec\to$230pc}\\
       &   &      &   & 0.057 &  13 42 08.333 &  35 39 15.40  &   $<$0.22   &                       &  &  500  &  0.92$\times$$<$0.32   &  167.5  &  0.049 &  0.31$\pm$0.05 &   \\
       &   &      &     & 0.057 &  13 42 08.350 &  35 39 16.76  & $<$0.23    &                       &  &  500  &  0.87$\times$0.22     &  159.3  &   0.049 &  0.28$\pm$0.05 &   \\
        &   &  0.26$\times$$<$0.20    &   150.50 & 0.057 & 13 42 08.354 &  35 39 15.92  &  0.24$\pm$0.06   &    0.33             &  &   500  & 0.92$\times$0.12     &   155.1 &  0.049 &  0.30$\pm$0.05 &   \\
        &   &      &   & 0.057 &  13 42 08.415&  35 39 16.35  &   $<$0.25  &                       &   &500    & 1.30$\times$0.22     &  140.3  &  0.049 &  0.32$\pm$0.05 &   \\
\cmidrule(rl){15-15}
    & Tot  &                     &          &        &               &              &                &   &   &     &                      &        &        &  2.6$\pm$0.2 &        \\
 \hline
NGC~5676  &    &       &       & 0.085  &  14 32 47.221 & 49 27 51.79 & $<$0.28      &       &  &200 & 0.38$\times$$<$1.0 & 14.5 & 0.091 & 0.38$\pm$0.09 &   \multirow{2}{1.8cm}{\centering  multi-components, 27.7$\arcsec$ offset}\\
         &    &        &       & 0.085  &  14 32 47.565 & 49 27 51.83 & $<$0.26      &       &  &200 & 0.46$\times$0.2    & 30.7 & 0.091 & 0.31$\pm$0.09 &   \\
\cmidrule(rl){15-15}
  & Tot  &                     &          &        &               &              &                &      &     &             &         &        &        &  0.7$\pm$0.2 &            \\
\hline
NGC~5907 &      &  0.19$\times$$<$0.13  & 22.2  & 0.097 &   15 15 49.124 & 56 20 26.42 & 0.44$\pm$0.10 & 0.72 & &750 &   0.16$\times$$<$0.19 & 31.9 & 0.090   &  0.41$\pm$0.09  &\multirow{3}{1.8cm}{\centering  multi-components, 55.7$\arcsec$ offset}\\
       &      &  1.90$\times$0.33  &   29.5 & 0.097 &  15 15 49.300 & 56 20 25.01 &    0.30$\pm$0.10 & 0.53 & &750 &  1.90$\times$0.15 & 69.9 & 0.090   &  0.33$\pm$0.09  &\\
       &      &  1.63$\times$0.22  &   157.6 & 0.097 &  15 15 49.372 & 56 20 24.33 &  0.34$\pm$0.10  & 0.52 & &750 &  0.63$\times$$<$0.35 & 145.8 & 0.090  &  0.42$\pm$0.09 &  \\
\cmidrule(rl){15-15}
      & Tot  &                     &          &        &               &              &                &      &     &        &              &        &        &  2.1$\pm$0.2 &            \\
\hline
NGC~6015 &      &      &    &  0.082 &   15 51 23.775  &    62 18 44.72 & $<$0.32     &           &  & 500  &   0.07$\times$0.18  &   57.6  &  0.079  & 0.38$\pm$0.10  &   \multirow{3}{1.8cm}{\centering  multi-components, 11.6$\arcsec$ offset}\\
       &      &      &    &  0.082 &    15 51 23.911 &  62 18 44.45   &  $<$0.35    &           &   &500  &   1.03$\times$0.36  &   4.5   &  0.079  & 0.50$\pm$0.09  &  \\
        &      &      &    &  0.082 &  15 51 23.974 & 62 18 43.70  &    $<$0.34    &           &  & 500  &   $<$0.33$\times$$<$0.42 &  78.7 &  0.079  &  0.39$\pm$0.09  & \\
        &      &      &    &  0.082 &  15 51 24.037 &  62 18 42.17    &  $<$0.39   &           &  & 500  &    1.14$\times$$<$0.25 &  57.5  & 0.079  & 0.27$\pm$0.09  &   \\
       &      &      &    &  0.082 &  15 51 24.039  &   62 18 41.945  &  $<$0.38   &           &  &  500  & 0.38$\times$0.04    &   21.1    &   0.079  &  0.32$\pm$0.09  &  \\
\cmidrule(rl){15-15}
           & Tot  &                     &          &        &               &              &                &      &     &                      &        &      &   &  2.6$\pm$0.2 &           \\
\hline 
\hline
\end{longtable}
\end{center}
Column description: (1) galaxy name; (2) radio component: core, jet,
or lobe or unidentified component if not labeled (N or S stand for
North or South); (3) deconvolved FWHM dimensions (major $\times$ minor
axes, $\theta_{\rm M}\times\theta_{\rm m}$) of the fitted component,
determined from an ellitpical Gaussian fit from the full-resolution
radio map; (4) PA of the deconvolved component, PA$_{\rm d}$ from the
full-resolution radio map (degree); (5) rms of the radio map close to
the specific component from the full-resolution radio map (mJB, mJy
beam$^{-1}$); (6)-(7) radio position in epoch J2000; (8) peak flux
density in mJy beam$^{-1}$, F$_{\rm peak}$ from the full-resolution
radio map; (9) integrated flux density, $F_{rm tot}$ in mJy, from the
full-resolution radio map; (10) $uv$-taper scale of the low-resolution
radio map in k$\lambda$; (11) deconvolved FWHM dimensions (major
$\times$ minor axes, $\theta_{\rm M}\times\theta_{\rm m}$) of the
fitted component, determined from an ellitpical Gaussian fit from the
low-resolution radio map; (12) PA of the deconvolved component,
PA$_{\rm d}$ from the low-resolution radio map (degree); (13) rms of
the radio map close to the specific component from the low-resolution
radio map (mJy beam$^{-1}$); (14) peak flux density in mJy
beam$^{-1}$, F$_{\rm peak}$ from the low-resolution radio map. At the
bottom of each target the total flux density of the radio source
associated with the galaxy is given in mJy, measured from the
low-resolution map; (15) radio morphology, size in arcsec and pc and
offset from the optical centre, if present.  \twocolumn
\end{landscape}
\begin{table*}
\begin{center}
\caption[]{Radio contour levels} 
\begin{tabular}{lccp{3cm}|C{0.05cm}cccl}
\hline \hline 
name     &  \multicolumn{3}{c|}{FR}                      &  \multicolumn{5}{c}{LR}  \\
          &   Beam         & PA         & levels       & &  k$\lambda$  & beam & PA &    levels    \\
\hline 
NGC~147   & 0.24$\times$0.14 & $-$63.2&    0.18$\times$($-$1,1) &  & 750 & 0.537$\times$0.25 & $-$41.6& 0.2$\times$($-$1,1,1.5,2,2.3) \\
NGC~266   &  0.48$\times$0.31 &  $-$40.0 & 0.21$\times$($-$1,1,2,4) & &750 &  0.57$\times$0.37 & $-$46.0& 0.25$\times$($-$1,1,2,4) \\
NGC~278   & 0.29$\times$0.14 &  $-$55.2 & 0.11$\times$($-$1,1,2) & &750 & 0.54$\times$0.25 & $-$42.9 & 0.13$\times$($-$1,1,1.5)  \\
NGC~315   &  0.12$\times$0.12 &  45.0 &  25.0$\times$($-$1,1,5,15) & &750 &  0.55$\times$0.36 & $-$45.7 &  10.0$\times$($-$1,1,5,15,30) \\
NGC~410   & 0.52$\times$0.32 & $-$46.85 &  0.28$\times$($-$1,1,2,4,7) &  &750 &0.62$\times$0.37 &  $-$49.9 & 0.28$\times$($-$1,1,2,4,7) \\ 
NGC~507   & 0.46$\times$0.35 &  $-$67.8 &  0.34$\times$($-$1,1,1.5,2,2.5) & &750 &0.60$\times$0.48 &  $-$64.6 &  0.34$\times$($-$1,1,1.5,2,2.5) \\
NGC~777   &  0.23$\times$0.15 &  7.6  & 0.165$\times$($-$1,1,2,3,4)    & & 600    &  0.29$\times$0.26 &  58.7 & 0.21$\times$($-$1,1,1.42,2,3,3.9)               \\
NGC~891   &  0.20$\times$0.17 & 65.6  & 0.15$\times$($-$1,1,2)         & &400      & 0.59$\times$0.53 & $-$22.9  & 0.12$\times$($-$1,1,1.5,2,2.4,3)          \\              
NGC~972   &  0.24$\times$0.17 & 37.2  & 0.13$\times$($-$1,1,1.5,2,2.7) &  &750      &  0.31$\times$0.29 & 66.8  & 0.13$\times$($-$1,1,1.5,2,3,4.2)            \\
NGC~2273  &  0.20$\times$0.16 &  48.1 & 0.20$\times$($-$1,1,2,4,8,13) &  &750  &    0.33$\times$0.27 & 85.1 & 0.20$\times$($-$1,1,2,4,8,16,22)  \\
NGC~2342  &  0.18$\times$0.18 &    0.0 &  0.15$\times$($-$1,1,1.5,2,2.5) &  & 750 &   0.34$\times$0.34 &  0.0 & 0.1$\times$($-$1,1,1.5,2,2.5,3,3.5,3.9,4.15)  \\
NGC~2342$^{*}$  &  0.18$\times$0.18 &    0.0 &  0.15$\times$($-$1,1,2) &  & 750 &   0.34$\times$0.34 &  0.0 & 0.17$\times$($-$1,1,1.9,2.4,3)  \\
UGC~3828  &  0.19$\times$0.16 & 50.3 &   0.15$\times$($-$1,1,1.5,1.85) &  &750 & 0.33$\times$0.26&  89.3 &  0.15$\times$($-$1,1,2,3)   \\
NGC~2300  & 0.20$\times$0.20 & 0.0 & 0.11$\times$($-$1,1,2,3,5,8) & &750 &  0.40$\times$0.40 & 0.0  & 0.13$\times$($-$1,1,2,3,5,8)\\
UGC~4028  & 0.20$\times$0.20 & 0.0  & 0.17$\times$($-$1,1,1.5) & &750  & 0.40$\times$0.40 & 0.0  &  0.13$\times$($-$1,1,1.5,2.5,3.2) \\
NGC~2639  &  0.20$\times$0.19 & 68.0 &  0.40$\times$($-$1,1,2,4,8, 16,29,40) &  & 750 &  0.34$\times$0.26 & $-$75.3& 1.5$\times$($-$1,1,2,4, 8,11.2,15)   \\
NGC~2634  &  0.20$\times$0.20 & 0.0 & 0.18$\times$($-$1,1,2)  &  &750 & 0.40$\times$0.40 & 0.0  & 0.16$\times$($-$1,1,2,2.5) \\
NGC~2681  &  0.19$\times$0.16 &  51.8 &  0.17$\times$($-$1,1,2,3,5)  &  &750  &  0.34$\times$0.26 & $-$76.4 &  0.20$\times$($-$1,1,1.7,2.5,3.5,5)     \\
NGC~2655  & 0.20$\times$0.20 & 0.0  &  0.20$\times$($-$1,1,2,4,8, 16,32,64,128)& &750 & 0.40$\times$0.40 & 0.0  & 0.40$\times$($-$1,1,2,4,8, 16,32,64,128)\\ 
NGC~2841  &  0.19$\times$0.16 & 52.7 &  0.16$\times$($-$1,1,1.5,2)   & &600 & 0.50$\times$0.32 &  $-$53.6&  0.17$\times$($-$1,1,1.3,1.77,2.5,3)  \\
NGC~3198 &  0.20$\times$0.20 &   0.00 &    0.21$\times$($-$1,1,1.5,2) & &750 &  0.46$\times$0.43 & $-$25.4 &    0.15$\times$($-$1,1,1.5,2,2.7)   \\
NGC~3198$^{*}$  &  0.20$\times$0.20 &     0.00 &   0.3$\times$($-$1,1,1.5)  & &750 &  0.46$\times$0.43 & $-$25.4 &   0.15$\times$($-$1,1,1.5,3,4)     \\         
NGC~3414 &  0.22$\times$0.14 &  20.60  &  0.11$\times$($-$1,1,2,4,8)  &  &750 & 0.47$\times$0.34 &  $-$37.6  &  0.21$\times$($-$1,1,2,4,8)     \\
NGC~3430 &   0.20$\times$0.20 &    0.0 &  0.17$\times$($-$1,1,1.5,2)  &  &750  &   0.48$\times$0.37  &  $-$30.9    &    0.16$\times$($-$1,1,1.5,2,2.5)   \\
NGC~3432  &   0.47$\times$0.28 &  $-$16.7 & 0.21$\times$($-$1,1,2,3) &  &750  &  0.59$\times$0.55 &  35.1 & 0.21$\times$($-$1,1,2) \\
\multirow{2}{*}{NGC~3665} &\multirow{2}{*}{0.20$\times$0.20} &   \multirow{2}{*}{0.00} &    \multirow{2}{3cm}{\centering 0.23$\times$($-$1,1,1.5, 2,5,12)} &  \rdelim\{{2}{20pt} & 750  & 0.45$\times$0.36 &  $-$34.6 & 0.90$\times$($-$0.80,1,1.5,2,4) \\
         &                                  &                        &                                                    &  & 500  &  0.57$\times$0.50 &  $-$18.0 & 1.0$\times$($-$0.80,1,1.5,2,4) \\
NGC~3675  & 0.17$\times$0.14  & $-$63.3  &   0.16$\times$($-$1,1,1.5)  &  & 750  &   0.24$\times$0.19& $-$54.0&0.14$\times$($-$1,1,1.5,2)  \\
NGC~3938  &  0.19$\times$0.14& $-$54.2 & 0.15$\times$($-$1,1,1.5)  &  &750& 0.24$\times$0.19&$-$47.0& 0.15$\times$($-$1,1,1.5,2)  \\
NGC~4013  & 0.19$\times$0.14& $-$56.4 &  0.14$\times$($-$1,1) & & 500 &  0.24$\times$ 0.20   &  $-$51.0 & 0.09$\times$($-$1,1,2,3) \\
NGC~4051  &   0.18$\times$0.14 &  $-$58.6  & 0.19$\times$($-$1,1,1.5,2, 2.5,3,3.5,4,5)   &  & 750  &   0.24$\times$0.19 &  $-$53.13 & 3.0$\times$($-$1,1,1.5,2,2.5, 3,3.5,4,5)  \\
\multirow{2}{*}{NGC~5005}  &   \multirow{2}{*}{0.21$\times$0.19}  &  \multirow{2}{*}{27.0}  &  \multirow{2}{3cm}{\centering 0.19$\times$($-$1,1,2,3,3.7, 4.5,7,8.5,10.75,12)} &  \rdelim\{{2}{20pt} & 750   &  0.34$\times$0.30 &  25.8 & 0.23$\times$($-$1,1,2,7,16) \\
         &                                  &                        &                                                                       &   & 200 &  1.57$\times$1.28 & 11.9 & 0.70$\times$($-$1,1,2,3,3.7,10,20) \\
\multirow{2}{*}{NGC~5194}  &   \multirow{2}{*}{0.18$\times$0.17} &  \multirow{2}{*}{82.7}  & \multirow{2}{*}{0.21$\times$($-$1,1,2,3)} &  \rdelim\{{2}{20pt} & 750 &  0.30$\times$0.27  & $-$53.6 & 0.18$\times$($-$1,1,2,3,6)  \\
         &                                  &                        &                                               &  &500 &  0.47$\times$0.42 &  14.7 & 0.18$\times$($-$1,1,2,2.5,3.3,5,8) \\
NGC~5195  &    0.20$\times$0.18 &  82.8 & 0.17$\times$($-$1,1,1.5,2,2.8) &  & 750  &  0.30$\times$0.29 & $-$59.9 & 0.21$\times$($-$1,1,1.5,2,2.8)  \\
NGC~5273  &  0.25$\times$0.17 &    32.4 & 0.11$\times$($-$1,1,1.5)      &  &500  &   0.46$\times$0.39 &  76.2  & 0.11$\times$($-$1,1,1.5,2,2.4)     \\ 
NGC~5353  &  0.18$\times$0.14 & 29.6  & 0.90$\times$($-$1,1,2,4,8,13)   &  &500  &  0.24$\times$0.22 & 66.3 & 0.70$\times$($-$1,1,2,4,8,16)       \\ 
NGC~5377  &    0.20$\times$0.18 & $-$86.5 &  0.18$\times$($-$1,1,1.5) &  &750 &  0.51$\times$0.39 & 4.3 & 0.19$\times$($-$1,1,1.85,2.5)    \\
NGC~5448  &     0.19$\times$0.18 &  $-$86.6 &  0.19$\times$($-$1,1,1.3, 1.6,1.85)  & &750  &  0.30$\times$0.28 & $-$51.8 &  0.19$\times$($-$1,1,1.5,1.88,2.3) \\
NGC~5676  &    0.35$\times$0.19  & 23.2 &   0.19$\times$($-$1,1,1.5)  & &200  & 1.06$\times$0.89 & 13.2 & 0.15$\times$($-$1,1,1.5,2)                   \\
NGC~5866  &    0.18$\times$0.18  & 0.0   & 0.40$\times$($-$1,1,2,4,8,16)  & &200 & 1.00$\times$0.96 & $-$50.8  & 0.25$\times$($-$1,1,2,3,6,15,30)          \\
NGC~5879  &   0.18$\times$0.18 & 0.0  &  0.15$\times$($-$1,1,1.5,2,)  & &750 &  0.41$\times$0.30 &  32.0 & 0.15$\times$($-$1,1,1.5,2,2.5)       \\
NGC~5907  &   0.28$\times$0.19  &   18.6 & 0.20$\times$($-$1,1,1.5)       & &750 &   0.42$\times$0.30 & 29.8 &  0.19$\times$($-$1,1,1.5,2)        \\
\multirow{2}{*}{NGC~5985}  &    \multirow{2}{*}{0.26$\times$0.19} &  \multirow{2}{*}{22.1} & \multirow{2}{*}{0.13$\times$($-$1,1,1.5)} &   \rdelim\{{2}{20pt} & 380   & 0.58$\times$0.55 & 78.3  &  0.11$\times$($-$1,1,1.5,2)  \\
          &                                  &                        &                                               &  & 200 &  1.13$\times$0.96 & $-$81.03 & 0.11$\times$($-$1,1,1.5,2,2.5) \\
NGC~6015  &   0.23$\times$0.18  &   25.4 &  0.15$\times$($-$1,1,1.5,2)   &  & 500  &   0.48$\times$0.42 & 67.5 & 0.15$\times$($-$1,1,1.5,2,3)     \\ 
NGC~6702  &  0.19$\times$0.16 & 56.7 & 0.14($-$1,1,2)     &  &600  &  0.57$\times$0.30 & $-$51.2 & 0.12$\times$($-$1,1,1.7,2.1,2.6,3.2)      \\
NGC~6703  &  0.19$\times$0.15 & 53.6  &   0.12$\times$($-$1,1,1.5,2) &  &750  & 0.52$\times$0.25 & $-$51.51 &  0.10$\times$($-$1,1,1.5,2.5)                     \\
NGC~6946  &   0.19$\times$0.15 & $-$84.1 &  0.13$\times$($-$1,1,2,3,4,5.5)&  &700 &  0.52$\times$0.27 & $-$42.73 & 0.13$\times$($-$1,1,2,3,4,8,11)                 \\
NGC~6951  &   0.20$\times$0.15 & 88.5  &  0.13$\times$(1,1,1.5,2.5,3.5) &  &750 &  0.48$\times$0.26 &  $-$39.3 &  0.13$\times$($-$1,1,1.5,2.5,3.5,4.2,5)            \\
NGC~7217  &    0.18$\times$0.18 &  0.0  & 1.5$\times$($-$1,1,1.5,2, 3,4,5) & &700  &  0.40$\times$0.4 &  0.0  & 1.5$\times$($-$1,1,2,3,4,5)                 \\
NGC~7798  & 0.20$\times$0.14 & 35.9  & 0.07$\times$($-$1,1,2,3,4)  & & 750& 0.50$\times$0.23 &  $-$41.0  &  0.1$\times$($-$1,1,2,3,4)                 \\                   
\hline
\hline
\end{tabular}
\label{contours}
\begin{flushleft}
  Column description: (1) source name; (2) FWHM of the elliptical
  Gaussian restoring beam in arcsec of the full-resolution maps;
  (3) PA of the restoring beam (degree) of the full-resolution maps;
  (4) radio contour levels (mJy beam$^{-1}$) of the full-resolution maps
  (Fig.~\ref{maps1} and \ref{maps2}); (5) uvtaper scale parameter in
  k$\lambda$ of the low-resolution radio maps; (6) FWHM of the
  elliptical Gaussian restoring beam in arcsec of the
  low-resolution maps; (7) PA of the restoring beam (degree) of the
  low-resolution maps; (8) radio contour levels (mJy beam$^{-1}$) of the
  low-resolution maps (Fig.~\ref{maps1} and
  \ref{maps2}). $*$ identifies the secondary radio source
  detected in the field, probably a background AGN.
\end{flushleft}
\end{center}
\end{table*}



%%%%%%%%%%%%%%%%%%%%%%%%%%%%%%%%%%%%%%%%%%%%%%%%%%

% Don't change these lines
\bsp	% typesetting comment
\label{lastpage}
\end{document}